\documentclass[12pt]{article}
\usepackage[T1]{fontenc}
\usepackage[utf8]{inputenc}
\usepackage{lmodern}
\usepackage{amsfonts}
\usepackage{amssymb}
\usepackage{amsmath}
\usepackage{amsthm}
\usepackage{mathtools}
\usepackage{bm}
\usepackage{xcolor}
\usepackage{natbib}
\usepackage{bibunits}
\usepackage{rotating}
\usepackage{booktabs}
\usepackage{tabularx,array,threeparttable,multirow}
\usepackage{graphicx}
\usepackage{caption}
\usepackage{subcaption}
\usepackage{float}
\usepackage{placeins}
\usepackage{enumitem}
\usepackage{microtype}
\usepackage{hyperref}
\usepackage[nameinlink,noabbrev]{cleveref}
\usepackage{tikz}
\usetikzlibrary{arrows.meta,positioning,calc,shapes.geometric}

\makeatletter
\let\bu@natlinkstart\hyper@natlinkstart
\let\bu@natlinkbreak\hyper@natlinkbreak
\let\bu@natanchorstart\hyper@natanchorstart
\def\hyper@natlinkstart#1{\bu@natlinkstart{\@bibunitname.#1}}
\def\hyper@natlinkbreak#1#2{\bu@natlinkbreak{#1}{\@bibunitname.#2}}
\def\hyper@natanchorstart#1{\bu@natanchorstart{\@bibunitname.#1}}
\makeatother

\IfFileExists{econometrica.bst}{\defaultbibliographystyle{econometrica}}{\defaultbibliographystyle{ecta}}
\defaultbibliography{references}

\hypersetup{
  colorlinks=true,
  linkcolor=blue!50!black,
  citecolor=blue!50!black,
  urlcolor=blue!50!black,
  bookmarksdepth=3
}

\newcommand{\draftdate}{\today}
\renewcommand{\baselinestretch}{1.3}
\numberwithin{equation}{section}
\setlist[itemize]{topsep=0.15em,itemsep=0.15em,leftmargin=1.5em}
\setlist[enumerate]{topsep=0.15em,itemsep=0.15em,leftmargin=1.5em}

\newcommand{\bc}{\begin{center}}
\newcommand{\ec}{\end{center}}
\newcommand{\be}{\begin{equation}}
\newcommand{\ee}{\end{equation}}
\newcommand{\bea}{\begin{eqnarray}}
\newcommand{\eea}{\end{eqnarray}}
\newcommand{\bean}{\begin{eqnarray*}}
\newcommand{\eean}{\end{eqnarray*}}

\newtheorem{theorem}{Theorem}[section]
\newtheorem{proposition}[theorem]{Proposition}

\theoremstyle{definition}
\newtheorem{assumption}[theorem]{Assumption}

\newtheorem{remark}[theorem]{Remark}

\crefname{assumption}{Assumption}{Assumptions}
\Crefname{assumption}{Assumption}{Assumptions}
\crefname{proposition}{Proposition}{Propositions}
\Crefname{proposition}{Proposition}{Propositions}
\crefname{theorem}{Theorem}{Theorems}
\Crefname{theorem}{Theorem}{Theorems}
\crefname{lemma}{Lemma}{Lemmas}
\Crefname{lemma}{Lemma}{Lemmas}
\crefname{corollary}{Corollary}{Corollaries}
\Crefname{corollary}{Corollary}{Corollaries}
\crefname{example}{Example}{Examples}
\Crefname{example}{Example}{Examples}
\crefname{remark}{Remark}{Remarks}
\Crefname{remark}{Remark}{Remarks}

\newcounter{pkt}

\newcommand{\E}{\mathbb E}
\newcommand{\Pbb}{\mathbb P}

\newcommand{\TV}{\operatorname{TV}}
\newcommand{\R}{\mathbb R}
\newcommand{\A}{\mathcal A}

\newcommand{\Pclass}{\Pi}

\newcommand{\Ical}{\mathcal I}

\newcommand{\ind}{\mathbf 1}

\newcommand{\argmax}{\operatornamewithlimits{argmax}}

\newcommand{\eps}{\varepsilon}

\newcommand{\logistic}{\operatorname{logit}^{-1}}

\newcounter{saveeqn}

\begin{document}
\begin{bibunit}
\hypersetup{pageanchor=false}

\title{\vspace*{-0.5 in} \textbf{Policy Learning with Network-Shaping Assignments}}

\author{Zhenxiao Chen \\
{\em \small University of Pennsylvania}
\and Zhiheng You\thanks{\setlength{\baselineskip}{4mm} Correspondence: Z. You and Z. Chen: Department of Economics, University of Pennsylvania, Philadelphia, PA 19104-6297. Email: zhyou@sas.upenn.edu (You) and zxchen@sas.upenn.edu (Chen).} \\
{\em \small University of Pennsylvania}}

\date{\draftdate}
\maketitle

\begin{abstract}
Many policies determine social arrangements whose effects depend on the relationships formed afterwards. 
We study how to learn such policies when a constrained joint action changes network formation and outcomes depend on exposures to the resulting network. 
Policy value integrates an outcome response over the distribution of post-assignment networks. 
The large number of feasible arrangements makes it difficult to evaluate policies directly from historical data. We therefore adopt a model-based approach that pools information across assignments to evaluate alternative arrangements.
We characterize identification of policy values and construct plug-in estimators. Our regret bounds account for sampling variation, first-stage estimation error, and computational errors. 
Our applications to classroom seating and business groups illustrate predicted welfare gains from learned rules relative to random assignment.
\end{abstract}

\noindent {\footnotesize {\em JEL codes:} C14, C18, C21, C44, D85, I21.}

\noindent {\footnotesize {\em Key words:} policy learning, potential outcomes, network interference, network formation, matching.}

\thispagestyle{empty}

\clearpage
\hypersetup{pageanchor=true}
\setcounter{page}{1}

\section{Introduction}

Schools, firms, and other institutions shape opportunities for interaction through the assignments they make. Teachers choose seating charts, organizations assign mentors, and business programs form meeting groups. These decisions allocate access to peers subject to restrictions on seating, eligibility, and group size. How they affect outcomes depends on the relationships and interactions that follow the assignment. We study how to use data from historical assignments to learn rules that can improve welfare.

Consider a teacher choosing a seating chart to improve average academic achievement. Seating two students together can change the probability that they become friends; the resulting friendship, in turn, may contribute to both students' achievement through peer effects. Under an additive benchmark, the expected gain in total achievement from a seating chart is the sum of the gains from its selected deskmate pairs. For each pair, this gain equals the seating-induced change in its friendship probability multiplied by the combined achievement contribution of that friendship. Students who would become friends even when seated apart may have little to gain from sitting together, while friendships with large potential benefits may be difficult to induce. Because each student can have only one deskmate, the teacher must evaluate these gains jointly.

These assignments are a practical policy instrument in education and business development. Studies of student peer groups examine academic outcomes under alternative group compositions \citep{CarrellSacerdoteWest2013,Garlick2018}, while classroom experiments study the effects of deskmate assignments on achievement and friendship formation \citep{WuZhangWang2023,RohrerKellerElwert2021,KellerElwert2023}. In business development, \citet{CaiSzeidl2018} organize owner-managers into recurring meeting groups to study learning, business relationships, and firm performance. By changing who receives access to which peers, reassignment can affect both aggregate outcomes and the distribution of benefits across these settings.

Learning an assignment rule requires comparing many feasible arrangements using limited data. For a classroom with an even number $M$ of students, there are $(M-1)!!$ complete pairings before imposing additional restrictions. Historical data typically cover only a small fraction of these assignments, and many feasible arrangements may receive negligible or zero probability under the historical design, making the overlap conditions commonly imposed in policy learning difficult to maintain. We therefore use models of network formation and outcomes to pool information across assignments. Evidence on how seating affects friendships for pairs with particular characteristics can inform the evaluation of many charts when combined with estimates of how those friendships contribute to achievement. Likewise, common responses to peer composition allow information from observed groups to inform the evaluation of other feasible groupings.

Our framework treats assignment as a joint action within a cluster. Using baseline characteristics and institutional constraints, a policy selects a feasible action, which induces a distribution of post-assignment networks. Individual mean outcomes depend on baseline covariates and specified network exposures; these exposures can also record assignment characteristics that directly affect outcomes. We evaluate each action by integrating the outcome response over the network distribution it induces and averaging across participants. Policy value then averages these expected outcomes across baseline environments.

We first study unrestricted optimal assignments at the population level and their approximation by score-based rules. When welfare is additive across selected pairs or groups, the seating and mentor assignment problems reduce to maximum-weight matching and set partitioning, respectively, allowing us to use established matching and integer-programming algorithms from operations research \citep{Edmonds1965,Galil1986,BalasPadberg1976}. With nonadditive welfare, exact optimization may be impractical, in which case heuristic search can provide an approximate solution. A score-based rule uses a common coefficient vector to score feasible pairs or groups from baseline characteristics and selects an arrangement subject to the institutional constraints. The gap between the unrestricted optimum and the best score-based rule measures policy-class approximation error.

We then characterize the policy values consistent with the observed distribution under the maintained formation and outcome models. Even when some model components remain unidentified, a policy value can be point identified. For a given additive outcome response, marginal link probabilities determine expected welfare even when the dependence among links is unknown. When outcomes include interactions between links, the corresponding joint probabilities also enter welfare. The welfare criterion therefore determines which aspects of network formation need to be learned. This distinction also clarifies the assignment variation relevant for policy evaluation under the maintained models. We also bound the welfare loss from choosing policies using misspecified formation or outcome models.

We construct plug-in estimates of action values by integrating the fitted outcome response over the estimated network distribution, using simulation when the integral cannot be evaluated analytically. We then choose score coefficients to maximize average estimated welfare across observed environments. We derive regret bounds relative to the best policy in the specified class, controlling value-estimation error uniformly over candidate policies. The bounds account for sampling variation, errors in estimating the formation and outcome models, numerical integration, and optimization error. For score-based policies, sampling error depends on the number of independent clusters, the score dimension, and the number of feasible assignments. We further bound first-stage errors by estimation rates multiplied by sensitivity factors determined by policy-induced exposures. In the running examples, matching constraints can limit the accumulation of link-probability errors, while larger mentor groups can amplify errors involving peer relationships. We also extend the first-stage analysis to nonregular formation and outcome responses.

We apply the framework to randomized classroom seating in Hungary \citep{RohrerKellerElwert2021,KellerElwert2023} and business meeting groups \citep{CaiSzeidl2018}. The seating application combines estimated friendship probabilities with objectives measuring mutual friendship count and lower-achieving students' access to higher-achieving friends. The business application uses exposures determined by group assignment to evaluate mean log revenue, with either equal welfare weights or weights placed on small firms. Under common assignment constraints, we compare \emph{Random assignment}, a \emph{Score-based policy}, and an \emph{Unrestricted policy}. In the fitted seating results, the learned friendship-count rule increases friendships but reduces academic access relative to random assignment, whereas the learned access rule increases both. The business objectives also select different groupings and distribute predicted revenue gains differently between small and large firms.

This paper builds on statistical treatment choice and policy learning \citep{Manski2004,HiranoPorter2009,KitagawaTetenov2018,AtheyWager2021}. Recent work extends policy learning to settings with spillovers.
\citet{Viviano2025} develops semiparametric welfare estimators and treatment-targeting rules under network interference, while \citet{ZhangImai2023} exploit additive outcome structure to evaluate and learn individualized treatment rules under clustered interference. \citet{HanOwusuShin2022} study the
choice of treatment fractions under anonymous social
interaction. \citet{KitagawaWang2023} develop
individualized vaccine allocation under epidemic spillovers, and \citet{KitagawaWang2026} optimize treatment allocation using the stationary distribution of outcomes in sequential network games. Our policy instrument is instead a feasible pairing or grouping that changes opportunities for interaction and the relationships that subsequently form. We evaluate these assignments by integrating an outcome
response over the network distribution induced by each
assignment, and study how estimation of both formation and outcomes affects the welfare of learned assignment rules.

Two contemporaneous papers are especially close. \citet{HazardKitagawa2025} learn a welfare-optimal two-sided matching from estimated pair costs. \citet{XuZhou2025} use successive batches of pair outputs to learn latent-type complementarities and choose later matches. Their one-period welfare criteria correspond to pair-additive special cases of our framework. Because our general formulation also permits individual outcomes to depend on exposures involving multiple relationships, welfare need not reduce to a sum of selected-pair outputs. We study identification of the integrated policy value and distinguish the contributions of formation and outcome estimation errors to regret. Their entropy-regularized transport and sequential exploration methods address complementary computational and learning problems.

Our framework draws on causal inference under interference, which develops exposure-based estimands and methods for evaluating effects when outcomes depend on others' assignments \citep{HudgensHalloran2008,AronowSamii2017,BasseFeller2018,Leung2022,AuerbachGuoTabordMeehan2026}. Models of network formation characterize how observed characteristics, unobserved heterogeneity, and strategic interactions shape relationships \citep{Graham2017,dePaula2020,HoffRafteryHandcock2002,Mele2017,Snijders2001}, while the peer-effects literature studies how social interactions affect outcomes and the identification of these effects \citep{Manski1993,BramoulleDjebbariFortin2009,Sacerdote2011,BallesterCalvoArmengolZenou2006,BramoulleDjebbariFortin2020}. Our framework combines these elements to evaluate assignments that change the distribution of outcome-relevant relationships. The post-assignment network serves as an intermediate outcome, connecting our representation to causal mediation analysis \citep{ImaiKeeleYamamoto2010}. Rather than decomposing effects into direct and indirect components, we focus on identifying policy values and learning assignment rules.

The remainder of the paper is organized as follows. \Cref{sec:framework} introduces the setup, and \Cref{sec:examples} develops the running examples. \Cref{sec:population-choice} studies optimal assignments and score-based policy classes. \Cref{sec:identification} establishes the identification results. \Cref{sec:learning} develops value estimation and regret bounds. \Cref{sec:applications} presents the empirical applications, and \Cref{sec:conclusion} concludes.

\section{Framework}
\label{sec:framework}

This section presents the general framework. In Section~\ref{subsec:setup}, we define the assignment problem, specify the network-formation and outcome models, and introduce the policy-value criterion. We then describe the observed data and sampling assumption in Section~\ref{subsec:data-sampling}.

\subsection{Setup}
\label{subsec:setup}

Clusters are indexed by $c=1,\ldots,C$, with $M_c$ individuals indexed by $i$. The baseline environment $H_c$ contains the information available before the planner acts. It includes individual covariates and prior relationships, together with the institutional conditions that determine feasible assignments. Given $H_c$, the feasible action set $\A(H_c)$ imposes the relevant constraints. Write $a$ for a candidate action and $D_c\in\A(H_c)$ for the action observed in cluster $c$.

A policy is a measurable rule $\pi$ satisfying $\pi(H_c)\in\A(H_c)$ for every baseline environment in its domain. The policy class $\Pclass$ is a prespecified collection of such rules. Our main specification is the parametric score-based class $\Pclass_\Theta=\{\pi_\theta:\theta\in\Theta\}$ with $\Theta\subseteq\R^d$, but the general framework also allows classes defined by an assignment algorithm. The set of environment--action pairs that these policies may select is
\[
\mathcal E_{\Pclass}=\{(h,a):a=\pi(h)\text{ for some }\pi\in\Pclass\}.
\]

For each feasible action $a\in\A(H_c)$, let $G_c(a)$ be the network that would form after implementing $a$. The variable $G_c(a)$ also records any deterministic characteristics of the assignment that affect outcomes. Thus it may contain both realized relationships and recorded assignment characteristics. The observed network is $G_c=G_c(D_c)$. The network-formation law is the conditional distribution of the post-action network under a candidate action. For a measurable set $B$ of post-action networks,
\[
K_0(B\mid h,a)=\Pbb[G_c(a)\in B\mid H_c=h].
\]
A parametric formation model represents this law by $K_{\eta_0}(dg\mid h,a)$, where $\eta_0$ indexes the true parameter.

For each action $a$, let $Y_{ic}(a)$ denote individual $i$'s potential outcome, with observed outcome $Y_{ic}=Y_{ic}(D_c)$.
We define the cluster-average potential outcome as
\[
\bar Y_c(a)=\frac{1}{M_c}\sum_{i=1}^{M_c}Y_{ic}(a).
\]
More generally, nonnegative individual welfare weights $w_i(H_c)$ can be used to define
\[
\bar Y_{c,w}(a)=
\frac{\sum_iw_i(H_c)Y_{ic}(a)}{\sum_iw_i(H_c)},
\qquad \sum_iw_i(H_c)>0.
\]
To evaluate these welfare criteria under alternative assignments, we specify how the conditional mean of individual outcomes depends on baseline covariates and the post-assignment network.
Write $M_h$ for the number of individuals in baseline environment $h$. For each $i\le M_h$, let $W_i(h)$ collect individual baseline covariates and let $T_i(h,g)$ summarize the characteristics of the
realized network $g$ and assignment characteristics relevant to individual $i$'s outcome.
The following assumption specifies the outcome response.

\begin{assumption}[Exposure-based outcome response]
\label{ass:outcome-response}
For every policy-relevant pair $(h,a)\in\mathcal E_{\Pclass}$, every individual $i\le M_h$, and $K_0(\cdot\mid h,a)$-almost every network realization $g$,
\[
\E[Y_{ic}(a)\mid H_c=h,G_c(a)=g]
=
m_{\gamma_0}\{W_i(h),T_i(h,g)\}.
\]
\end{assumption}

For each network realization $g$, write
\[
\mu_{0,i}(g,h)=m_{\gamma_0}\{W_i(h),T_i(h,g)\},
\qquad
\mu_0(h,g)=\frac1{M_h}\sum_{i=1}^{M_h}\mu_{0,i}(g,h).
\]
Combining the formation law and the outcome response, the conditional
value of action $a$ in environment $h$, under parameters
$(\eta,\gamma)$, is
\[
v_{\eta,\gamma}(h,a)
=
\int
\left[
\frac1{M_h}\sum_{i=1}^{M_h}
m_\gamma\{W_i(h),T_i(h,g)\}
\right]
K_\eta(dg\mid h,a).
\]
Under Assumption~\ref{ass:outcome-response}, iterated expectations give
\[
v_0(h,a)
:=
\E[\bar Y_c(a)\mid H_c=h]
=
\int \mu_0(h,g)K_0(dg\mid h,a)
=
v_{\eta_0,\gamma_0}(h,a).
\]
The value of a policy $\pi$ is
\[
V_0(\pi)=\E\left[v_0\{H_c,\pi(H_c)\}\right],
\]
where the expectation is taken over the distribution of baseline
environments. The optimal policy in the class $\Pclass$ is thus
\[
\pi_\Pclass^*\in\argmax_{\pi\in\Pclass}V_0(\pi).
\]

\begin{remark}[Assignment complexity and information pooling]
\label{rem:dimension-support}
For an even cluster of size $M$, there are $(M-1)!!$ perfect matchings, while an undirected binary post-action network has $2^{\binom{M}{2}}$ possible realizations. Under uniform assignment over perfect matchings at a fixed baseline environment, a given matching appears on average $C/(M-1)!!$ times in $C$ independent assignments. The precision of separate action-specific estimates can therefore deteriorate rapidly with cluster size. Common formation covariates and exposure responses allow information to be shared across assignments. Section~\ref{sec:identification} characterizes when the maintained models and historical design determine their values.
\end{remark}

\begin{remark}[Deterministic post-action network]
\label{rem:deterministic-network}
If there is a measurable map $g^\dagger$ such that $K_0(dg\mid h,a)=\delta_{g^\dagger(h,a)}(dg)$ for every policy-relevant $(h,a)$, then
\[
v_0(h,a)=\mu_0\{h,g^\dagger(h,a)\},
\]
so no integration over random networks is required. If, in addition, $M_h=1$, $\A(h)=\{0,1\}$, and $g^\dagger(h,a)$ is equivalent to the treatment label $a$, the framework reduces to the standard binary-treatment policy-learning problem.
\end{remark}

\begin{remark}[Direct match costs as a special case]
\label{rem:hazard-kitagawa-objective}
\citet{HazardKitagawa2025} study two-sided one-to-one matching: a planner pairs units with observed characteristics $x_i$ and $z_j$ to minimize a direct pair cost $c(x_i,z_j)$. In our notation, $h=\{(x_i)_{i=1}^n,(z_j)_{j=1}^n\}$ collects the covariates on both sides of the market, $\sigma(i)$ is unit $i$'s partner, $\rho$ is a randomized policy over permutations, and $g_\sigma$ is the realized matching, with
\[
K_0(dg\mid h,\rho)
=
\sum_{\sigma\in\mathfrak S_n}\rho(\sigma)\delta_{g_\sigma}(dg).
\]
Set $M_h=n$, $W_i(h)=x_i$, $T_i(h,g_\sigma)=z_{\sigma(i)}$, and $m_{\gamma_0}\{x_i,z_j\}=-c(x_i,z_j)$. Then
\[
v_0(h,\rho)
=
-\frac1n\sum_{\sigma\in\mathfrak S_n}\rho(\sigma)
\sum_{i=1}^n c\{x_i,z_{\sigma(i)}\},
\]
their negative expected average match cost. Our welfare functional contains their pair-additive criterion. Their entropy-regularized estimator and specialized regret analysis address a distinct learning procedure.
\end{remark}

\begin{remark}[Pairwise network output as a special case]
\label{rem:xu-zhou-objective}
\citet{XuZhou2025} study batched online network formation, repeatedly pairing agents and using observed pair outputs to learn latent types and complementarities for later batches. For one batch, let $h$ denote the current agent pool and pre-batch information, $a$ the selected edge set, and $Z_{ij}(a)$ pair $\{i,j\}$'s output, set to zero if unselected. Then $G_c(a)=\{Z_{ij}(a):i<j\}$ is the weighted output network, whose distribution follows their weighted stochastic block model. With $M_h$ agents, allocate each edge payoff equally to its endpoints:
\[
T_i(h,g)=\frac12\sum_{j\ne i}g_{ij},
\qquad
m_{\gamma_0}\{W_i(h),t\}=t.
\]
Thus
\[
v_0(h,a)
=
\frac1{M_h}\E\!\left[\left.\sum_{i<j}Z_{ij}(a)\,\right|H_c=h\right],
\]
their expected total network output up to normalization, with the same one-period optimal assignment. Their sequential learning procedure also accounts for the value of information in later batches.
\end{remark}

\subsection{Sampling}
\label{subsec:data-sampling}

Collect the observed variables as
\[
O_c=(H_c,D_c,G_c,Y_{1c},\ldots,Y_{M_c,c}),\qquad c=1,\ldots,C,
\]
where $G_c$ is the realized post-action network and $Y_{ic}$ are observed outcomes. The cluster-average observed outcome is $\bar Y_c=M_c^{-1}\sum_iY_{ic}$.

\begin{assumption}[Independent clusters and bounded outcomes]
\label{ass:iid-clusters}
The clusters $O_1,\ldots,O_C$ are independent draws from a common population. Within-cluster dependence is unrestricted unless stated otherwise. There is a finite constant $\bar B$ such that $|\bar Y_c(a)|\le \bar B$ for every feasible action $a$ almost surely.
\end{assumption}

The primary sample size in the general theory is the number of independent clusters $C$. Shared shocks and overlapping interactions can make the observations within a cluster dependent. The cluster average therefore need not have the precision of an average of $M_c$ independent observations. Under additional application-specific conditions, larger within-cluster samples can improve estimation of the formation and outcome models.

\begin{remark}[Normalization]
\label{rem:normalization}
The use of $\bar Y_c$ makes policy values comparable across clusters of different size. If the planner instead values total outcomes in a fixed population, the same theory applies after replacing $\bar Y_c$ by the desired cluster-level payoff and adjusting the boundedness condition. For growing cluster sizes, a normalization is usually needed for stable regret statements.
\end{remark}

\section{Running Examples}
\label{sec:examples}

This section develops deskmate seating and mentor assignments as running examples of the framework. In each example, we specify the feasible assignments, the relationships that follow them, and the outcome response. These specifications provide the inputs for the optimal assignment problems in Section~\ref{sec:population-choice}. Section~\ref{subsec:formation-response-specs} introduces the network-formation and outcome specifications. Sections~\ref{subsec:deskmates} and~\ref{subsec:mentors} develop the classroom seating and mentor assignment examples. Section~\ref{sec:design-examples} discusses experimental designs for these settings. Appendix~\ref{app:network-introductions} considers introductions between agents, where welfare can depend on interactions among an agent's realized contacts.

\subsection{Formation and Response}
\label{subsec:formation-response-specs}

The outcome response determines which features of the formation law are needed to evaluate an assignment. When the response is linear in individual link indicators, their marginal probabilities suffice. For a binary network characteristic indexed by $\ell$, a convenient marginal specification is
\[
\Pbb_\eta[G_{\ell c}(a)=1\mid H_c=h,a]
=\logistic\{\eta'Z_{\ell c}(h,a)\},
\qquad
Z_{\ell c}(h,a)=\big(Z_{\ell c}^{0}(h),Z_{\ell c}^{act}(h,a)\big),
\]
where $\logistic(u)=\{1+\exp(-u)\}^{-1}$. The index $\ell$ identifies the network characteristic being modeled, and $\eta$ contains the formation coefficients. The baseline covariates $Z_{\ell c}^{0}(h)$ capture its pre-assignment determinants. The covariates $Z_{\ell c}^{act}(h,a)$ describe how the action changes contact opportunities. This specification is useful when assignment affects the probability that a relationship forms; see \citet{Graham2017} and \citet{dePaula2020}.

More general outcome responses can depend on which links form together, making their joint distribution relevant for welfare. Strategic formation models provide one way to specify this joint law by generating the network as an equilibrium outcome \citep{JacksonWolinsky1996,BalaGoyal2000,BlochJackson2006,dePaulaRichardsShubikTamer2018,Mele2017,Sheng2020}. For an undirected network, consider the payoff
\[
u_i(g;h,a,\varepsilon)
=
\sum_{j\neq i}g_{ij}\{\beta'W_{ij}(h,a)+\varepsilon_{ij}\}
+\rho\sum_{\substack{j<k\\ j,k\neq i}}g_{ij}g_{ik}g_{jk}
-\kappa d_i(g)^2 .
\]
Here $g_{ij}$ indicates a link between $i$ and $j$, whose observed determinants are collected in $W_{ij}(h,a)$ and whose unobserved shock is $\varepsilon_{ij}$. Degree is $d_i(g)=\sum_{j\neq i}g_{ij}$. The parameter $\rho$ governs the payoff from triangles, and $\kappa$ governs the cost of maintaining many links. Pairwise stability requires each existing link to be acceptable to both agents and each absent link to be rejected by at least one. If $\mathcal S(h,a,\varepsilon)$ is the set of stable networks and $\pi_\eta(g\mid h,a,\varepsilon)$ is an equilibrium-selection rule, then
\[
K_\eta(g\mid h,a)=
\int \pi_\eta(g\mid h,a,\varepsilon)
\ind[g\in\mathcal S(h,a,\varepsilon)]\,dF_{\eta,\varepsilon}(\varepsilon).
\]
Here $F_{\eta,\varepsilon}$ is the distribution of formation shocks. This model allows a change in one link to affect the incentives to form other links. The dyadic specifications below may therefore describe either independent Bernoulli links or the marginal probabilities of a dependent network model.

The exposure map connects network formation to outcomes by recording the network and assignment characteristics used by the response. For a generic post-action network, an illustrative exposure map and the corresponding average response are
\[
\begin{aligned}
T_i(h,g)&=\big(d_i(g),\overline X_{N_i(g)}(h),T_{\triangle,i}(g),\mathrm{Cent}_i(g)\big),\\
\overline m_\gamma(h,g)&=
\frac1{M_h}\sum_{i=1}^{M_h}m_\gamma\{W_i(h),T_i(h,g)\},
\end{aligned}
\]
where $N_i(g)$ is the set of individual $i$'s neighbors, $d_i(g)$ is network degree, and $\overline X_{N_i(g)}(h)$ summarizes neighbors' baseline covariates. The other two coordinates summarize higher-order network characteristics, such as the number of triangles involving individual \(i\) and measures of $i$'s centrality in the network. The exposure map is chosen so that it and the baseline covariates determine the action-specific conditional mean. Any direct effect of the administrative assignment must also be represented in $T_i$.

\subsection{Deskmate Seating}
\label{subsec:deskmates}

In the first example, a teacher chooses deskmate pairs to improve average academic achievement. The seating chart changes opportunities for friendship, and the resulting friendships enter students' outcome responses.

Let $I_h$ be the students in a classroom and let $E_h^{desk}\subseteq \binom{I_h}{2}$ be the pair set allowed by school restrictions. A deskmate action is a matching,
\begin{equation*}
\A^{desk}(h)=\left\{a\subseteq E_h^{desk}:\sum_{j:\{i,j\}\in E_h^{desk}}x_{ij}(a)\le1\ \text{for all }i\in I_h\right\},
\qquad
x_{ij}(a)=\ind[\{i,j\}\in a].
\end{equation*}
The notation $\binom{I_h}{2}$ denotes unordered student pairs. The indicator $x_{ij}(a)$ records whether students $i$ and $j$ are selected as deskmates, and the matching constraint gives each student at most one partner. The set $E_h^{desk}$ excludes pairings that violate school restrictions. A school may require language support for some students or prohibit pairings because of conflicts. Studies of peer assignment motivate modeling the seating chart as an action that affects friendship formation \citep{CarrellSacerdoteWest2013,Garlick2018,WuZhangWang2023,KellerElwert2023}.

The post-action network is $G^{desk}=(a,F^+)$, where $F_{ij}^+\in\{0,1\}$ records whether students $i$ and $j$ are friends after the seating action. For pairs that are not seated together, set $x_{ij}(a)=0$. A dyadic friendship formation model is
\begin{equation}
\label{eq:deskmate-link-probability}
\begin{aligned}
&\Pbb_\eta(F_{ij}^+(a)=1\mid H=h,a)=p_{ij}^{F,x_{ij}(a)}(h),\\
&p_{ij}^{F,x}(h)=\logistic\{\eta_0^F+\eta_1^F x+\eta_2^F F_{ij}^0
+\eta_3^F|\mathrm{Ach}_i-\mathrm{Ach}_j| +\eta_4^F\mathrm{Cl}_{ij}^0+\eta_5^F xF_{ij}^0\}.
\end{aligned}
\end{equation}
Here $F_{ij}^0$ indicates baseline friendship and $\mathrm{Ach}_i$ is baseline achievement. The variable $\mathrm{Cl}_{ij}^0$ summarizes common friends before assignment and may be scaled by classroom size. The coefficient $\eta_1^F$ gives the seating effect for pairs who were initially unconnected, while $\eta_5^F$ allows this effect to differ for existing friends. Write $p_{ij}^{F,x}(h)$ for the friendship probability when the seating indicator is $x\in\{0,1\}$. The change in this probability from seating the pair together is
\begin{equation}
\label{eq:deskmate-probability-increment}
\delta_{ij}^F(h)=p_{ij}^{F,1}(h)-p_{ij}^{F,0}(h).
\end{equation}
Conditional independence gives the formation law
\begin{equation*}
K_\eta^{desk}(dF^+\mid h,a)=
\prod_{\{i,j\}\subset I_h}
\big[p_{ij}^{F,x_{ij}(a)}(h)\big]^{F_{ij}^+}
\big[1-p_{ij}^{F,x_{ij}(a)}(h)\big]^{1-F_{ij}^+}.
\end{equation*}
Independent Bernoulli links provide a convenient baseline. A joint formation model can instead allow dependence across friendships, including dependence induced by shared classroom shocks or strategic interaction.

We next specify how post-assignment friendships enter the outcome response for academic achievement. Let $W_i(h)$ contain baseline student and classroom covariates, and let $T_i^{desk}(h,G^{desk})$ record the realized friendship links incident to $i$ together with the baseline characteristics of the corresponding pairs. A linear conditional mean for achievement is
\begin{equation*}
m_\gamma^{desk}\{W_i(h),T_i^{desk}(h,G^{desk})\}
=
\alpha_{\gamma,i}^{desk}(h)
+\sum_{j\in I_h\setminus\{i\}}b_{\gamma,ij}^F(h)F_{ij}^+.
\end{equation*}
Here $\alpha_{\gamma,i}^{desk}(h)$ is student $i$'s baseline response and $b_{\gamma,ij}^F(h)$ measures the contribution of friendship with student $j$. The contribution may differ between the two students. Define
\[
\alpha_\gamma^{desk}(h)
=\frac1{|I_h|}\sum_{i\in I_h}\alpha_{\gamma,i}^{desk}(h),
\qquad
\omega_{\gamma,ij}^F(h)
=b_{\gamma,ij}^F(h)+b_{\gamma,ji}^F(h).
\]
Averaging over students gives
\begin{equation}
\label{eq:deskmate-outcome-response}
\frac1{|I_h|}\sum_{i\in I_h}m_\gamma^{desk}\{W_i(h),T_i^{desk}(h,G^{desk})\}
=
\alpha_\gamma^{desk}(h)+\frac1{|I_h|}\sum_{\{i,j\}\subset I_h}\omega_{\gamma,ij}^F(h)F_{ij}^+ .
\end{equation}
The pair coefficient $\omega_{\gamma,ij}^F(h)$ measures the combined contribution of a friendship to the two students' conditional mean outcomes and may vary with their baseline characteristics. At the true response, write the coefficient as $\omega_{ij}^F(h)$. Taking expectations in \eqref{eq:deskmate-outcome-response} replaces each friendship indicator by its formation probability. Combined with \eqref{eq:deskmate-link-probability}, this gives the pair-additive action value in Section~\ref{subsec:oracle-reductions}.

\subsection{Mentor Assignments}
\label{subsec:mentors}

In the second example, the planner assigns juniors to mentors to improve retention or productivity. The assignment determines direct mentoring relationships and creates opportunities for peer relationships among juniors assigned to the same mentor. Mentor load and group composition affect how these relationships contribute to outcomes.

Let $J_h$ be junior workers and $\mathcal M_h$ mentors. Mentor $j$ has capacity $b_j(h)$, and eligibility restrictions determine which junior--mentor matches are allowed. A feasible assignment is a binary matrix $a=(a_{ij})$ satisfying
\begin{equation*}
a_{ij}\in\{0,1\},
\qquad
\sum_{j\in\mathcal M_h}a_{ij}=1,
\qquad
\sum_{i\in J_h}a_{ij}\le b_j(h),
\qquad
a_{ij}=0\ \text{for ineligible pairs}.
\end{equation*}
Here $a_{ij}=1$ indicates that junior $i$ is assigned to mentor $j$. Each junior receives one mentor, and an outside option can be represented by a dummy mentor when nonassignment is allowed. The capacity constraint limits the number of juniors assigned to mentor $j$. Write $n_j(a)=\sum_{i\in J_h}a_{ij}$ for the mentor's load and $I_j(a)=\{i\in J_h:a_{ij}=1\}$ for the assigned group.

The post-action network is $G^{ment}=(a,\mathrm{Co})$. Assignment determines direct junior--mentor relationships through $a_{ij}$, while $\mathrm{Co}_{ii'}\in\{0,1\}$ records whether juniors $i$ and $i'$ form a peer relationship while sharing a mentor. Conditional on $h$ and $a$, these links have probabilities
\begin{equation}
\label{eq:mentor-comentee-probability}
\begin{aligned}
\Pbb_\eta(\mathrm{Co}_{ii'}(a)=1\mid h,a)&=\sum_{j\in\mathcal M_h}a_{ij}a_{i'j}\ p_{ii'j}^{\mathrm{Co}}(h,n_j(a)),\\
p_{ii'j}^{\mathrm{Co}}(h,n)&=\logistic\{\eta_{\mathrm{Co}}^{\prime}  Z_{ii'j}^{\mathrm{Co}}(h,n)\}.
\end{aligned}
\end{equation}
The product $a_{ij}a_{i'j}$ permits a peer link only when both juniors share mentor $j$. The covariates $Z_{ii'j}^{\mathrm{Co}}(h,n)$ describe mentor load and characteristics of the two juniors. For the linear outcome response below, these marginal peer-link probabilities suffice to compute expected welfare, even if peer relationships are dependent.

We next specify how mentoring and peer relationships enter the outcome response for retention or productivity. Mentor $j$ has baseline quality $q_j^0(h)>0$, with effective quality at load $n$ given by
\begin{equation}
\label{eq:mentor-load-quality}
q_j(n;h)=q_j^0(h)\psi_j(n),
\qquad
\psi_j(n)=\{1+\chi_j(n-1)\}^{-1},
\qquad \chi_j\ge0 .
\end{equation}
Thus adding juniors to a mentor can dilute the mentoring quality received by those already assigned to that mentor. Let $\lambda_{ij}(h)$ measure the compatibility between junior $i$ and mentor $j$, and let $S_{ii'}(h)=S_{i'i}(h)$ measure the spillover from a peer relationship between juniors $i$ and $i'$, which may be positive or negative. For junior $i$ assigned to mentor $j$, a linear conditional mean for retention or productivity is
\begin{equation*}
\begin{aligned}
m_\gamma^{ment}\{W_i(h),T_i^{ment}(h,G^{ment})\}
&=\alpha_{\gamma,i}^{ment}(h)
+\beta_q\ q_j(n_j(a);h)+\beta_\lambda\ \lambda_{ij}(h)\\
&+\beta_{q\lambda}\ q_j(n_j(a);h)\ \lambda_{ij}(h)+\frac{\beta_{\mathrm{Co}}}{2}
\sum_{i'\in I_j(a)\setminus\{i\}}
\mathrm{Co}_{ii'}\ S_{ii'}(h),
\end{aligned}
\end{equation*}
where $\alpha_{\gamma,i}^{ment}(h)$ is junior $i$'s baseline response. Each peer relationship contributes $\beta_{\mathrm{Co}}S_{ii'}(h)/2$ to each junior's conditional mean. Writing $\alpha_\gamma^{ment}(h)=|J_h|^{-1}\sum_{i\in J_h}\alpha_{\gamma,i}^{ment}(h)$, averaging over juniors gives
\begin{equation}
\label{eq:mentor-outcome-response}
\begin{aligned}
&\frac1{|J_h|}\sum_{i\in J_h}m_\gamma^{ment}\{W_i(h),T_i^{ment}(h,G^{ment})\} \\
&\quad=
\alpha_\gamma^{ment}(h)+\frac1{|J_h|}\Bigg[
\sum_{j\in\mathcal M_h}\sum_{i\in I_j(a)}
\Big\{\beta_q\ q_j(n_j(a);h)+\beta_\lambda\ \lambda_{ij}(h)
+\beta_{q\lambda}\ q_j(n_j(a);h)\ \lambda_{ij}(h)\Big\} \\
&\qquad\qquad\qquad
+
\sum_{j\in\mathcal M_h}\sum_{\{i,i'\}\subset I_j(a)}
\beta_{\mathrm{Co}}\ \mathrm{Co}_{ii'}\ S_{ii'}(h)
\Bigg].
\end{aligned}
\end{equation}
The first sum gives the contribution from mentoring, including the interaction between mentor quality and junior--mentor fit. The second sum gives spillovers from peer relationships within each assigned group. Taking expectations replaces the realized peer-link indicators with their formation probabilities. Since mentoring quality and peer spillovers depend on group membership, the resulting value is additive across mentor groups rather than across individual junior--mentor matches, as used in Section~\ref{subsec:oracle-reductions}.

\subsection{Experimental Design Examples}
\label{sec:design-examples}

The running examples also illustrate how experiments could provide information on the formation and outcome specifications. The designs below introduce randomized variation in seating or mentor assignments, subject to institutional constraints, and measure the resulting relationships and outcomes.

\paragraph{Classroom seating experiment.}

A classroom seating experiment can randomize pairings within the school's restrictions, using baseline attributes such as achievement differences or prior friendship to define pair types. Variation in seating status across these types informs the heterogeneity in friendship responses. For example, \citet{WuZhangWang2023} randomize deskmate exposure between previously high- and low-achieving students, while \citet{KellerElwert2023} randomize full seating charts. Measurements of post-assignment friendships and subsequent achievement connect the assigned pairs to the relationships through which seating affects outcomes.

\paragraph{Mentor assignment experiment.}

A mentor assignment experiment could randomize assignments of juniors to mentors, subject to eligibility and capacity constraints. The randomization could vary mentor loads and, holding load fixed, the composition of juniors assigned to each mentor. Variation in load provides information about how mentoring quality changes with the number of assigned juniors, while variation in group composition informs junior--mentor compatibility and peer spillovers. Observing the assignments, subsequent peer relationships, and outcomes such as worker retention or productivity provides data for estimating the formation and outcome response models under the maintained specifications.

\section{Population-Optimal Policies and Score-Based Classes}
\label{sec:population-choice}

This section studies policy choice at the population level, taking the conditional value function $v_0$ as given. We distinguish unrestricted optimal assignments from the best policies within a restricted class. If the planner observes a baseline environment $h$ and can choose any feasible action in $\A(h)$, the unrestricted optimal action is
\[
a_0^\star(h)\in\argmax_{a\in\A(h)}v_0(h,a),
\]
and the associated unrestricted policy is $\pi_{\mathrm{unr}}^\star(h)=a_0^\star(h)$. For a restricted policy class $\Pclass$, the best rule is
\[
\pi_\Pclass^*\in\argmax_{\pi\in\Pclass}V_0(\pi),
\]
using the policy value and best-in-class notation defined in
Section~\ref{subsec:setup}.
The unrestricted optimum allows a separate choice in every baseline environment. The restricted optimum chooses among the rules in $\Pclass$. The difference between their values measures the cost of restricting the policy class. Section~\ref{subsec:oracle-reductions} first characterizes optimal actions in the running examples. Section~\ref{subsec:score-based-design} then uses their structure to construct score-based policy classes.

\subsection{Population Oracles: Actions and Policies}
\label{subsec:oracle-reductions}

When the integrated action value is additive across selected pairs or groups, the unrestricted assignment problem can be formulated as a standard combinatorial optimization problem in operations research. For classroom seating, pair additivity leads to a maximum-weight matching problem, while for mentor assignment, additivity across mentor groups leads to a set-partitioning problem. Both problems admit established optimization algorithms. In both formulations, the objective coefficients represent the expected welfare contributions of the selected pairs or groups, derived from the network-formation and outcome models. The matching formulation is related to the pair-surplus approach of \citet{HazardKitagawa2025}, with surplus here incorporating the effects of assignment-induced relationships.

\paragraph{Deskmate seating.}
Taking expectations in \eqref{eq:deskmate-outcome-response}
under the formation model in \eqref{eq:deskmate-link-probability}
gives a pair-additive value. Allowing for known seating costs,
\begin{equation}
\label{eq:deskmate-additive}
v_0(h,a)=\alpha(h)+\sum_{\{i,j\}\in E_h^{desk}}\Delta_{ij}^{desk}(h)x_{ij}(a),
\end{equation}
where $\alpha(h)$ collects the assignment-invariant terms and
\[
\Delta_{ij}^{desk}(h)
=
\frac1{|I_h|}\omega_{ij}^F(h)\delta_{ij}^F(h)-c_{ij}(h).
\]
Here $c_{ij}(h)$ is the known cost of seating the pair together,
set to zero when costs are excluded. Since $\alpha(h)$ does not
depend on the assignment, maximizing conditional welfare is
equivalent to solving the maximum-weight matching problem
\begin{equation}
\label{eq:deskmate-matching-ip}
\begin{array}{ll}
\max_x
&\displaystyle \sum_{\{i,j\}\in E_h^{desk}}\Delta_{ij}^{desk}(h)x_{ij}\\[0.4em]
\mathrm{s.t.}
&\displaystyle \sum_{j:\{i,j\}\in E_h^{desk}}x_{ij}\le1,\qquad i\in I_h,\\[0.7em]
&x_{ij}\in\{0,1\},\qquad \{i,j\}\in E_h^{desk}.
\end{array}
\end{equation}
The feasible binary vectors correspond exactly to the matchings
in $\A^{desk}(h)$. Standard maximum-weight matching algorithms solve this problem directly \citep{Edmonds1965,Galil1986}.

\paragraph{Mentor assignments.}
Taking expectations in \eqref{eq:mentor-outcome-response}
replaces each peer-link indicator by its probability from
\eqref{eq:mentor-comentee-probability}. Allowing for known
assignment costs, the resulting value is additive across mentor groups:
\begin{equation}
\label{eq:mentor-cell-additive-main}
v_0(h,a)=\alpha(h)+\sum_{j\in\mathcal M_h}\mathcal W_j(I_j(a);h).
\end{equation}
Here $\alpha(h)$ is the assignment-invariant baseline. For each
mentor $j$, the feasible groups are
\[
\mathcal B_j(h)=
\{B\subseteq J_h:\ B\ \text{contains only juniors eligible for mentor }j,\ |B|\le b_j(h)\},
\]
including the empty group. The net contribution of
$B\in\mathcal B_j(h)$ is
\begin{equation}
\label{eq:mentor-cell-value-main}
\begin{aligned}
\mathcal W_j(B;h)
=&\ \frac1{|J_h|}\Bigg[
\sum_{i\in B}\left\{
\beta_q q_j(|B|;h)+\beta_\lambda\lambda_{ij}(h)
+\beta_{q\lambda}q_j(|B|;h)\lambda_{ij}(h)
\right\} \\
&\qquad+
\sum_{\{i,i'\}\subset B}
\beta_{\mathrm{Co}}p_{ii'j}^{\mathrm{Co}}(h,|B|)S_{ii'}(h)
\Bigg]-c_j^{ment}(B;h).
\end{aligned}
\end{equation}
Here $c_j^{ment}(B;h)$ is a known group-assignment cost, set to
zero when costs are excluded, and $\mathcal W_j(\emptyset;h)=0$.
With $y_{jB}$ indicating that mentor $j$ receives group $B$, the
unrestricted assignment problem has the set-partitioning formulation
\begin{equation}
\label{eq:mentor-set-partition-ip}
\begin{array}{ll}
\max_y
&\displaystyle \sum_{j\in\mathcal M_h}\sum_{B\in\mathcal B_j(h)}
\mathcal W_j(B;h)y_{jB}\\[0.6em]
\mathrm{s.t.}
&\displaystyle \sum_{B\in\mathcal B_j(h)}y_{jB}=1,\qquad j\in\mathcal M_h,\\[0.8em]
&\displaystyle \sum_{j\in\mathcal M_h}\sum_{B\in\mathcal B_j(h):\,i\in B}y_{jB}=1,\qquad i\in J_h,\\[0.8em]
&y_{jB}\in\{0,1\},\qquad j\in\mathcal M_h,\ B\in\mathcal B_j(h).
\end{array}
\end{equation}
The first constraint assigns each mentor one feasible group, with the empty group representing an unused mentor. The second assigns each junior to exactly one of the selected groups. The feasible binary solutions correspond exactly to feasible
mentor assignments, and the objective equals $v_0(h,a)-\alpha(h)$.
Since $\alpha(h)$ is independent of the assignment, any optimum
of \eqref{eq:mentor-set-partition-ip} yields an unrestricted
optimal assignment. Replacing $y_{jB}\in\{0,1\}$ by $0\le y_{jB}\le1$ gives a linear-programming relaxation whose solutions can be fractional. Integer solutions can be obtained by cutting-plane or branch-and-bound methods \citep{BalasPadberg1976}. When there are many feasible groups, column-generation methods construct additional groups as needed during optimization \citep{BarnhartJohnsonNemhauserSavelsberghVance1998}.

\paragraph{Failure of additivity.} The additive structure need not hold when welfare depends on interactions between realized contacts. Consider the network-introduction example in Appendix~\ref{app:network-introductions}, where two contacts may be complements or substitutes. Suppose the contribution of contacts with $j$ and $k$ to agent $i$'s payoff is \[ \nu_{ij}R^+_{ij}+\nu_{ik}R^+_{ik} +\kappa_i R^+_{ij}R^+_{ik}. \] Here $R^+_{ij}$ indicates a realized contact. When $\kappa_i\ne0$, the contribution of contact $j$ depends on whether contact $k$ also forms. The expected contribution of this interaction is \[ \kappa_i\,\Pbb(R^+_{ij}=1,R^+_{ik}=1\mid h,a), \] which may depend on whether both introductions are selected. Consequently, the welfare contribution of one introduction may depend on the other, and the integrated action value need not be additive across selected pairs.
When exact maximization is computationally impractical, heuristic search can compare feasible assignments using their integrated welfare. For example, a local search can consider adding, removing, or swapping introductions,
subject to the assignment constraints, and retain changes that increase welfare. Such a search need not attain the unrestricted optimum, and incomplete maximization of estimated welfare is accounted for by the optimization-error term in Section~\ref{subsec:empirical-maximization}.

\subsection{Score-Based Rules as Approximation}
\label{subsec:score-based-design}

The preceding optimization formulations motivate a restricted class of score-based assignment rules. These rules assign scores to feasible pairs or groups using a common coefficient vector and select an assignment by maximizing the total score. 
Let $\mathcal R(h)$ be the set of components scored by the rule, such as feasible pairs or mentor groups. The indicator $T_r(h,a)$ records whether action $a$ selects component $r$. Write the conditional value as
\[
v_0(h,a)=C_0(h)+\sum_{r\in\mathcal R(h)}T_r(h,a)\Delta_r(h)+R_0(h,a),
\]
where $C_0(h)$ is the action-invariant baseline, $\Delta_r(h)$ is the welfare contribution of component $r$, and $R_0(h,a)$ is a remainder term that captures departures from the additive representation. 

To define the score-based rule, we give component $r$ the score $\theta'\xi_r(h)$, where $\xi_r(h)$ contains baseline features. Consider
\[
\Pclass_\Theta=\{\pi_\theta:\theta\in\Theta\},
\qquad
\pi_\theta(h)\in\argmax_{a\in\A(h)}
\sum_{r\in\mathcal R(h)}T_r(h,a)\theta'\xi_r(h),
\]
where $\Theta\subseteq\R^d$ may be continuous. Given $\theta$, the maximization selects a feasible assignment
in each baseline environment using the same coefficient vector across environments. 
The resulting policy is evaluated using its population value $V_0(\pi_\theta)$, which incorporates any remainder $R_0$ in the true action value. The best policy in this restricted class is
\[
\pi_\Theta^*\in\argmax_{\pi\in\Pclass_\Theta}V_0(\pi),
\]
and it generally differs from the unrestricted optimum $\pi_{\mathrm{unr}}^\star$. The gap $V_0(\pi_{\mathrm{unr}}^\star)-V_0(\pi_\Theta^*)$ is the policy-class approximation error; the statistical regret bounds in Section~\ref{sec:learning} control the additional loss from estimating and selecting a rule within the class.

The structure of the welfare function can guide the choice of score features. In the running examples, the formation and outcome specifications indicate which characteristics and interactions enter surplus. 

\paragraph{Deskmate seating.} The exact pair surplus in the additive benchmark is
\[
\Delta_{ij}^{\mathrm{desk}}(h)
=
\frac1{|I_h|}\omega_{ij}^F(h)\delta_{ij}^F(h)
-c_{ij}(h).
\]
The response term $\omega_{ij}^F(h)$ is the value of the induced friendship link, while $\delta_{ij}^F(h)$ is the seating-induced increment defined in \eqref{eq:deskmate-probability-increment}. If linear functions of baseline features approximate the friendship contribution and probability increment,
\[
\omega_{ij}^F(h)\approx \beta_\omega'z_{ij}^{\omega}(h),
\qquad
\delta_{ij}^F(h)\approx \beta_p'z_{ij}^{p}(h),
\]
then multiplying these approximations gives:
\[
\omega_{ij}^F(h)\delta_{ij}^F(h)
\approx
(\beta_\omega\otimes\beta_p)'\left[z_{ij}^{\omega}(h)\otimes z_{ij}^{p}(h)\right].
\]
Here $\otimes$ denotes the Kronecker product. This product structure motivates including baseline characteristics and their interactions in the score. For the deskmate example in Section~\ref{subsec:deskmates}, suppose prior friendship $F^0_{ij}$, achievement distance $|\mathrm{Ach}_i-\mathrm{Ach}_j|$, and common-friend measure $\mathrm{Cl}^0_{ij}$ enter both approximations. One parsimonious choice of score features is
\[
\begin{aligned}
\xi_{ij}^{desk}(h)=\big(&
F^0_{ij},\ \mathrm{Cl}^0_{ij},\ |\mathrm{Ach}_i-\mathrm{Ach}_j|, |\mathrm{Ach}_i-\mathrm{Ach}_j|^2,F^0_{ij}|\mathrm{Ach}_i-\mathrm{Ach}_j|, \mathrm{Cl}^0_{ij}|\mathrm{Ach}_i-\mathrm{Ach}_j|
\big).
\end{aligned}
\]

\paragraph{Mentor assignments.}  We construct score features at the group level. For a group $B$ assigned to mentor $j$, the score approximates
\[
\theta'\xi_j^{ment}(B,h)
\approx
\mathcal W_j(B;h),
\]
so $\xi_j^{ment}(B,h)$ should contain covariates corresponding to the terms in \eqref{eq:mentor-cell-value-main}. Let $R_{ij}^{JM}(h,|B|)$ collect baseline characteristics and functions of group size relevant to mentor quality and junior--mentor compatibility, including their interactions. Let $R_{ii'j}^{\mathrm{Co}}(h,|B|)$ contain covariates for pairs of juniors in the group. The remaining group characteristics are collected in $Z_j(B,h)$. One specification could be
\[
\begin{aligned}
\xi_j^{ment}(B,h)=\Big(
&\sum_{i\in B} R_{ij}^{JM}(h,|B|),\quad
\sum_{\{i,i'\}\subset B}R_{ii'j}^{\mathrm{Co}}(h,|B|),\quad
Z_j(B,h)
\Big).
\end{aligned}
\]
These features depend on the baseline environment $h$ and the proposed group $B$. Constructing score features at the group level allows the score to reflect how adding a junior affects mentoring benefits and peer spillovers for the other juniors assigned to the same mentor.

\section{Identification and Model Misspecification}
\label{sec:identification}

This section studies what the observed data identify about policy values and how model misspecification can affect policy choice. Section~\ref{subsec:policy-value-identification} characterizes the identified sets for action and policy values. Section~\ref{subsec:approximation-bias} bounds the welfare loss from using misspecified or approximate values.

\subsection{Policy-Value Identification}
\label{subsec:policy-value-identification}

We study identification of policy value while allowing the formation and outcome parameters to be partially identified. Let $P_O$ be the observed distribution of $O_c$ in Section~\ref{subsec:data-sampling}. Let $\Ical(P_O)$ be the set of parameter pairs $(\eta,\gamma)$ that can generate $P_O$ under a full-data distribution of observed variables and potential networks and outcomes satisfying the maintained setup. Observationally equivalent specifications may imply different network distributions or outcome responses under a candidate policy. The historical design and maintained model restrictions jointly determine $\Ical(P_O)$.

Let $Q_{\eta,i,a}(\cdot\mid h)$ denote the distribution of the
exposure $T_i(h,G)$ when $G\sim K_\eta(\cdot\mid h,a)$.
For a policy-relevant action and a policy, define
\begin{align}
\mathcal V_{P_O}(h,a)
&=
\left\{
\frac1{M_h}\sum_{i=1}^{M_h}
\int m_\gamma\{W_i(h),t\}Q_{\eta,i,a}(dt\mid h)
:
(\eta,\gamma)\in\Ical(P_O)
\right\},
\label{eq:conditional-identified-set}
\\
\mathcal V_{P_O}(\pi)
&=
\left\{
\E_{P_O}\left[v_{\eta,\gamma}\{H_c,\pi(H_c)\}\right]
:
(\eta,\gamma)\in\Ical(P_O)
\right\}.
\label{eq:policy-identified-set}
\end{align}
The expectation in \eqref{eq:policy-identified-set} uses the observed marginal distribution of baseline environments.

\begin{proposition}[Policy-value identification]
\label{prop:identification}
Suppose the setup in Section~\ref{subsec:setup} holds and $\Ical(P_O)$ is nonempty. Then \eqref{eq:conditional-identified-set} and \eqref{eq:policy-identified-set} are the sharp identified sets for $v_0(h,a)$ and $V_0(\pi)$, respectively. In particular, $v_0(h,a)$ is point identified if and only if, for every $(\eta,\gamma),(\widetilde\eta,\widetilde\gamma)\in\Ical(P_O)$,
\begin{equation}
\label{eq:exact-action-identification}
\frac1{M_h}\sum_{i=1}^{M_h}
\int m_\gamma\{W_i(h),t\}Q_{\eta,i,a}(dt\mid h)
=
\frac1{M_h}\sum_{i=1}^{M_h}
\int m_{\widetilde\gamma}\{W_i(h),t\}Q_{\widetilde\eta,i,a}(dt\mid h).
\end{equation}
Likewise, $V_0(\pi)$ is point identified if and only if
\begin{equation}
\label{eq:exact-policy-identification}
\E_{P_O}\left[v_{\eta,\gamma}\{H_c,\pi(H_c)\}\right]
=
\E_{P_O}\left[v_{\widetilde\eta,\widetilde\gamma}\{H_c,\pi(H_c)\}\right]
\end{equation}
for every two such pairs. Whenever \eqref{eq:exact-action-identification} holds, its common value can be computed using any $(\eta,\gamma)\in\Ical(P_O)$:
\begin{equation}
\label{eq:identification-value}
v_0(h,a)
=
\frac1{M_h}\sum_{i=1}^{M_h}
\int m_\gamma\{W_i(h),t\}Q_{\eta,i,a}(dt\mid h).
\end{equation}
\end{proposition}

Point identification of $\eta$ and $\gamma$ implies the displayed equalities. More generally, policy values are identified when the remaining ambiguity in formation and outcomes leaves their integrated values unchanged. For the additive responses in the seating and mentor examples, taking expectations replaces each realized link indicator with its probability of forming. Thus formation models that agree on these probabilities under each candidate assignment imply the same policy values. Their predictions about which links form together can differ without affecting expected welfare. When the outcome response includes interactions between links, the probabilities of those links forming together also enter policy value.

\begin{remark}[Implications for experimental design]
\label{rem:experimental-design}
Proposition~\ref{prop:identification} suggests directing experimental variation toward assignment margins that affect welfare comparisons among candidate policies.
For classroom seating, randomizing deskmate status across pair types generates variation in friendship formation and, consequently, in the friendship exposures that enter academic achievement. Under the maintained models, this variation informs both the seating-induced
friendship increments and the contributions of friendships to achievement, which jointly determine pair surplus in \eqref{eq:deskmate-additive}. 
For mentor assignments, variation in mentor loads informs how mentoring quality and peer formation change with load, while variation in group composition at a given load informs junior--mentor compatibility and peer spillovers. These components jointly determine the group values in \eqref{eq:mentor-cell-value-main}.
\end{remark}

\subsection{Model Misspecification and Value Approximation}
\label{subsec:approximation-bias}

Policy learning uses values computed from the fitted formation and outcome models. Misspecification of either model can change the value assigned to an arrangement and its ranking among feasible alternatives. In this section, we derive bounds on the resulting welfare
loss, first in terms of errors in policy values and then in terms of errors in the formation law and outcome response.

Let $\widetilde v(h,a)$ denote a model-implied or computationally approximate value, and let $v_0(h,a)$ be the true value. Define
\[
\Delta(\Pclass)=
\sup_{\pi\in\Pclass}
\left|
\E\left[\widetilde v(H_c,\pi(H_c))-v_0(H_c,\pi(H_c))\right]
\right|,
\qquad
\widetilde V(\pi)=\E[\widetilde v(H_c,\pi(H_c))],
\]
and let $\widetilde\pi^*\in\argmax_{\pi\in\Pclass}\widetilde V(\pi)$.

\begin{proposition}[Value approximation and model misspecification]
\label{prop:approx-bias}
\label{prop:formation-target-shift}
\label{prop:outcome-target-shift}
Under the setup in Section~\ref{subsec:setup}, the following hold.
\begin{enumerate}[label=(\roman*)]
\item For any learned policy $\widehat\pi\in\Pclass$,
\[
V_0(\pi_\Pclass^*)-V_0(\widehat\pi)
\le
[\widetilde V(\widetilde\pi^*)-\widetilde V(\widehat\pi)]
+2\Delta(\Pclass).
\]

\item Let $\widetilde K(\cdot\mid h,a)$ be an approximate formation law. Write $\TV(P,Q)=\sup_B|P(B)-Q(B)|$ for total variation distance and define
\[
\widetilde v_K(h,a)=\int \mu_0(h,g)\widetilde K(dg\mid h,a),
\qquad
\widetilde V_K(\pi)=\E[\widetilde v_K(H_c,\pi(H_c))].
\]
If $\widetilde\pi_K^*\in\argmax_{\pi\in\Pclass}\widetilde V_K(\pi)$ and
$\mathcal D_K(\Pclass)=\sup_{\pi\in\Pclass}|V_0(\pi)-\widetilde V_K(\pi)|$, then
\[
V_0(\pi_\Pclass^*)-V_0(\widetilde\pi_K^*)\le 2\mathcal D_K(\Pclass).
\]
If $|\mu_0(h,g)|\le \bar B$ and $\TV\{K_0(\cdot\mid h,a),\widetilde K(\cdot\mid h,a)\}\le\delta_K$ uniformly over policy-relevant $(h,a)$, then
\[
V_0(\pi_\Pclass^*)-V_0(\widetilde\pi_K^*)\le 4\bar B\delta_K.
\]

\item Let $\widetilde m$ be an approximate outcome response and define
\[
\widetilde v_m(h,a)=
\int
\left[\frac1{M_h}\sum_{i=1}^{M_h}
\widetilde m\{W_i(h),T_i(h,g)\}\right]
K_0(dg\mid h,a),
\qquad
\widetilde V_m(\pi)=\E[\widetilde v_m(H_c,\pi(H_c))].
\]
If $\widetilde\pi_m^*\in\argmax_{\pi\in\Pclass}\widetilde V_m(\pi)$ and
$\mathcal D_m(\Pclass)=\sup_{\pi\in\Pclass}|V_0(\pi)-\widetilde V_m(\pi)|$, then
\[
V_0(\pi_\Pclass^*)-V_0(\widetilde\pi_m^*)\le 2\mathcal D_m(\Pclass).
\]
If the average response error is at most $\delta_m$ uniformly over policy-relevant actions,
\[
\sup_{(h,a)\in\mathcal E_{\Pclass}}
\left|
\int \frac1{M_h}\sum_i
\left[\widetilde m\{W_i(h),T_i(h,g)\}-m_{\gamma_0}\{W_i(h),T_i(h,g)\}\right]
K_0(dg\mid h,a)
\right|
\le \delta_m,
\]
then $V_0(\pi_\Pclass^*)-V_0(\widetilde\pi_m^*)\le 2\delta_m$.
\end{enumerate}
\end{proposition}

The first bound separates regret under the approximate value from its discrepancy with the true value. Increasing the sample size can reduce estimation error for $\widetilde v$ while leaving this discrepancy unchanged. The remaining bounds consider misspecification of each model separately. Formation misspecification changes the exposure distribution over which outcomes are averaged. Outcome misspecification changes the value assigned to those exposures.

In the deskmate example, an inaccurate friendship model can misrank seating charts. An outcome response that omits nonlinear effects of friendships can also assign incorrect values to charts even with accurate formation probabilities. In the mentor example, misspecifying how mentor load affects peer formation can distort the expected spillovers within assigned groups. Even with accurate peer-link probabilities, an outcome response that omits the decline in mentoring quality with load or the interaction between mentor quality and junior--mentor compatibility can misstate the direct benefits of mentoring. These errors can change the estimated welfare of mentor groups and their rankings across feasible assignments.

\begin{remark}[Policy choice under model uncertainty] \label{rem:robust-policy-choice} 
The preceding bounds assess the welfare consequences of choosing a policy using a single approximate value function. An alternative is to account for model uncertainty directly in policy choice. Let $\mathcal U$ be a nonempty set of plausible conditional value functions, each generated by a candidate formation law and outcome response. A maximin policy solves 
\[ \pi_{\mathcal U}^{\mathrm{rob}} \in \argmax_{\pi\in\Pclass} \inf_{\widetilde v\in\mathcal U} \E\!\left[ \widetilde v\{H_c,\pi(H_c)\} \right]. \] 
This criterion selects the policy with the highest worst-case welfare over $\mathcal U$. The set can be generated by the observationally compatible specifications in Section~\ref{subsec:policy-value-identification}, or by specified departures from a reference formation or outcome model. If the true value function belongs to $\mathcal U$, the worst-case value is a lower bound on the selected policy's true welfare. 
\end{remark}

\section{Policy Learning}
\label{sec:learning}

This section studies policy selection by empirical value maximization and bounds regret relative to the best rule in the policy class. Section~\ref{subsec:general-estimation-recipe} constructs the plug-in action-value estimates. Section~\ref{subsec:empirical-maximization} defines the selection rule, decomposes regret, and gives sampling and simulation rates for score-based policies. Section~\ref{subsec:first-stage-regret-terms} relates the first-stage terms to estimation accuracy and policy-value sensitivity. Section~\ref{subsec:running-example-first-stage} specializes the bounds to the running examples.

\subsection{Plug-in Value Estimation}
\label{subsec:general-estimation-recipe}

The estimator is a sample analogue of the identification formula. Using the observed clusters $O_1,\ldots,O_C$, estimate the formation law by $\widehat K$ and the outcome response by $\widehat m$. These first-stage estimates determine the fitted value of each candidate assignment.

For a baseline environment $h$ and network realization $g$, define the predicted cluster mean by
\[
\widehat\mu(h,g)
=
\frac1{M_h}\sum_{i=1}^{M_h}
\widehat m\{W_i(h),T_i(h,g)\}.
\]
The estimated value of an action $a\in\A(H_c)$ is
\begin{equation}
\label{eq:estimated-value}
\widehat v_c(a)
=
\int \widehat\mu(H_c,g)\widehat K(dg\mid H_c,a).
\end{equation}
Thus the fitted formation law supplies the distribution of post-action networks, while the fitted outcome response assigns a mean outcome to the exposures generated by each network. The same fitted models are used for every cluster and candidate action.

If the individual response is linear in a feature vector $\phi_i(h,g)$,
\[
\widehat m\{W_i(H_c),T_i(H_c,g)\}=\widehat\beta'\phi_i(H_c,g),
\]
then the integral reduces to
\[
\widehat v_c(a)
=
\widehat\beta'\widehat{\bar\Phi}_a(H_c),
\qquad
\widehat{\bar\Phi}_a(H_c)
=
\frac1{M_c}\sum_i\int \phi_i(H_c,g)\widehat K(dg\mid H_c,a).
\]
For a linear response, only the expected exposure features are needed to compute $\widehat v_c(a)$. When these expectations are available in closed form, the value can be evaluated analytically.

When the integral in \eqref{eq:estimated-value} cannot be evaluated analytically, draw $g_c^{(1)},\ldots,g_c^{(S)}$ independently from $\widehat K(\cdot\mid H_c,a)$, conditional on the observed data, and compute
\[
\widehat v_{c,S}(a)
=
\frac1S\sum_{s=1}^S
\frac1{M_c}\sum_{i=1}^{M_c}
\widehat m\{W_i(H_c),T_i(H_c,g_c^{(s)})\}.
\]
Simulation approximates the same value integral. Its numerical error enters the regret bound in Section~\ref{subsec:empirical-maximization}. Repeated evaluations of a cluster--action pair use the same simulated value. Appendix~\ref{app:implementation-map} illustrates analytic value calculations and simulation for the running examples.

\subsection{Empirical Welfare Maximization and Regret}
\label{subsec:empirical-maximization}

We choose a policy $\widehat\pi\in\Pclass$ to maximize estimated welfare and study its regret $V_0(\pi_\Pclass^*)-V_0(\widehat\pi)$ relative to the best policy in $\Pclass$. Let $\widehat v_c(a)$ be the plug-in estimate of action value from Section~\ref{subsec:general-estimation-recipe}, with the integral evaluated exactly or approximated numerically. We hold the estimated formation and outcome models fixed when maximizing
\begin{equation}
\label{eq:ewm-criterion}
\widehat V_C(\pi)
\equiv
\widehat V_C^{est}(\pi)
=
\frac1C\sum_{c=1}^C \widehat v_c\{\pi(H_c)\}.
\end{equation}

For the score-based class $\Pclass_\Theta$, each coefficient vector $\theta$ determines an assignment by maximizing the score in Section~\ref{subsec:score-based-design}. With a fixed rule for breaking ties, the selected assignments and their estimated welfare are
\begin{align*}
a_c(\theta)=\pi_\theta(H_c)
&\in\argmax_{a\in\A(H_c)}
\sum_{r\in\mathcal R(H_c)}T_r(H_c,a)\theta'\xi_r(H_c),\\
\widehat V_C(\pi_\theta)
&=\frac1C\sum_{c=1}^C\widehat v_c\{a_c(\theta)\}.
\end{align*}
We choose $\widehat\theta$ to maximize $\widehat V_C(\pi_\theta)$ and set $\widehat\pi=\pi_{\widehat\theta}$. In the running examples, maximizing the additive score is a matching or set-partitioning problem. The unrestricted policy instead chooses assignments to maximize estimated welfare. When estimated welfare has the additive form described in Section~\ref{subsec:oracle-reductions}, the objective coefficients are the estimated pair or group welfare contributions. For nonadditive values, the procedure computes estimated welfare for each assignment considered.

An approximate maximizer satisfies
\[
\widehat V_C(\widehat\pi)
\ge \sup_{\pi\in\Pclass}\widehat V_C(\pi)-\eps_{opt,C},
\qquad \eps_{opt,C}\ge0.
\]
Here $\eps_{opt,C}$ bounds the difference between the largest estimated welfare in $\Pclass$ and that of the computed policy. The objective $\widehat V_C(\pi_\theta)$ can be nonsmooth because small changes in $\theta$ can change the selected assignments. A numerical algorithm may provide an upper bound on its optimization error; otherwise, $\eps_{opt,C}$ remains an unknown error term in the regret bound. Simulated policy comparisons can use the same random draws. Appendix~\ref{app:implementation-map} gives the value calculations for the running examples.

To bound regret, define the value-estimation error and sampling error by
\begin{align*}
r_{v,C}
&=\sup_{\pi\in\Pclass}
\left|
\frac1C\sum_{c=1}^C
\left[\widehat v_c(\pi(H_c))-v_0(H_c,\pi(H_c))\right]
\right|,\\
s_C
&=\sup_{\pi\in\Pclass}
\left|
\frac1C\sum_{c=1}^C v_0(H_c,\pi(H_c))-V_0(\pi)
\right|.
\end{align*}
The term $r_{v,C}$ measures the error from replacing true action values with their estimates in the sample average. The term $s_C$ measures the difference between the sample average of true action values and population welfare. The approximate maximization condition implies
\[
V_0(\pi_\Pclass^*)-V_0(\widehat\pi)
\le 2s_C+2r_{v,C}+\eps_{opt,C}.
\]
The bound separates value-estimation error, sampling error, and optimization error. The term $r_{v,C}$ includes numerical integration error when the integral defining $\widehat v_c$ is approximated.

For the plug-in estimator, value-estimation error comes from estimating the outcome response and formation distribution and from approximating the integral. Let
\[
\widehat v_c^{int}(a)=
\int \widehat\mu(H_c,g)\widehat K(dg\mid H_c,a)
\]
denote the exact plug-in integral. Define the three error terms as
\begin{align}
\mathcal E_{\mu,C}
&=
\sup_{\pi\in\Pclass}
\frac1C\sum_{c=1}^C
\int
\left|
\widehat\mu(H_c,g)-\mu_0(H_c,g)
\right|
\widehat K(dg\mid H_c,\pi(H_c)),
\label{eq:E-mu-main}\\
\mathcal E_{K,C}
&=
\sup_{\pi\in\Pclass}
\left|
\frac1C\sum_{c=1}^C
\int
\mu_0(H_c,g)
(\widehat K-K_0)(dg\mid H_c,\pi(H_c))
\right|,
\label{eq:E-K-main}\\
\mathcal E_{MC,C}
&=
\sup_{\pi\in\Pclass}
\left|
\frac1C\sum_{c=1}^C
\left[
\widehat v_c(\pi(H_c))
-
\widehat v_c^{int}(\pi(H_c))
\right]
\right|.
\label{eq:E-MC-main}
\end{align}

The triangle inequality gives
\[
r_{v,C}\le
\mathcal E_{\mu,C}+\mathcal E_{K,C}+\mathcal E_{MC,C}.
\]
The term $\mathcal E_{\mu,C}$ averages absolute errors in the estimated outcome response over the estimated network distribution. The term $\mathcal E_{K,C}$ measures the effect of formation estimation error on welfare, holding the outcome response at its true value. The numerical integration error $\mathcal E_{MC,C}$ is zero when the integral is evaluated exactly.

The preceding inequalities hold for any prespecified policy class. For the score-based class in Section~\ref{subsec:score-based-design}, let $\Theta_C\subseteq\R^{d_C}$, $d_C\ge1$, and $K_C=\sup_h|\A(h)|<\infty$. Scores use prespecified features, and ties are resolved by a fixed measurable ordering of feasible actions. The dimension $d_C$ counts score coefficients, and $K_C$ bounds the number of feasible assignments in any environment. Define the induced class of conditional policy-value functions by
\[
\mathcal F_C=\{h\mapsto v_0(h,\pi_\theta(h)):\theta\in\Theta_C\}.
\]
We assume that $\mathcal F_C$ is pointwise measurable: it contains a countable subclass $\mathcal F_C^0$ such that every $f\in\mathcal F_C$ is the pointwise limit of a sequence in $\mathcal F_C^0$.

\begin{theorem}[Plug-in regret bound]
\label{thm:main-regret-decomposition}
Let $\Pclass=\Pclass_{\Theta_C}$ satisfy the conditions above, and let $\widehat\pi\in\Pclass$ maximize \eqref{eq:ewm-criterion} up to optimization error $\eps_{opt,C}$, using the plug-in value estimates defined above. Then
\begin{equation}
\label{eq:main-regret}
V_0(\pi_\Pclass^*)-V_0(\widehat\pi)
\le
2s_C+2\mathcal E_{\mu,C}+2\mathcal E_{K,C}
+2\mathcal E_{MC,C}+\eps_{opt,C}.
\end{equation}
Under Assumption~\ref{ass:iid-clusters},
\begin{equation}
\label{eq:uniform-bound}
s_C=O_p\!\left(\bar B\sqrt{\frac{d_C\log(eK_C)}{C}}\right).
\end{equation}
If $\widehat v_c=\widehat v_{c,S}$ as defined in Section~\ref{subsec:general-estimation-recipe} and the fitted cluster response is bounded in absolute value by $\bar B$ under the fitted formation laws, then
\begin{equation}
\label{eq:simulation-rate}
\mathcal E_{MC,C}
=O_p\!\left(\bar B\sqrt{\frac{\log(eK_C)}{S}}\right).
\end{equation}
With exact integration, $\mathcal E_{MC,C}=0$.
\end{theorem}

The sampling bound in \eqref{eq:uniform-bound} decreases with the number of independent clusters but grows with the score dimension and the number of feasible assignments. In particular, sampling error vanishes when $d_C\log(eK_C)/C\to0$. The simulation bound in \eqref{eq:simulation-rate} depends on the number of network draws $S$ and the number of feasible assignments. This error vanishes when $\log(eK_C)/S\to0$ and is zero under exact integration. Section~\ref{subsec:first-stage-regret-terms} relates the remaining formation- and outcome-estimation errors to first-stage estimation rates and policy-value sensitivity.

\begin{remark}[Policy-class approximation and sampling error]
\label{rem:policy-class-tradeoff}
The choice of policy class involves a trade-off between approximation error and sampling error. Let $\pi_{\mathrm{unr}}^\star$ denote the unrestricted optimal policy and $\pi_\Pclass^*$ the best policy in $\Pclass$. Define
$\mathcal E_{\mathrm{app}}(\Pclass)
=V_0(\pi_{\mathrm{unr}}^\star)-V_0(\pi_\Pclass^*)$.
Then
\[
\begin{aligned}
V_0(\pi_{\mathrm{unr}}^\star)-V_0(\widehat\pi)
&=\mathcal E_{\mathrm{app}}(\Pclass)
+[V_0(\pi_\Pclass^*)-V_0(\widehat\pi)]\\
&\le \mathcal E_{\mathrm{app}}(\Pclass)+2s_C
+2(\mathcal E_{\mu,C}+\mathcal E_{K,C}
+\mathcal E_{MC,C})+\eps_{opt,C}.
\end{aligned}
\]
For nested policy classes, adding score features can reduce policy-class approximation error, but can increase the complexity of the policy class. In particular, the sampling bound in \eqref{eq:uniform-bound} depends on the score dimension $d_C$.
Section~\ref{subsec:score-based-design} discusses policy-class approximation, while Theorem~\ref{thm:main-regret-decomposition} quantifies the sampling error associated with score-based rules.
\end{remark}

\subsection{What Determines the First-Stage Regret Terms}
\label{subsec:first-stage-regret-terms}

Theorem~\ref{thm:main-regret-decomposition} bounds regret in terms of the outcome and formation errors $\mathcal E_{\mu,C}$ and $\mathcal E_{K,C}$. We now relate these errors to the accuracy of the first-stage estimates. Their contribution depends on how changes in each model alter values under candidate assignments.

Suppose the outcome and formation models are locally parameterized by $\gamma$ and $\eta$, with true values $\gamma_0$ and $\eta_0$. We state the conditions in parameter norms. Corresponding nonparametric conditions can instead use distances between the estimated functions or distributions.

\begin{assumption}[Local Lipschitz bounds]
\label{ass:policy-relevant-sensitivity}
Write $\mu_\gamma(h,g)=M_h^{-1}\sum_i m_\gamma\{W_i(h),T_i(h,g)\}$, $\widehat\mu=\mu_{\widehat\gamma}$, and $\widehat K=K_{\widehat\eta}$. There are nonnegative functions $L_\mu(h,g)$ and $L_K(h,a)$ such that, on a neighborhood containing the true and estimated parameters,
\begin{align}
|\mu_\gamma(h,g)-\mu_{\gamma_0}(h,g)|
&\le L_\mu(h,g)\|\gamma-\gamma_0\|,
\label{eq:outcome-local-lipschitz}\\
\left|
\int \mu_0(h,g)\{K_\eta-K_{\eta_0}\}(dg\mid h,a)
\right|
&\le L_K(h,a)\|\eta-\eta_0\|.
\label{eq:formation-local-lipschitz}
\end{align}
The displayed quantities are finite on the set of actions and network realizations relevant to the policy class.
\end{assumption}

Define the cluster-action sensitivity factors
\begin{equation*}
\Gamma_{\mu,c}(a)
=\int L_\mu(H_c,g)\widehat K(dg\mid H_c,a),
\qquad
\Gamma_{K,c}(a)=L_K(H_c,a),
\end{equation*}
and their largest averages over candidate policies
\begin{equation*}
\Lambda_{\mu,C}
=\sup_{\pi\in\Pclass}\frac1C\sum_{c=1}^C
\Gamma_{\mu,c}(\pi(H_c)),
\qquad
\Lambda_{K,C}
=\sup_{\pi\in\Pclass}\frac1C\sum_{c=1}^C
\Gamma_{K,c}(\pi(H_c)).
\end{equation*}
Also let $r_{\gamma,C}=\|\widehat\gamma-\gamma_0\|$ and $r_{\eta,C}=\|\widehat\eta-\eta_0\|$.

\begin{proposition}[First-stage sensitivity]
\label{prop:first-stage-sensitivity}
Under Assumption~\ref{ass:policy-relevant-sensitivity},
\begin{align}
\mathcal E_{\mu,C}
&\le
\sup_{\pi\in\Pclass}\frac1C\sum_{c=1}^C
\Gamma_{\mu,c}(\pi(H_c))
\|\widehat\gamma-\gamma_0\|
= \Lambda_{\mu,C}r_{\gamma,C},
\label{eq:E-mu-sensitivity}\\
\mathcal E_{K,C}
&\le
\sup_{\pi\in\Pclass}\frac1C\sum_{c=1}^C
\Gamma_{K,c}(\pi(H_c))
\|\widehat\eta-\eta_0\|
= \Lambda_{K,C}r_{\eta,C}.
\label{eq:E-K-sensitivity}
\end{align}
Consequently, the policy selected using plug-in value estimates obeys
\[
V_0(\pi_\Pclass^*)-V_0(\widehat\pi)
\le
2s_C+2\Lambda_{\mu,C}r_{\gamma,C}
+2\Lambda_{K,C}r_{\eta,C}
+2\mathcal E_{MC,C}+\eps_{opt,C}.
\]
\end{proposition}

The plug-in value estimator is generally sensitive to first-stage estimation errors at first order. The rates $r_{\gamma,C}$ and $r_{\eta,C}$ measure estimation error in the outcome and formation models. The factors $\Lambda_{\mu,C}$ and $\Lambda_{K,C}$ measure how strongly candidate policies amplify that error. For example, an assignment that concentrates on large values of an outcome regressor can magnify a small coefficient error. High mentor loads have a similar effect when many peer relationships contribute to welfare. Good average fit can therefore coexist with inaccurate values at the assignments favored by optimization. Conversely, uncertainty about a feature that contributes almost equally under all candidate assignments has little effect on their ranking. Both estimation accuracy and the exposures induced by candidate policies determine the relevance of a first-stage error.

For a linear outcome response $\mu_\gamma(h,g)=\gamma'\Phi(h,g)$, one can take $L_\mu(h,g)=\|\Phi(h,g)\|_*$, where $\|z\|_* = \sup_{\|u\|\le1}|z^\prime u|$ is the dual of the parameter norm. The factor $\Gamma_{\mu,c}(a)$ is then the expected norm of the exposure vector under the fitted network law for action $a$. For a differentiable formation model, a local choice for $L_K(h,a)$ is
\[
L_K(h,a)
=
\sup_{\eta\in\mathcal N}
\left\|\frac{\partial}{\partial\eta}
\int \mu_0(h,g)K_\eta(dg\mid h,a)
\right\|_*,
\]
where $\mathcal N$ contains $\eta_0$ and the fitted values. This derivative measures how changes in the formation parameter affect value after accounting for the outcome response.

\begin{remark}[Nonregular formation and response models]
\label{rem:nonregular-module-sensitivity}
Assumption~\ref{ass:policy-relevant-sensitivity} may fail when small parameter changes lead to discontinuous or sharply varying policy values. In pairwise-stable formation, a link whose surplus is near zero can appear or disappear after a small parameter change \citep{JacksonWolinsky1996}. A linear-quadratic response can also become highly sensitive near the boundary of equilibrium stability \citep{BallesterCalvoArmengolZenou2006}. With multiple equilibria, small parameter changes can also change which equilibrium is selected. The regret decomposition in Theorem~\ref{thm:main-regret-decomposition} continues to apply, but the linear sensitivity bounds in Proposition~\ref{prop:first-stage-sensitivity} need not hold. Appendix~\ref{app:nonregular-first-stage} instead uses moduli of continuity to bound the same formation and outcome value errors. Margin or tail conditions control how often candidate policies encounter link-formation thresholds, near-singular network responses, or equilibrium-selection ties. The resulting rates depend on estimation accuracy and the frequency of these cases. When ties persist and the competing equilibria imply different welfare, additional restrictions or evaluation at the lowest value across admissible equilibria may be needed.
\end{remark}

\subsection{First-Stage Terms in the Running Examples}
\label{subsec:running-example-first-stage}

The first-stage terms depend on estimation accuracy and the exposures generated by feasible assignments. We use the notation of Section~\ref{sec:examples}, with hats denoting estimated functions. Action-invariant value components are omitted because they cancel from policy comparisons. Appendices~\ref{app:deskmate-first-stage}--\ref{app:intro-first-stage} give the derivations and bounds for outcome-estimation error.

\paragraph{Deskmate seating.}
Under the pair-additive value in \eqref{eq:deskmate-additive}, the policy-relevant coefficient on a selected pair combines its friendship value $\omega_{ij}^F(h)$ and the seating-induced probability increment $\delta_{ij}^F(h)$. Let $\widehat\omega_{ij}^F(h)$ be the fitted friendship value and let
\[
\widehat\delta_{ij}^F(h)
=
\widehat p_{ij}^{F,1}(h)-\widehat p_{ij}^{F,0}(h)
\]
be the fitted increment. The pair-level error has the exact decomposition
\[
\widehat\omega_{ij}^F\widehat\delta_{ij}^F
-
\omega_{ij}^F\delta_{ij}^F
=
(\widehat\omega_{ij}^F-\omega_{ij}^F)\widehat\delta_{ij}^F
+
\omega_{ij}^F(\widehat\delta_{ij}^F-\delta_{ij}^F).
\]
Let $\mathcal E_{K,C}^{desk}$ denote the formation error after removing assignment-invariant value components. Then
\begin{equation}
\label{eq:deskmate-EK-specialization}
\mathcal E_{K,C}^{desk}
\le
\sup_{\pi\in\Pclass}\frac1C\sum_{c=1}^C
\frac1{|I_{H_c}|}
\sum_{\{i,j\}\in\pi(H_c)}
|\omega_{ij}^F(H_c)|
|\widehat\delta_{ij}^F(H_c)-\delta_{ij}^F(H_c)|.
\end{equation}
The outcome contribution has the same form with $|\widehat\omega_{ij}^F-\omega_{ij}^F|\,|\widehat\delta_{ij}^F|$. Formation error enters through the estimated change in friendship probability; errors common to the seated and unseated probabilities cancel. With bounded regressors and friendship contributions, the logistic specification bounds \eqref{eq:deskmate-EK-specialization} by a constant times the formation-parameter estimation error. The matching constraint limits the sum to at most $|I_h|/2$ selected pairs.

\paragraph{Mentor assignments.}
The group value in \eqref{eq:mentor-cell-value-main} shows that formation error enters through the peer-link probabilities of pairs assigned to the same mentor. Let $\mathcal E_{K,C}^{ment}$ denote this component of $\mathcal E_{K,C}$. Let $\delta_{\mathrm{Co},C}$ be the largest fitted probability error over observed clusters and feasible mentor groups, let $b_{\max,C}$ be the largest mentor capacity, and suppose $|S_{ii'}(h)|\le\bar S$. Then
\begin{equation}
\label{eq:mentor-capacity-EK}
\mathcal E_{K,C}^{ment}
\le
\frac{|\beta_{\mathrm{Co}}|\bar S}{2}
(b_{\max,C}-1)\delta_{\mathrm{Co},C}.
\end{equation}
The factor $b_{\max,C}-1$ bounds the number of peers sharing a mentor with each junior. Mentor load also amplifies outcome-model error. If $\widehat\chi_j$ is the estimate of the dilution parameter in \eqref{eq:mentor-load-quality}, then
\[
|q_j(n;h,\widehat\chi_j)-q_j(n;h,\chi_j)|
\le q_j^0(h)(n-1)|\widehat\chi_j-\chi_j|.
\]
Consequently, error in the mentor-quality parameter is weighted by the selected mentor loads and by the coefficients multiplying mentor quality. The exact definitions and the integrated outcome bound are given in Appendix~\ref{app:mentor-first-stage}.

\paragraph{Network introductions.}
In the introduction example, contacts can be complements or substitutes. Formation error must therefore account for their joint distribution. Appendix~\ref{app:intro-first-stage} bounds this error through the distribution of the exposure vector entering outcomes.

\section{Empirical Applications}
\label{sec:applications}\label{sec:workflow}

We apply our framework to randomized experiments on classroom seating and business meeting groups. In the seating application, we consider assignments that encourage friendships or connect lower-achieving students with higher-achieving classmates. In the business application, we consider how grouping firms can improve revenue, either for all firms or specifically for small firms.
We compare three assignment approaches under common assignment constraints: \emph{Random assignment}, which draws uniformly from feasible assignments; a \emph{Score-based policy}, which implements the score-based approach in Section~\ref{subsec:empirical-maximization}; and an \emph{Unrestricted policy}, which implements unrestricted empirical maximization from Section~\ref{subsec:empirical-maximization}.

\subsection{Classroom Seating}
\label{sec:hungary-application}

For our first application, we study classroom seating policies that aim to increase mutual friendships or improve lower-achieving students' access to higher-achieving classmates. We use data from the randomized seating experiment studied by \citet{RohrerKellerElwert2021} and \citet{KellerElwert2023}. The experiment randomly assigned students to deskmates within classrooms for one semester and measured their friendships afterward. Our sample includes 3,109 students in 182 classrooms across 40 schools, with 26,840 observed within-class pairs. Each classroom is a separate decision environment. Feasible charts pair as many students as possible within each classroom, leaving one unmatched when the number of students is odd.

\paragraph{Friendship formation and welfare.}
The post-assignment network records mutual friendships at the end of the semester. For an assignment $a$, let $x_{ij}(a)$ indicate whether students $i$ and $j$ are deskmates. We model each pair's friendship probability as
\begin{equation}
\label{eq:hungary-formation}
\Pr\{F^+_{ij}=1\mid H_c,x_{ij}(a)\}
=\logistic\{\eta_1x_{ij}(a)+\eta_2'Z_{ij}(H_c)+\eta_3'x_{ij}(a)Z^{int}_{ij}(H_c)\},
\end{equation}
where $Z_{ij}(H_c)$ includes gender agreement, pair ethnicity, mean baseline grade, absolute grade difference, and log classroom size, while $Z^{int}_{ij}(H_c)$ contains the baseline covariates selected for interaction with deskmate status $x_{ij}(a)$: gender agreement and absolute grade difference.

Estimating \eqref{eq:hungary-formation} gives friendship probabilities $\widehat p^0_{c,ij}$ for students seated apart and $\widehat p^1_{c,ij}$ for deskmates in classroom $c$. Their difference, $\widehat\delta_{c,ij}=\widehat p^1_{c,ij}-\widehat p^0_{c,ij}$, is the predicted increment from seating the pair together. Writing $n_c$ for classroom size, our first objective, friendship count, measures the expected number
of mutual friends per student under assignment $a$:
\begin{equation}
\label{eq:hungary-value}
\widehat v_{c,T}(a)=\frac{2}{n_c}\left[
\sum_{i<j}\widehat p^0_{c,ij}+\sum_{\{i,j\}\in a}\widehat\delta_{c,ij}\right].
\end{equation}
The first sum evaluates all pairs as seated apart, and the second adds the selected deskmates' increments. Each friendship link is counted for both students.

The second objective, academic access, values friendships with higher-achieving classmates. Let $Ach_i$ denote student $i$'s baseline achievement, measured by the average of teacher-assigned grades recorded before the experiment.
Among students with observed baseline achievement, let $L_c$ contain those at or below the classroom median and write $\ell_c=|L_c|$. These sets contain 1,711 target students in total. The access weight increases with the baseline achievement gap, with $x_+=\max\{x,0\}$ denoting the positive part:
\[
\omega^A_{c,ij}=\mathbf1\{i\in L_c\}(Ach_j-Ach_i)_+
+\mathbf1\{j\in L_c\}(Ach_i-Ach_j)_+.
\]
Expected academic access per target student is
\[
\widehat v_{c,A}(a)=\frac{1}{\ell_c}\left[
\sum_{i<j}\omega^A_{c,ij}\widehat p^0_{c,ij}
+\sum_{\{i,j\}\in a}\omega^A_{c,ij}\widehat\delta_{c,ij}\right].
\]
This payoff measures access to higher-achieving friends in grade-gap units. Students with missing grades remain eligible for assignment, with zero access weight on their pairs.

\paragraph{Assignment rules.}
For the friendship-count objective, the score uses three pair-level features: gender agreement, mixed ethnicity, and the negative absolute difference in baseline grades. For academic access, the score uses
the access weight and its interactions with these features. We select the coefficients from a finite grid to maximize average fitted welfare (see Appendix~\ref{app:hungary-application-details} for details). Both score-based and unrestricted policies use maximum-weight matching.

Panel A of Table~\ref{tab:revised-hungary-headline} reports the selected score coefficients. The friendship-count rule favors pairs of the same gender and with similar grades and assigns a negative coefficient to mixed ethnicity. The access rule places its largest coefficient on the access weight, with positive coefficients on its interactions with gender agreement and negative grade distance. Random assignment draws uniformly from maximum-cardinality matchings within each classroom.

\begin{table}[!t]\centering
\caption{Classroom seating: score rules and fitted welfare}
\label{tab:revised-hungary-headline}
\begin{minipage}{\textwidth}
\renewcommand{\baselinestretch}{1.0}\small
\setlength{\tabcolsep}{4pt}
\captionsetup[subtable]{font=small,justification=centering,singlelinecheck=false,skip=3pt}
\begin{subtable}[t]{\linewidth}\centering
\caption*{\textit{Panel A: Normalized score coefficients}}
\begin{tabularx}{\linewidth}{@{}>{\raggedright\arraybackslash}Xcc@{}}
\toprule
Score feature & Friendship count & Academic access \\
\midrule
Gender agreement & 0.82 & -- \\
Mixed ethnicity & -0.41 & -- \\
Negative grade distance & 0.41 & -- \\
Access weight & -- & 0.82 \\
Access $\times$ gender agreement & -- & 0.41 \\
Access $\times$ mixed ethnicity & -- & 0.00 \\
Access $\times$ negative grade distance & -- & 0.41 \\
\bottomrule
\end{tabularx}
\end{subtable}\par\vspace{5pt}
\begin{subtable}[t]{\linewidth}\centering
\caption*{\textit{Panel B: Fitted welfare levels}}
\begin{tabularx}{\linewidth}{@{}>{\raggedright\arraybackslash}Xccc@{}}
\toprule
Policy & Objective & \shortstack{Mutual friends\\per student} & \shortstack{Academic access\\per target} \\
\midrule
Random assignment & -- & 2.67 & 1.38 \\
Score-based policy & Friendship count & 2.71 & 1.36 \\
Score-based policy & Academic access & 2.70 & 1.44 \\
Unrestricted policy & Friendship count & 2.71 & 1.38 \\
Unrestricted policy & Academic access & 2.70 & 1.44 \\
\bottomrule
\end{tabularx}
\end{subtable}\par\vspace{5pt}
\begin{subtable}[t]{\linewidth}\centering
\caption*{\textit{Panel C: Improvements over random assignment}}
\begin{tabularx}{\linewidth}{@{}>{\raggedright\arraybackslash}Xccc@{}}
\toprule
Policy & Objective & \shortstack{Gain per 100\\students or targets} & \shortstack{90\%\\interval} \\
\midrule
Score-based policy & Friendship count & 3.93 & [-0.51, 8.38] \\
Score-based policy & Academic access & 5.99 & [2.33, 9.64] \\
Unrestricted policy & Friendship count & 4.17 & -- \\
Unrestricted policy & Academic access & 6.09 & -- \\
\bottomrule
\end{tabularx}
\end{subtable}\par\vspace{5pt}
{\footnotesize\raggedright\noindent Notes: Panel C reports gains relative to random assignment, per 100 students for friendship count and per 100 target students for academic access. Brackets give nominal 90\% bootstrap confidence intervals. Appendix~\ref{app:empirical-inference} describes the inference procedure.\par}
\end{minipage}\end{table}

\paragraph{Welfare comparisons.}
Panels B and C of Table~\ref{tab:revised-hungary-headline} report the welfare under three policies. Relative to random assignment, the score-based policies yield fitted gains of 3.93 mutual friends per 100 students for friendship count and 5.99 grade-gap units per 100 target students for academic access. The nominal 90\% interval lies above zero for the access gain and includes zero for the friendship-count gain. Unrestricted policies yield slightly larger fitted gains on both objectives. For academic access, the gain is 6.09 grade-gap units per 100 target students, compared with 5.99 under the score-based policy.

A rule learned to improve one objective may either raise or lower welfare under the other. The score-based friendship-count rule lowers fitted academic access by 2.24 grade-gap units per 100 target students relative to random assignment; the interval for this contrast lies below zero. The access rule raises both fitted outcomes, although its friendship-count gain is smaller than that of the count rule. The intervals for both gains lie above zero. Appendix Table~\ref{tab:seating-cross-improvements} reports the gains and intervals for both outcomes.

\paragraph{Sensitivity to the formation specification.}
The alternative formation models imply similar fitted gains but can favor substantially different seating charts. Under the quadratic-grade logit, the reselected access rule shares 47.20\% of its selected pairs with the main rule. Evaluating both rules under this specification gives fitted gains of 6.01 grade-gap units per 100 target students for the main rule and 6.10 for the reselected rule. Appendix Table~\ref{tab:seating-functional-form} compares the gains and assignments across specifications.

\paragraph{Implementation costs.}
Reassigning students may improve friendships, but breaking existing deskmate pairs can impose implementation costs. 
To study the effect of implementation costs, we therefore evaluate reassignment relative to the recorded chart, holding its pair count fixed in each classroom. Let $a_c^0$ be that chart, $m_c^0=|a_c^0|$, and $d_c(a)=|a_c^0\setminus a|/m_c^0$ the fraction of recorded pairs broken. We model the net welfare as
\[
\widehat v_{c,T}(a)-\kappa d_c(a),\qquad |a|=m_c^0,
\]
where $\kappa$ is measured in friendships per student. We augment the pair score with an indicator for retaining a recorded pair and choose the coefficients on baseline pair characteristics and pair retention jointly to maximize fitted net welfare at each value of $\kappa$. 

\begin{figure}[t!]
\centering
\caption{Friendship gains and seating disruption}
\label{fig:hungary-implementation-frontier}
\includegraphics[width=0.72\linewidth]{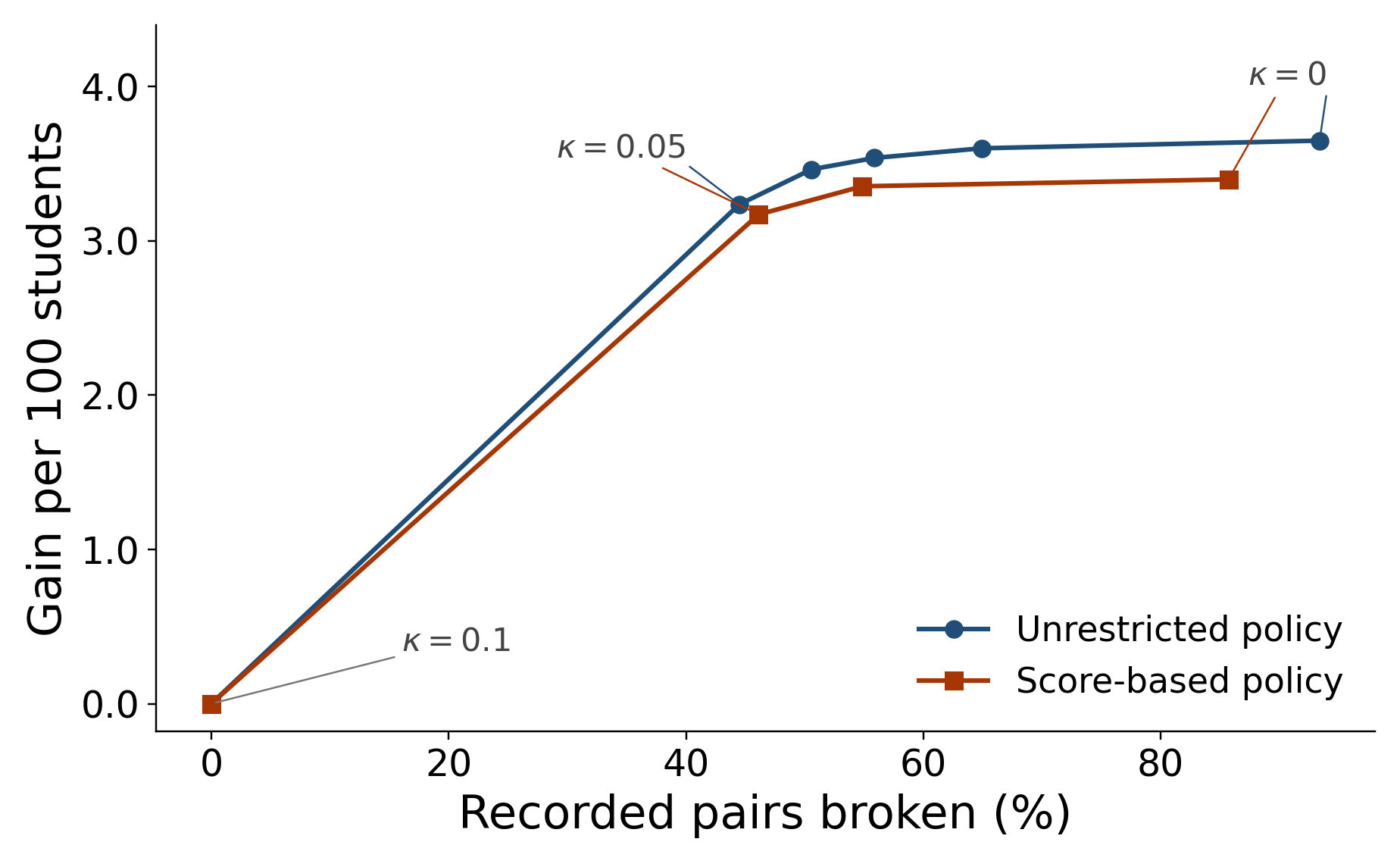}
\par\vspace{4pt}
{\footnotesize\raggedright\noindent Notes: Gains are mutual friends per 100 students relative to the recorded chart. \par}
\end{figure}

Figure~\ref{fig:hungary-implementation-frontier} plots the percentage of recorded pairs broken against the fitted gross friendship gain relative to the recorded chart. Each point represents a policy selected at a given implementation cost, with separate curves for score-based and unrestricted policies. Much of the fitted friendship gain can be retained while reducing seating disruption. At $\kappa=0.05$, the score-based policy preserves more than half of the recorded pairs while retaining 93.31\% of its zero-cost gain. The unrestricted policy exhibits a similar trade-off. At $\kappa\geq 0.10$, both policies preserve the recorded chart. Appendix~\ref{app:hungary-application-details} reports the full results and selected coefficients across implementation costs.

\subsection{Business Groups}
\label{sec:cs-application}

For our second application, we study how assigning firms to business meeting groups can improve revenue. We use the experiment of \citet{CaiSzeidl2018}, which organized firm owner-managers into recurring meeting groups to study business relationships and firm performance. The assignment sample contains 1,409 firms in 141 groups across 21 counties. We reallocate firms within counties, preserving the number of groups of each observed size, so each county forms a separate decision environment.

\paragraph{Outcome response and welfare.}
Let $a$ partition the firms in county $c$ into meeting groups, with $G_i(a)$ denoting firm $i$'s group. Let $Y_{ic}(a)$ be follow-up log revenue under $a$. Given baseline county information $H_c$, we specify
\begin{equation}
\label{eq:cs-outcome-response}
\E[Y_{ic}(a)\mid H_c]
=\alpha_{s(i)}+X_i'\gamma+x_i\{G_i(a)\}'\beta.
\end{equation}
Here $\alpha_{s(i)}$ is the fixed effect for firm $i$'s
randomization stratum in the original experiment, defined by county, firm-size category, and sector. Firm controls $X_i$ include log baseline employment and revenue, owner age, and indicators for a male owner, a college degree or higher, and government work experience. The vector $x_i(G)$ contains firm $i$'s group-composition exposures in group $G$.

To define $x_i(G)$, we first classify firms by baseline
employment size. Small firms have employment at or below the county median among firms originally assigned to meeting groups; large firms lie above it.
For firm $i$ in group $G$, peers are the other members of $G$. The six components of $x_i(G)$ are the shares of large peers, peers in the same sector, and peers whose owners have government work experience; the within-group standard deviation of log baseline employment; the log of mean baseline employment among peers; and group size.
These exposures are determined by assignment, so no separate relationship-formation model is needed.

We consider two welfare objectives: average expected log
revenue across all firms and among small firms only.
Let $w_i$ denote the welfare weight for firm $i$.
All-firm welfare sets $w_i=1$ for every firm, while
small-firm welfare sets $w_i=1$ for small firms and zero
otherwise. Under assignment $a$, county welfare is
\begin{equation}
\label{eq:cs-group-value}
v_{c,w}(a)
=\frac{1}{\sum_i w_i}
\sum_{G\in a}\sum_{i\in G}
w_i\{\alpha_{s(i)}+X_i'\gamma+x_i(G)'\beta\}.
\end{equation}
We evaluate county welfare using the estimated response
coefficients and then average equally across counties.
Appendix Table~\ref{tab:cs-response-slopes} reports
the coefficient estimates.

\paragraph{Assignment rules.}
For all-firm welfare, the score uses three group-composition features: same-sector peer share, employment dispersion, and log mean peer employment. The county averages of the other three exposures are invariant to reassignment and therefore do not affect welfare comparisons between assignments under the linear outcome model. For small-firm welfare, all six features enter the score because their averages among small firms can vary with reassignment. We standardize the features and select coefficients from a finite grid to maximize average fitted welfare. Both score-based and unrestricted policies use exhaustive enumeration for small two-group problems and mixed-integer nonlinear programming otherwise. Appendix~\ref{app:cs-application-details} provides the computational details.

Panel A of Table~\ref{tab:revised-china-business-headline} describes the resulting score coefficients. The all-firm rule favors groups with similar employment levels: its coefficient on employment dispersion is negative, while the other two are zero. The small-firm rule places its largest absolute coefficient on government-experience peer share, with a negative sign, and assigns a positive coefficient to log mean peer employment. Random assignment provides the benchmark by assigning equal probability to all feasible within-county groupings with the observed group sizes.

\begin{table}[!t]\centering
\caption{Business groups: score rules and fitted welfare}
\label{tab:revised-china-business-headline}
\begin{minipage}{\textwidth}
\renewcommand{\baselinestretch}{1.0}\small
\setlength{\tabcolsep}{4pt}
\captionsetup[subtable]{font=small,justification=centering,singlelinecheck=false,skip=3pt}
\begin{subtable}[t]{\linewidth}\centering
\caption*{\textit{Panel A: Normalized score coefficients}}
\begin{tabularx}{\linewidth}{@{}>{\raggedright\arraybackslash}Xcc@{}}
\toprule
Score feature & All firms & Small firms \\
\midrule
Large-peer share & Invariant & -0.22 \\
Same-sector peer share & 0.00 & -0.14 \\
Group SD of log employment & -1.00 & -0.04 \\
Log mean peer employment & 0.00 & 0.56 \\
Government-experience peer share & Invariant & -0.77 \\
Group size & Invariant & 0.18 \\
\bottomrule
\end{tabularx}
\end{subtable}\par\vspace{5pt}
\begin{subtable}[t]{\linewidth}\centering
\caption*{\textit{Panel B: Fitted welfare levels}}
\begin{tabularx}{\linewidth}{@{}>{\raggedright\arraybackslash}Xcccc@{}}
\toprule
Policy & Objective & All firms & Small firms & Large firms \\
\midrule
Random assignment & -- & 5.81 & 5.06 & 6.67 \\
Score-based policy & All firms & 5.92 & 5.04 & 6.91 \\
Score-based policy & Small firms & 5.86 & 5.25 & 6.55 \\
Unrestricted policy & All firms & 5.97 & 5.20 & 6.83 \\
Unrestricted policy & Small firms & 5.89 & 5.31 & 6.55 \\
\bottomrule
\end{tabularx}
\end{subtable}\par\vspace{5pt}
\begin{subtable}[t]{\linewidth}\centering
\caption*{\textit{Panel C: Improvements over random assignment}}
\begin{tabularx}{\linewidth}{@{}>{\raggedright\arraybackslash}Xccc@{}}
\toprule
Policy & Objective & \shortstack{Geometric-mean\\revenue gain (\%)} & \shortstack{90\%\\interval} \\
\midrule
Score-based policy & All firms & 10.65 & [-15.53, 44.93] \\
Score-based policy & Small firms & 21.80 & [-7.24, 59.95] \\
Unrestricted policy & All firms & 16.59 & -- \\
Unrestricted policy & Small firms & 28.35 & -- \\
\bottomrule
\end{tabularx}
\end{subtable}\par\vspace{5pt}
{\footnotesize\raggedright\noindent Notes: In Panel A, `Invariant' denotes an exposure whose all-firm county average is unchanged by reassignment. Panel B reports mean log revenue; Panel C reports percentage changes in geometric-mean revenue relative to random assignment. Brackets give nominal 90\% bootstrap confidence intervals. Appendix~\ref{app:empirical-inference} describes the inference procedure.\par}
\end{minipage}\end{table}

\paragraph{Welfare comparisons.}
Table~\ref{tab:revised-china-business-headline} reports fitted welfare levels in Panel B and gains relative to random assignment in Panel C. The score-based gains are 10.65\% for all firms and 21.80\% for small firms under their respective objectives. Both nominal 90\% intervals include zero. Unrestricted policies attain larger fitted gains, including 16.59\% for all-firm welfare.

To understand the sources of the fitted welfare gains, we decompose each rule's gain into contributions from changes in the six group-composition exposures relative to random assignment. Each contribution is the change in average exposure multiplied by its estimated response coefficient. Appendix Table~\ref{tab:cs-score-decomposition} reports this decomposition. For the all-firm rule, lower employment dispersion increases fitted welfare, but the accompanying reduction in mean peer employment partly offsets this gain. For the small-firm rule, the largest positive contributions come from reduced exposure to large peers and peers with government work experience.

The welfare effects of assignment policies can differ substantially across firm sizes. The score-based rule for the all-firm objective lowers fitted geometric-mean revenue for small firms by 1.87\% relative to random assignment, but raises it for large firms by 26.79\%. Targeting small firms instead raises fitted geometric-mean revenue for small firms by 21.80\%, but lowers it for large firms by 11.44\%. Appendix Table~\ref{tab:business-cross-improvements} reports the complete comparisons. These findings suggest that policymakers may also wish to consider distributional consequences when evaluating assignment policies.

\paragraph{Sensitivity to the response specification.}
The alternative response specifications yield positive fitted welfare gains, although their magnitudes and the selected assignments can differ. Under the additive quadratic specification, which adds squared exposure terms to \eqref{eq:cs-outcome-response}, the all-firm assignments remain unchanged, with a fitted gain of 43.53\% compared with 10.65\% under the linear model. For the small-firm objective, reselecting the rule under the quadratic specification yields different groupings with higher fitted welfare under that model. Appendix Table~\ref{tab:business-functional-form} provides the full comparisons.
\FloatBarrier

\section{Conclusion}
\label{sec:conclusion}

Network-shaping policies determine assignments whose effects depend on the relationships that form afterward. This paper evaluates those assignments by integrating an outcome response over a network-formation distribution. We characterize policy-value identification under the maintained models and construct a plug-in value estimator. The regret analysis relates policy performance to first-stage estimation error and sampling variation, with separate contributions from numerical integration and optimization. Policy-class restrictions and model misspecification have distinct effects on welfare. Applications to classroom seating and business groups illustrate how the planner's objective changes the preferred assignments and the distribution of predicted gains.

The framework suggests several directions for future research. First, policy learning could be extended to settings where policy values are partially identified. Observationally equivalent formation and outcome models may imply different welfare values and rankings across feasible assignments. \citet{ChristensenMoonSchorfheide2026} develop asymptotically optimal statistical decision rules for discrete choices with partially identified payoffs. Building on this approach, future work could characterize when optimal assignment policies remain identified and develop minimax risk or regret learning procedures for constrained, network-shaping assignments.

Second, one could study experimental design for learning welfare-maximizing assignment rules. Choosing randomization probabilities over feasible pairings or groupings to reduce the expected regret of the subsequently selected policy would provide a decision-theoretic basis for assignment experiments (see, e.g., \citealp{You2026} for decision-theoretic experimental design for the subsequent policy choice). Such designs could concentrate limited experimental variation on pair characteristics and group compositions that distinguish promising policies, subject to the implementation constraints.

Third, a dynamic extension could study repeated assignments when relationships persist across periods. Maintaining existing ties and creating new contacts would then have different implications for current welfare and future opportunities for interaction. The planner would account for these continuation values and, when formation and outcome responses are unknown, the information generated by earlier assignments. Central questions are when continuation values preserve the additive structure that permits optimization, and how estimation errors affect welfare over dynamic decisions.

\putbib
\end{bibunit}

\clearpage
\renewcommand{\thepage}{A.\arabic{page}}
\setcounter{page}{1}

\begin{bibunit}
\begin{appendix}
\markright{Online Appendix -- This Version: \draftdate }

\renewcommand{\theequation}{\thesection.\arabic{equation}}
\setcounter{equation}{0}
\renewcommand*\thetable{A-\arabic{table}}
\setcounter{table}{0}
\renewcommand*\thefigure{A-\arabic{figure}}
\setcounter{figure}{0}
\renewcommand*\thetheorem{\thesection.\arabic{theorem}}
\setcounter{theorem}{0}

\begin{center}
{\large {\bf Online Appendix: Policy Learning with Network-Shaping Assignments}}

{\bf Zhenxiao Chen and Zhiheng You}
\end{center}

\thispagestyle{empty}

\noindent This appendix consists of the following sections:

\begin{itemize}
\item[A.] Proofs
\item[B.] Network Introductions
\item[C.] Policy Learning: Computation and Error Bounds
\item[D.] Empirical Applications: Estimation and Supplementary Results
\end{itemize}

\newpage
\section{Proofs}
\label{app:proofs}

\subsection{Proof of \texorpdfstring{Proposition~\ref{prop:identification}}{Proposition~\getrefnumber{prop:identification}}}

\begin{proof}
Fix a full-data law compatible with $P_O$ and let $(\eta,\gamma)\in\Ical(P_O)$ be its formation and outcome parameters. Here $Q_{\eta,i,a}(\cdot\mid h)$ denotes the distribution of the exposure $T_i(h,G)$ induced by $G\sim K_\eta(\cdot\mid h,a)$. By the maintained outcome-response condition and iterated expectations,
\[
\begin{aligned}
v_0(h,a)
&=\frac1{M_h}\sum_{i=1}^{M_h}
\E[Y_{ic}(a)\mid H_c=h]\\
&=\int \frac1{M_h}\sum_{i=1}^{M_h}
 m_\gamma\{W_i(h),T_i(h,g)\}K_\eta(dg\mid h,a)\\
&=\frac1{M_h}\sum_{i=1}^{M_h}
\int m_\gamma\{W_i(h),t\}Q_{\eta,i,a}(dt\mid h).
\end{aligned}
\]
Every compatible full-data law therefore implies an action value in \eqref{eq:conditional-identified-set}. Conversely, the definition of $\Ical(P_O)$ ensures that each parameter pair in this set is compatible with at least one such law, so every value in the displayed set is attainable. This establishes sharpness. Because the marginal distribution of $H_c$ is observed, evaluating each compatible model at $a=\pi(H_c)$ and averaging gives the sharp policy-value set in \eqref{eq:policy-identified-set}. A value is point identified exactly when its identified set contains a single element, which gives \eqref{eq:exact-action-identification}, \eqref{eq:exact-policy-identification}, and \eqref{eq:identification-value}.
\end{proof}

\subsection{Proof of \texorpdfstring{Proposition~\ref{prop:approx-bias}}{Proposition~\getrefnumber{prop:approx-bias}}}

\begin{proof}
For part (i), add and subtract the policy values computed using the approximate model:
\[
\begin{aligned}
V_0(\pi_\Pclass^*)-V_0(\widehat\pi)
=&[V_0(\pi_\Pclass^*)-\widetilde V(\pi_\Pclass^*)]
+[\widetilde V(\pi_\Pclass^*)-\widetilde V(\widetilde\pi^*)]\\
&+[\widetilde V(\widetilde\pi^*)-\widetilde V(\widehat\pi)]
+[\widetilde V(\widehat\pi)-V_0(\widehat\pi)].
\end{aligned}
\]
The second term is nonpositive because $\widetilde\pi^*$ maximizes $\widetilde V$ over $\Pclass$, and the first and fourth terms are each bounded above by $\Delta(\Pclass)$.

Part (ii) follows by applying part (i) with $\widetilde v=\widetilde v_K$ and $\widehat\pi=\widetilde\pi_K^*$, so the regret under the approximate value is zero and $\Delta(\Pclass)=\mathcal D_K(\Pclass)$. Under the total-variation condition, for any $\pi\in\Pclass$,
\[
V_0(\pi)-\widetilde V_K(\pi)
=
\E\left[
\int \mu_0(H_c,g)\{K_0-\widetilde K\}(dg\mid H_c,\pi(H_c))
\right],
\]
and, using $|\int f\,d(P-Q)|\le 2\|f\|_\infty\TV(P,Q)$,
\[
\left|\int \mu_0(h,g)(K_0-\widetilde K)(dg\mid h,a)\right|
\le
2\bar B\TV\{K_0(\cdot\mid h,a),\widetilde K(\cdot\mid h,a)\}
\le2\bar B\delta_K.
\]
Thus $\mathcal D_K(\Pclass)\le2\bar B\delta_K$, which gives the displayed $4\bar B\delta_K$ bound.

Part (iii) similarly applies part (i) with $\widetilde v=\widetilde v_m$ and $\widehat\pi=\widetilde\pi_m^*$, yielding the $2\mathcal D_m(\Pclass)$ bound. For any $\pi\in\Pclass$,
\[
\begin{aligned}
V_0(\pi)-\widetilde V_m(\pi)
=\E\Bigg[\int \frac1{M_{H_c}}\sum_i
&\{m_{\gamma_0}(W_i(H_c),T_i(H_c,g))\\
&\quad-\widetilde m(W_i(H_c),T_i(H_c,g))\}
K_0(dg\mid H_c,\pi(H_c))\Bigg].
\end{aligned}
\]
The uniform average response-error condition implies $\mathcal D_m(\Pclass)\le\delta_m$, and hence $V_0(\pi_\Pclass^*)-V_0(\widetilde\pi_m^*)\le2\delta_m$.
\end{proof}

\subsection{Proof of \texorpdfstring{Theorem~\ref{thm:main-regret-decomposition}}{Theorem~\getrefnumber{thm:main-regret-decomposition}}}

\begin{proof}
\allowdisplaybreaks[1]
The proof first bounds regret by sampling and value-estimation errors, then bounds the first-stage and simulation contributions to value estimation. It concludes with the sampling and simulation rates. Write $v_\pi(h)=v_0(h,\pi(h))$, $P_Cf=C^{-1}\sum_{c=1}^C f(H_c)$, and $Pf=\E[f(H_c)]$, so that $Pv_\pi=V_0(\pi)$.

\emph{Regret inequality.}
The approximate maximization condition gives
\[
\widehat V_C(\pi_\Pclass^*)-\widehat V_C(\widehat\pi)
\le \sup_{\pi\in\Pclass}\widehat V_C(\pi)-\widehat V_C(\widehat\pi)
\le \eps_{opt,C}.
\]
By the definitions of $s_C$ and $r_{v,C}$,
\begin{align*}
V_0(\pi_\Pclass^*)-V_0(\widehat\pi)
={}&[V_0(\pi_\Pclass^*)-P_Cv_{\pi_\Pclass^*}]
   +[P_Cv_{\pi_\Pclass^*}-\widehat V_C(\pi_\Pclass^*)]\\
 &+[\widehat V_C(\pi_\Pclass^*)-\widehat V_C(\widehat\pi)]\\
 &+[\widehat V_C(\widehat\pi)-P_Cv_{\widehat\pi}]
   +[P_Cv_{\widehat\pi}-V_0(\widehat\pi)]\\
\le{}&2s_C+2r_{v,C}+\eps_{opt,C}.
\end{align*}

\emph{Value-estimation error.}
For each cluster $c$ and action $a$,
\begin{align*}
\widehat v_c(a)-v_0(H_c,a)
={}&\widehat v_c(a)-\widehat v_c^{int}(a)\\
 &+\int[\widehat\mu(H_c,g)-\mu_0(H_c,g)]\widehat K(dg\mid H_c,a)\\
 &+\int\mu_0(H_c,g)(\widehat K-K_0)(dg\mid H_c,a).
\end{align*}
Writing $a_c=\pi(H_c)$, the triangle inequality yields
\begin{align*}
r_{v,C}
\le{}&\sup_{\pi\in\Pclass}
 \left|\frac1C\sum_{c=1}^C[\widehat v_c(a_c)-\widehat v_c^{int}(a_c)]\right|\\
 &+\sup_{\pi\in\Pclass}\frac1C\sum_{c=1}^C
 \int|\widehat\mu(H_c,g)-\mu_0(H_c,g)|\widehat K(dg\mid H_c,a_c)\\
 &+\sup_{\pi\in\Pclass}\left|\frac1C\sum_{c=1}^C
 \int\mu_0(H_c,g)(\widehat K-K_0)(dg\mid H_c,a_c)\right|\\
={}&\mathcal E_{MC,C}+\mathcal E_{\mu,C}+\mathcal E_{K,C}.
\end{align*}
Consequently,
\[
V_0(\pi_\Pclass^*)-V_0(\widehat\pi)
\le 2s_C+2\mathcal E_{\mu,C}+2\mathcal E_{K,C}
      +2\mathcal E_{MC,C}+\eps_{opt,C},
\]
which proves \eqref{eq:main-regret}.

\emph{Sampling error.}
We first bound the complexity of the induced value-function class by counting how many different action choices the scores can generate. The score in Section~\ref{subsec:score-based-design} can be written as $\theta'\varphi_C(h,a)$, where
\[
\varphi_C(h,a)=\sum_{r\in\mathcal R(h)}T_r(h,a)\xi_r(h).
\]
Thus $\pi_\theta(h)\in\argmax_{a\in\A(h)}\theta'\varphi_C(h,a)$, with the stated rule for breaking ties.

Fix $n$ baseline environments $h_1,\ldots,h_n$ and thresholds $y_1,\ldots,y_n$. To bound the VC-subgraph dimension, consider the indicator that the selected action's value is at least the threshold:
\[
\ind\{v_0(h_\ell,\pi_\theta(h_\ell))\ge y_\ell\}.
\]
For a fixed $h_\ell$, the selected action can change only when two actions tie in score, that is, when
\[
\theta'[\varphi_C(h_\ell,a)-\varphi_C(h_\ell,a')]=0
\]
for some pair $a,a'\in\A(h_\ell)$. Across the $n$ environments the nonzero score differences define at most $N=nK_C(K_C-1)/2$ hyperplanes in $\R^{d_C}$. The signs of the score differences, including zero signs, determine the maximizing actions and the fixed ordering determines the selected action. Hence all subgraph labels are constant on each face of the hyperplane arrangement, including its lower-dimensional faces.

An arrangement of $N$ hyperplanes in $\R^{d_C}$ has at most
\[
\sum_{j=0}^{d_C}2^j\binom Nj
\]
faces, with $\binom Nj=0$ for $j>N$. This bound follows by adding hyperplanes successively: the number of faces is at most the previous count plus twice the number induced on the new hyperplane, and induction gives the displayed sum. If $N\ge d_C$, then
\[
\sum_{j=0}^{d_C}2^j\binom Nj
\le 2^{d_C}\sum_{j=0}^{d_C}\binom Nj
\le \left(\frac{2eN}{d_C}\right)^{d_C}
\le \left(\frac{enK_C^2}{d_C}\right)^{d_C}.
\]
Shattering $n$ observations therefore requires
\[
2^n\le\left(\frac{enK_C^2}{d_C}\right)^{d_C}.
\]
Set $x=n/d_C$. Taking logarithms gives $x\log2\le1+\log x+2\log K_C$. Since $\log x\le x\log2/2$ for $x\ge4$, this implies $x\le 2(1+2\log K_C)/\log2$ whenever $x\ge4$. Thus $n$ is bounded by a universal constant times $d_C\log(eK_C)$. If $K_C\ge2$ and $N<d_C$, then $n<2d_C/[K_C(K_C-1)]\le d_C$, which satisfies the same bound. If $K_C=1$, the value class is a singleton. These cases imply that the VC-subgraph index is at most $A_0d_C\log(eK_C)$ for a universal constant $A_0$.

Assumption~\ref{ass:iid-clusters} and the value representation imply that the functions in $\mathcal F_C$ are bounded by $\bar B$.
If $\bar B=0$, then $s_C=0$; suppose $\bar B>0$.
Pointwise measurability and boundedness imply
\[
s_C=\sup_{f\in\mathcal F_C}|(P_C-P)f|
   =\sup_{f\in\mathcal F_C^0}|(P_C-P)f|.
\]
Let $H_1',\ldots,H_C'$ be an independent copy of the sample, and let
$\varepsilon_1,\ldots,\varepsilon_C$ be independent Rademacher signs,
independent of both samples. Symmetrization gives
\begin{align*}
\E s_C
&\le \E\sup_{f\in\mathcal F_C}
 \left|\frac1C\sum_{c=1}^C\{f(H_c)-f(H_c')\}\right|\\
&=\E\sup_{f\in\mathcal F_C}
 \left|\frac1C\sum_{c=1}^C\varepsilon_c\{f(H_c)-f(H_c')\}\right|\\
&\le 2\E_H\E_\varepsilon\sup_{f\in\mathcal F_C}
 \left|\frac1C\sum_{c=1}^C\varepsilon_cf(H_c)\right|.
\end{align*}
Here $\E_\varepsilon$ conditions on the sample. Let $N(t,\mathcal F_C,L_2(Q))$ be the covering number: the minimum number of $L_2(Q)$ balls of radius $t$ needed to cover $\mathcal F_C$.
The VC-subgraph index bound above and the entropy bound for VC-subgraph classes \citep{vanDerVaartWellner1996} imply
\[
\sup_Q\log N(u\bar B,\mathcal F_C,L_2(Q))
\le A_2d_C\log(eK_C)\log(A_1/u),\qquad 0<u\le1,
\]
where $Q$ ranges over finitely supported probability measures and
$A_1\ge e$ and $A_2>0$ are independent of $C$.
The conditional Rademacher maximal inequality and this entropy bound
\citep{vanDerVaartWellner1996} give
\begin{align*}
\E_\varepsilon\sup_{f\in\mathcal F_C}
 \left|\frac1C\sum_{c=1}^C\varepsilon_cf(H_c)\right|
&\le \frac{A_3\bar B}{\sqrt C}
 \int_0^1\sqrt{1+\log N(u\bar B,\mathcal F_C,L_2(P_C))}\,du\\
&\le A_4\bar B\sqrt{\frac{d_C\log(eK_C)}{C}}
 \int_0^1\sqrt{\log(A_1/u)}\,du\\
&\le A_4\bar B\sqrt{\frac{d_C\log(eK_C)}{C}}
 \left(\sqrt{\log A_1}+\frac{\sqrt\pi}{2}\right).
\end{align*}
The last inequality uses
$\sqrt{\log(A_1/u)}\le\sqrt{\log A_1}+\sqrt{\log(1/u)}$
and $\int_0^1\sqrt{\log(1/u)}\,du=\sqrt\pi/2$.
Thus, for a constant $A_5$ independent of $C$,
\[
\E s_C\le A_5\bar B\sqrt{\frac{d_C\log(eK_C)}{C}}.
\]
For every $t>0$, Markov's inequality gives
\[
\Pbb\!\left(s_C>t\bar B\sqrt{\frac{d_C\log(eK_C)}{C}}\right)
\le\frac{\E s_C}{t\bar B\sqrt{d_C\log(eK_C)/C}}
\le\frac{A_5}{t}.
\]
Hence $s_C=O_p(\bar B\sqrt{d_C\log(eK_C)/C})$, which proves \eqref{eq:uniform-bound}.

\emph{Simulation error.}
We bound the largest simulation error within each cluster and then average across clusters. This avoids requiring independence of simulation errors across clusters or actions. Let $\mathcal D_C=(O_1,\ldots,O_C)$. For $\widehat v_c=\widehat v_{c,S}$ as defined in Section~\ref{subsec:general-estimation-recipe}, write
\[
X_{c,a}^{(s)}=\widehat\mu(H_c,G_{c,a}^{(s)}),
\qquad
e_{c,S}(a)=\widehat v_{c,S}(a)-\widehat v_c^{int}(a).
\]
Here $G_{c,a}^{(s)}$ denotes the $s$th network draw for cluster $c$ and action $a$. Conditional on $\mathcal D_C$, the variables $X_{c,a}^{(s)}$ are independent across $s$, lie in $[-\bar B,\bar B]$, and have mean $\widehat v_c^{int}(a)$. Hoeffding's lemma gives, for every $\lambda\in\R$,
\begin{align*}
\E[e^{\lambda e_{c,S}(a)}\mid\mathcal D_C]
&=\prod_{s=1}^S
\E\!\left[
\exp\!\left\{\frac{\lambda}{S}
(X_{c,a}^{(s)}-\widehat v_c^{int}(a))\right\}
\,\middle|\,\mathcal D_C\right]\\
&\le \exp\!\left(\frac{\lambda^2\bar B^2}{2S}\right).
\end{align*}
Let $Z_c=\max_{a\in\A(H_c)}|e_{c,S}(a)|$. For $\lambda>0$, Jensen's inequality gives
\begin{align*}
\E[Z_c\mid\mathcal D_C]
&\le\frac1\lambda\log\E[e^{\lambda Z_c}\mid\mathcal D_C]\\
&\le\frac1\lambda\log\sum_{a\in\A(H_c)}
\E[e^{\lambda e_{c,S}(a)}+e^{-\lambda e_{c,S}(a)}\mid\mathcal D_C]\\
&\le\frac{\log(2K_C)}{\lambda}+\frac{\lambda\bar B^2}{2S}.
\end{align*}
If $\bar B=0$, every simulation error is zero. Otherwise, minimizing the last expression over $\lambda>0$ gives
\[
\E[Z_c\mid\mathcal D_C]
\le\bar B\sqrt{\frac{2\log(2K_C)}{S}}.
\]
The triangle inequality gives
\[
\mathcal E_{MC,C}
=\sup_{\pi\in\Pclass}
\left|\frac1C\sum_{c=1}^C e_{c,S}\{\pi(H_c)\}\right|
\le\frac1C\sum_{c=1}^C Z_c.
\]
Taking conditional expectations and averaging the preceding bound proves
\[
\E[\mathcal E_{MC,C}\mid\mathcal D_C]
\le\frac1C\sum_{c=1}^C\E[Z_c\mid\mathcal D_C]
\le\bar B\sqrt{\frac{2\log(2K_C)}{S}}.
\]
For $t>0$, conditional Markov's inequality and $\log(2K_C)\le\log(eK_C)$ imply
\[
\Pbb\!\left(
\mathcal E_{MC,C}>t\bar B\sqrt{\frac{\log(eK_C)}{S}}
\,\middle|\,\mathcal D_C\right)
\le\frac{\sqrt2}{t}.
\]
Taking expectations over $\mathcal D_C$ proves \eqref{eq:simulation-rate}. With exact integration, $\widehat v_c=\widehat v_c^{int}$, so \eqref{eq:E-MC-main} gives $\mathcal E_{MC,C}=0$.
\end{proof}

\subsection{Proof of \texorpdfstring{Proposition~\ref{prop:first-stage-sensitivity}}{Proposition~\getrefnumber{prop:first-stage-sensitivity}}}
\label{app:first-stage-sensitivity-proof}

\begin{proof}
Recall from \eqref{eq:E-mu-main} that the outcome-response error is an average of nonnegative integrated absolute errors. Fix any policy $\pi\in\Pclass$. Because $\widehat\mu=\mu_{\widehat\gamma}$ and both $\widehat\gamma$ and $\gamma_0$ lie in the neighborhood specified in Assumption~\ref{ass:policy-relevant-sensitivity}, \eqref{eq:outcome-local-lipschitz} implies, pointwise on the network realizations relevant to the policy class,
\[
\left|
\widehat\mu(H_c,g)-\mu_0(H_c,g)
\right|
\le
L_\mu(H_c,g)\|\widehat\gamma-\gamma_0\|.
\]
Integrating this inequality with respect to the probability measure $\widehat K(\cdot\mid H_c,\pi(H_c))$ and averaging over clusters yields
\[
\begin{aligned}
&\frac1C\sum_{c=1}^C
\int
\left|
\widehat\mu(H_c,g)-\mu_0(H_c,g)
\right|
\widehat K(dg\mid H_c,\pi(H_c))\\
&\qquad\le
\frac1C\sum_{c=1}^C
\|\widehat\gamma-\gamma_0\|
\int L_\mu(H_c,g)
\widehat K(dg\mid H_c,\pi(H_c))=
\frac1C\sum_{c=1}^C
\Gamma_{\mu,c}(\pi(H_c))
\|\widehat\gamma-\gamma_0\|.
\end{aligned}
\]
Taking the supremum over $\pi\in\Pclass$ proves the inequality in \eqref{eq:E-mu-sensitivity}. Since $\|\widehat\gamma-\gamma_0\|=r_{\gamma,C}$ does not depend on $c$,
\[
\begin{aligned}
\mathcal E_{\mu,C}
&\le
r_{\gamma,C}
\sup_{\pi\in\Pclass}\frac1C\sum_{c=1}^C
\Gamma_{\mu,c}(\pi(H_c))
=
\Lambda_{\mu,C}r_{\gamma,C},
\end{aligned}
\]
which gives the stated bound.

For the formation-law term, fix $\pi\in\Pclass$ and define
\[
d_{K,c}(\pi)
=
\int \mu_0(H_c,g)
(\widehat K-K_0)(dg\mid H_c,\pi(H_c)).
\]
The triangle inequality gives
\[
\left|\frac1C\sum_{c=1}^C d_{K,c}(\pi)\right|
\le
\frac1C\sum_{c=1}^C |d_{K,c}(\pi)|.
\]
Because $\widehat K=K_{\widehat\eta}$ and $K_0=K_{\eta_0}$, and because $\widehat\eta$ and $\eta_0$ lie in the neighborhood in Assumption~\ref{ass:policy-relevant-sensitivity}, \eqref{eq:formation-local-lipschitz} implies
\[
|d_{K,c}(\pi)|
\le
L_K(H_c,\pi(H_c))
\|\widehat\eta-\eta_0\|
=
\Gamma_{K,c}(\pi(H_c))
\|\widehat\eta-\eta_0\|.
\]
Taking the supremum over $\pi$ gives the inequality in \eqref{eq:E-K-sensitivity}. Factoring out $r_{\eta,C}=\|\widehat\eta-\eta_0\|$ yields
\[
\mathcal E_{K,C}
\le
r_{\eta,C}
\sup_{\pi\in\Pclass}\frac1C\sum_{c=1}^C
\Gamma_{K,c}(\pi(H_c))
=
\Lambda_{K,C}r_{\eta,C},
\]
which gives the stated bound in \eqref{eq:E-K-sensitivity}.

Finally, the regret decomposition in Section~\ref{subsec:empirical-maximization} gives
\[
V_0(\pi_\Pclass^*)-V_0(\widehat\pi)
\le
2s_C+2\mathcal E_{\mu,C}+2\mathcal E_{K,C}
+2\mathcal E_{MC,C}+\eps_{opt,C}.
\]
Substitution of the two bounds just proved gives the displayed regret inequality in Proposition~\ref{prop:first-stage-sensitivity}. When the action-value integral is evaluated exactly, $\mathcal E_{MC,C}=0$.
\end{proof}

\clearpage
\section{Network Introductions}
\label{app:network-introductions}
\label{subsec:introductions}

This appendix develops network introductions as a third example of the framework. A planner selects pairs of agents to create opportunities for contact. The realized contacts can affect outcomes both individually and through interactions with an agent's other contacts, so the example illustrates how pairwise assignments can generate nonadditive welfare.

Let $I_h$ be the set of agents. The planner chooses introductions subject to a total limit and a maximum number involving any one agent:
\begin{equation*}
\A^{intro}(h)=\left\{a\subseteq \binom{I_h}{2}: |a|\le \bar L_h,
\quad r_i(a)=\sum_{j\neq i}\ind[\{i,j\}\in a]\le \bar r_h\ \text{for all }i\in I_h\right\}.
\end{equation*}
Here $\bar L_h$ limits the total number of introductions, $\bar r_h$ limits the number involving each agent, and $r_i(a)$ records agent $i$'s assigned introductions. Experiments on social and business interactions motivate this setting \citep{FeigenbergFieldPande2013,CaiSzeidl2018,FafchampsQuinn2018,VasilakyLeonard2018}.

The post-action network is $G^{intro}=(a,R^+)$, where $R_{ij}^+\in\{0,1\}$ records whether agents $i$ and $j$ have contact after assignment. Let $\iota_{ij}(a)=\ind[\{i,j\}\in a]$. A model for contact formation is
\begin{equation}
\label{eq:intro-link-probability}
\begin{aligned}
\Pbb_\eta(R_{ij}^+(a)=1\mid h,a)&=p_{ij}^R(h,a),\\
p_{ij}^R(h,a)&=\logistic\{\eta_0^R+\eta_1^R\iota_{ij}(a)+\eta_2^R R_{ij}^0(h)+\eta_3^R\mathrm{Complement}_{ij}(h)\\
&\hspace{1.0in}-\eta_4^R\mathrm{Distance}_{ij}(h)-\eta_5^R r_i(a)-\eta_6^R r_j(a)\}.
\end{aligned}
\end{equation}
The variable $R_{ij}^0(h)$ records prior contact, $\mathrm{Complement}_{ij}(h)$ measures complementarities in resources or markets, and $\mathrm{Distance}_{ij}(h)$ measures social or geographic distance. The introduction indicator changes the contact probability, while $r_i(a)$ and $r_j(a)$ allow that probability to depend on how many introductions each agent receives. Contacts can also form between agents who are not introduced. Conditional independence gives
\begin{equation}
\label{eq:intro-formation-law}
K_\eta^{intro}(dR^+\mid h,a)
=
\prod_{\{i,j\}\subset I_h}
\big[p_{ij}^R(h,a)\big]^{R_{ij}^+}
\big[1-p_{ij}^R(h,a)\big]^{1-R_{ij}^+}.
\end{equation}
Independent Bernoulli contacts provide a convenient baseline. A joint formation law can instead allow common shocks or dependence among contacts involving the same agent.

The outcome response depends on both individual contacts and combinations of contacts. Let $X_{ij}(h)$ collect the relevant baseline pair characteristics, and define two weighted summaries of realized contacts:
\begin{equation}
\label{eq:intro-exposure-map}
\begin{aligned}
D_i^R&=\sum_{j\ne i}R_{ij}^+\nu_\gamma\{X_{ij}(h)\},
&
B_i^R&=\sum_{\{j,k\}\subset I_h\setminus\{i\}}R_{ij}^+R_{ik}^+\kappa_\gamma\{X_{ij}(h),X_{ik}(h)\}.
\end{aligned}
\end{equation}
The term $D_i^R$ sums individual contacts with weights $\nu_\gamma$ that depend on contact type. The term $B_i^R$ sums pairs of contacts, with $\kappa_\gamma$ specifying their interaction weights. Each product $R_{ij}^+R_{ik}^+$ equals one only when both contacts form.

An individual outcome response that incorporates contact benefits and the cost of receiving multiple introductions is
\begin{equation}
\label{eq:intro-outcome-response}
m_\gamma^{intro}\{W_i(h),T_i^{intro}(h,G^{intro})\}
=
\alpha_\gamma^{intro}(h)+\gamma_W'W_i(h)+\phi_\gamma(D_i^R,B_i^R)-\gamma_r r_i(a).
\end{equation}
Here $\phi_\gamma$ maps the two contact summaries into the outcome, and $\gamma_r$ measures the cost of additional introductions. At the true response, write the coefficients as $\gamma_{W,0}$ and $\gamma_{r,0}$ and the functions as $\nu_0$, $\kappa_0$, and $\phi_0$. Averaging over agents gives
\begin{equation*}
\begin{aligned}
&\frac1{|I_h|}\sum_{i\in I_h}
m_\gamma^{intro}\{W_i(h),T_i^{intro}(h,G^{intro})\}\\
&\qquad=
\alpha_\gamma^{intro}(h)
+\frac1{|I_h|}\sum_{i\in I_h}
\left[
\gamma_W'W_i(h)+\phi_\gamma(D_i^R,B_i^R)-\gamma_r r_i(a)
\right].
\end{aligned}
\end{equation*}

Integrating this average response over the contact-formation law gives the value of an introduction action. Two features can prevent additivity across selected introductions. First, $r_i(a)$ and $r_j(a)$ allow other assigned introductions to change the probability of contact between $i$ and $j$. Second, the interaction summary $B_i^R$ and nonlinearities in $\phi_\gamma$ allow the value of one contact to depend on other realized contacts. The assignment problem therefore need not reduce to maximum-weight matching, as discussed in Section~\ref{subsec:oracle-reductions}.

An introduction experiment can randomize feasible assignments to vary which types of pairs are introduced and how many introductions each agent receives. The resulting contacts and outcomes provide information for estimating the formation and outcome models used to evaluate candidate policies.

\clearpage
\section{Policy Learning: Computation and Error Bounds}
\label{app:policy-learning-details}

This section supplements Section~\ref{sec:learning}. Appendix~\ref{app:implementation-map} explains how to compute fitted action values. Appendices~\ref{app:deskmate-first-stage}--\ref{app:intro-first-stage} derive the first-stage bounds for the running examples. Appendix~\ref{app:nonregular-first-stage} then extends the analysis to formation and outcome models for which uniform Lipschitz bounds are unavailable.

\subsection{Value Calculations}
\label{app:implementation-map}

The specifications in Section~\ref{sec:examples} and Appendix~\ref{app:network-introductions} determine how to compute the plug-in values in \eqref{eq:estimated-value}. Terms that are constant across assignments can be omitted when selecting a policy and restored when reporting welfare levels.

\paragraph{Deskmate seating.}
Estimate the friendship probabilities in \eqref{eq:deskmate-link-probability} and the friendship contributions to outcomes in \eqref{eq:deskmate-outcome-response}. For each feasible pair, the estimated welfare gain from seating its members together is
\begin{equation*}
\widehat\Delta_{ij}^{desk}(h)
=
\frac{1}{|I_h|}\widehat\omega_{ij}^F(h)
\left\{\widehat p_{ij}^{F,1}(h)-\widehat p_{ij}^{F,0}(h)\right\}
-c_{ij}(h).
\end{equation*}
Using these weights in \eqref{eq:deskmate-matching-ip} gives the unrestricted assignment under the fitted additive value. When the response includes nonlinear network features, the simulation procedure in Section~\ref{subsec:general-estimation-recipe} can instead evaluate each candidate assignment.

\paragraph{Mentor assignments.}
For each feasible group $B$ and mentor $j$, substitute the estimated mentor-quality function, peer-link probabilities, and outcome coefficients into \eqref{eq:mentor-cell-value-main}, with the assignment cost included as specified there. The resulting $\widehat{\mathcal W}_j(B;h)$ is the objective coefficient in \eqref{eq:mentor-set-partition-ip}. Summing over the selected groups gives fitted welfare up to the assignment-invariant baseline. For responses nonlinear in the realized peer network, draw networks from the fitted joint formation law and average the predicted outcomes as in \eqref{eq:estimated-value}.

\paragraph{Network introductions.}
Estimate the formation and outcome models in \eqref{eq:intro-link-probability}--\eqref{eq:intro-outcome-response}. Under conditionally independent contacts and $\phi(d,b)=d+b$, the value integral is analytic: replace each $R_{ij}^+$ by its fitted probability and each $R_{ij}^+R_{ik}^+$ by the product of the fitted probabilities. For more general specifications, draw complete contact networks from the fitted joint formation law, compute $(D_i^R,B_i^R)$ and the predicted individual outcomes for each draw, and average over agents and draws. Drawing the full network preserves the dependence needed to evaluate interactions between contacts.

\subsection{Bounds for Deskmate Seating}
\label{app:deskmate-first-stage}

The seating calculation first removes the value common to all assignments and then bounds errors in the gains from selected pairs. Integrating the pair-additive response over friendship formation gives
\[
v_0^{desk}(h,a)
=
\alpha_0^{desk}(h)
+
\frac1{|I_h|}\sum_{\{i,j\}\subset I_h}
\omega_{ij}^F(h)p_{ij}^{F,x_{ij}(a)}(h).
\]
The assignment-invariant baseline is
\[
b_0^{desk}(h)
=
\alpha_0^{desk}(h)
+
\frac1{|I_h|}\sum_{\{i,j\}\subset I_h}
\omega_{ij}^F(h)p_{ij}^{F,0}(h),
\]
and define $\widehat b^{desk}(h)$ analogously with fitted quantities. Since $x_{ij}(a)=1$ exactly for selected pairs,
\[
v_0^{desk}(h,a)-b_0^{desk}(h)
=
\frac1{|I_h|}\sum_{\{i,j\}\in a}
\omega_{ij}^F(h)\delta_{ij}^F(h),
\]
The fitted value satisfies the corresponding identity. Adding and subtracting $\omega_{ij}^F\widehat\delta_{ij}^F$ in each selected pair's contribution attributes its estimation error to the friendship contribution and the probability increment. The formation component gives \eqref{eq:deskmate-EK-specialization}.

For the logistic formation model, the derivative bound converts coefficient error into probability error. Let $\boldsymbol\eta_{F,0}=(\eta_0^F,\ldots,\eta_5^F)'$ collect the true coefficients and let $\widehat{\boldsymbol\eta}_F$ be their estimate. The regressor associated with seating status $x$ is
\[
Z_{ij}^{F,x}(h)
=
\big(1,x,F_{ij}^0,|\mathrm{Ach}_i-\mathrm{Ach}_j|,
\mathrm{Cl}_{ij}^0,xF_{ij}^0\big)'.
\]
Set
\[
B_{Z,F}=\sup_{h,i,j,x\in\{0,1\}}\|Z_{ij}^{F,x}(h)\|_*,
\qquad
r_{\eta_F,C}=\|\widehat{\boldsymbol\eta}_F-\boldsymbol\eta_{F,0}\|,
\qquad
\bar\omega_F=\sup_{h,i,j}|\omega_{ij}^F(h)|.
\]
The mean-value theorem and $\sup_u\logistic(u)\{1-\logistic(u)\}=1/4$ imply
\[
|\widehat p_{ij}^{F,x}-p_{ij}^{F,x}|
\le
\frac14\|Z_{ij}^{F,x}(h)\|_*
\|\widehat{\boldsymbol\eta}_F-\boldsymbol\eta_{F,0}\|.
\]
Hence
\[
|\widehat\delta_{ij}^F-\delta_{ij}^F|
\le \frac{B_{Z,F}}2r_{\eta_F,C}.
\]
Because a matching contains at most $|I_h|/2$ pairs, the right-hand side of \eqref{eq:deskmate-EK-specialization} is at most $\bar\omega_FB_{Z,F}r_{\eta_F,C}/4$. Thus the matching constraint limits the number of probability errors that contribute to each assignment's welfare error.

\subsection{Bounds for Mentor Assignments}
\label{app:mentor-first-stage}

For mentor assignments, formation error affects the expected peer-spillover terms, whereas outcome-response error can also affect direct mentoring benefits. Conditional on $(h,a)$, the direct mentoring terms are deterministic and only the peer links are random. In particular,
\[
\E_K[\mathrm{Co}_{ii'}\mid h,a]
=
p_{ii'j}^{\mathrm{Co}}(h,n_j(a))
\quad\text{when }i,i'\in I_j(a).
\]
Let $\widehat p_{ii'j}^{\mathrm{Co}}(h,n)$ be the fitted probability and define
\begin{equation}
\label{eq:mentor-EK-specialization}
\Delta_K^{ment}(h,a)
=
\frac{\beta_{\mathrm{Co}}}{|J_h|}
\sum_{j\in\mathcal M_h}
\sum_{\{i,i'\}\subset I_j(a)}
S_{ii'}(h)
\left\{
\widehat p_{ii'j}^{\mathrm{Co}}(h,n_j(a))
-p_{ii'j}^{\mathrm{Co}}(h,n_j(a))
\right\}.
\end{equation}
Averaging this change in value across clusters and taking the largest absolute deviation over candidate policies gives the formation-error term in Section~\ref{subsec:running-example-first-stage}:
\[
\mathcal E_{K,C}^{ment}
=
\sup_{\pi\in\Pclass}
\left|\frac1C\sum_{c=1}^C
\Delta_K^{ment}\{H_c,\pi(H_c)\}\right|.
\]
Define
\begin{align*}
b_{\max,C}
&=\max_{1\le c\le C}\max_{j\in\mathcal M_{H_c}}b_j(H_c),
\quad
\bar S
=\sup_{h,i,i'}|S_{ii'}(h)|,\\
\delta_{\mathrm{Co},C}
&=\max_{1\le c\le C}
\max_{j\in\mathcal M_{H_c}}
\max_{1\le n\le b_j(H_c)}
\max_{\{i,i'\}\subset J_{H_c}}
\left|
\widehat p_{ii'j}^{\mathrm{Co}}(H_c,n)
-p_{ii'j}^{\mathrm{Co}}(H_c,n)
\right|.
\end{align*}
The absolute value of \eqref{eq:mentor-EK-specialization} is at most
\[
\frac{|\beta_{\mathrm{Co}}|\bar S\delta_{\mathrm{Co},C}}{|J_h|}
\sum_j\binom{n_j(a)}2 \le
\frac{|\beta_{\mathrm{Co}}|\bar S\delta_{\mathrm{Co},C}}{|J_h|}
\sum_j\frac{n_j(a)(b_{\max,C}-1)}2 =
\frac{|\beta_{\mathrm{Co}}|\bar S}{2}
(b_{\max,C}-1)\delta_{\mathrm{Co},C},
\]
because $\sum_jn_j(a)=|J_h|$. Taking the policy supremum and averaging over clusters proves \eqref{eq:mentor-capacity-EK}.

For outcome-response error, consider estimation of $\chi_j$, the parameter governing how mentoring quality declines with load. Its effect is controlled by
\[
\left|\frac{\partial q_j(n;h)}{\partial\chi_j}\right|
=
\frac{q_j^0(h)(n-1)}{\{1+\chi_j(n-1)\}^2}
\le q_j^0(h)(n-1).
\]
Let $\mathcal E_{\mu,C}^{ment,\chi}$ denote the part of the mentor outcome-response term due only to replacing $\chi_j$ by its estimate $\widehat\chi_j$. The mean-value theorem and the triangle inequality give
\begin{equation*}
\mathcal E_{\mu,C}^{ment,\chi}
\le
\sup_{\pi\in\Pclass}\frac1C\sum_{c=1}^C
\frac1{|J_{H_c}|}
\sum_{j\in\mathcal M_{H_c}}
\sum_{i\in I_j(\pi(H_c))}
|\beta_q+\beta_{q\lambda}\lambda_{ij}(H_c)|
q_j^0(H_c)
\{n_j(\pi(H_c))-1\}
|\widehat\chi_j-\chi_j|.
\end{equation*}
Errors in the remaining outcome coefficients can be bounded in the same way, using the magnitudes of the covariates and exposures multiplying them.

\subsection{Bounds for Network Introductions}
\label{app:intro-first-stage}

For network introductions, formation error changes the distribution of both individual contacts and pairs of contacts. The baseline $\alpha_0^{intro}(h)+\gamma_{W,0}'W_i(h)$ and the introduction count are fixed conditional on $(h,a)$. Holding the outcome model fixed, replacing the true formation law by its estimate therefore changes expected welfare only through $\phi_0(D_i^R,B_i^R)$. Define
\begin{align*}
\Delta_K^{intro}(h,a)
&=
\frac1{|I_h|}\sum_{i\in I_h}
\left[
\int \phi_0(D_i^R,B_i^R)\widehat K^{intro}(dR^+\mid h,a)
-
\int \phi_0(D_i^R,B_i^R)K_0^{intro}(dR^+\mid h,a)
\right],
\\
\mathcal E_{K,C}^{intro}
&=
\sup_{\pi\in\Pclass}
\left|\frac1C\sum_{c=1}^C
\Delta_K^{intro}\{H_c,\pi(H_c)\}\right|.
\end{align*}

For nonlinear $\phi_0$, compare draws $R'\sim\widehat K^{intro}(\cdot\mid h,a)$ and $R\sim K_0^{intro}(\cdot\mid h,a)$ under a coupling $\Gamma$, meaning a joint distribution with these fitted and true marginals. Define $D_i^R(r)$ and $B_i^R(r)$ using the true functions $\nu_0$ and $\kappa_0$ at contact vector $r$. If $\phi_0$ is $L_\phi$-Lipschitz under the $\ell_1$ metric, then
\begin{align*}
|\Delta_K^{intro}(h,a)|
&\le
\frac{L_\phi}{|I_h|}\sum_{i\in I_h}
\E_\Gamma\left[
|D_i^R(R')-D_i^R(R)|
+
|B_i^R(R')-B_i^R(R)|
\right].
\end{align*}
Minimizing over couplings gives the smallest expected difference in the outcome-relevant exposure vectors. This bound also applies when contact formation is dependent.

When $\phi_0(d,b)=d+b$, the response is linear in the two contact summaries but still includes interactions between realized contacts. The random part of average welfare is
\[
\frac1{|I_h|}\sum_{i\in I_h}
\left[
\sum_{j\ne i}\nu_0\{X_{ij}(h)\}R_{ij}^+
+
\sum_{\{j,k\}\subset I_h\setminus\{i\}}
\kappa_0\{X_{ij}(h),X_{ik}(h)\}R_{ij}^+R_{ik}^+
\right].
\]
In addition to the marginal probability $p_{ij}^R(h,a)=\E[R_{ij}^+\mid h,a]$, define the probability that both contacts form:
\begin{equation*}
q_{ijk}^R(h,a)
=
\E_{K_0^{intro}}[R_{ij}^+R_{ik}^+\mid h,a],
\qquad j<k,\quad j,k\ne i,
\end{equation*}
Let $\widehat q_{ijk}^R(h,a)$ be its fitted counterpart. Suppressing the common arguments $(h,a)$, the error in expected welfare due to formation estimation is
\begin{equation*}
\Delta_K^{intro}(h,a) =
\frac1{|I_h|}\sum_{i\in I_h}
\Bigg[
\sum_{j\ne i}\nu_0\{X_{ij}(h)\}
(\widehat p_{ij}^R-p_{ij}^R)
+
\sum_{\{j,k\}\subset I_h\setminus\{i\}}
\kappa_0\{X_{ij}(h),X_{ik}(h)\}
(\widehat q_{ijk}^R-q_{ijk}^R)
\Bigg].
\end{equation*}
If contacts are conditionally independent under both the true and fitted formation laws, then $q_{ijk}^R=p_{ij}^Rp_{ik}^R$ and
\[
|\widehat q_{ijk}^R-q_{ijk}^R|
\le
|\widehat p_{ij}^R-p_{ij}^R|
+
|\widehat p_{ik}^R-p_{ik}^R|.
\]
Consequently,
\begin{align*}
|\Delta_K^{intro}(h,a)|
&\le
\frac1{|I_h|}\sum_{i\in I_h}
\Bigg[
\sum_{j\ne i}|\nu_0\{X_{ij}(h)\}|\,e_{ij}^R(h,a)
\\
&\qquad+
\sum_{\{j,k\}\subset I_h\setminus\{i\}}
|\kappa_0\{X_{ij}(h),X_{ik}(h)\}|
\{e_{ij}^R(h,a)+e_{ik}^R(h,a)\}
\Bigg],
\end{align*}
Here $e_{ij}^R(h,a)=|\widehat p_{ij}^R(h,a)-p_{ij}^R(h,a)|$ is the error in a marginal contact probability. The introduction limit $r_i(a)\le\bar r_h$ does not by itself limit realized contacts, because contacts can form without an introduction. Reducing the numbers of individual-contact and contact-interaction terms to $O(\bar r_h)$ and $O(\bar r_h^2)$ therefore requires an additional sparsity restriction on the contacts entering outcomes.

For outcome-response error, let $\widehat D_i^R$ and $\widehat B_i^R$ use the estimated weights $\widehat\nu$ and $\widehat\kappa$ on a realized contact vector. The error in $\phi$ evaluated at these fitted summaries is
\[
\delta_\phi(h,a)
=
\frac1{|I_h|}\sum_{i\in I_h}
\int
\left|
\widehat\phi(\widehat D_i^R,\widehat B_i^R)
-
\phi_0(\widehat D_i^R,\widehat B_i^R)
\right|
\widehat K^{intro}(dR^+\mid h,a).
\]
After subtracting the assignment-invariant baseline, the remaining outcome error comes from estimating $\phi$, the contact weights, and the introduction cost. If $\phi_0$ is $L_\phi$-Lipschitz, its integrated absolute value at $(h,a)$ is bounded by
\begin{align*}
&\delta_\phi(h,a)
+
\frac{L_\phi}{|I_h|}\sum_{i\in I_h}
\int
\Bigg[
\sum_{j\ne i}R_{ij}^+
|\widehat\nu\{X_{ij}(h)\}-\nu_0\{X_{ij}(h)\}|
\\
&\qquad
+
\sum_{\{j,k\}\subset I_h\setminus\{i\}}
R_{ij}^+R_{ik}^+
|\widehat\kappa\{X_{ij}(h),X_{ik}(h)\}
-\kappa_0\{X_{ij}(h),X_{ik}(h)\}|
\Bigg]
\widehat K^{intro}(dR^+\mid h,a)
\\
&\qquad+
\bar r_h|\widehat\gamma_r-\gamma_{r,0}|.
\end{align*}
The bound weights errors in $\nu$ and $\kappa$ by the fitted probabilities of individual and joint contacts. The introduction limit controls the contribution of error in the cost coefficient $\gamma_r$.

\subsection{Regret Bounds under Weaker Regularity}
\label{app:nonregular-first-stage}

The preceding calculations bound how errors in formation and outcome inputs affect policy values. The regret decomposition in Section~\ref{subsec:empirical-maximization} also applies when uniform local Lipschitz bounds are unavailable. In that case, moduli of continuity measure how much policy value can change within a parameter neighborhood of radius $t$. Define
\begin{align}
\omega_{\mu,C}(t)
&=
\sup_{\pi\in\Pclass}\frac1C\sum_{c=1}^C
\int
\sup_{\|\gamma-\gamma_0\|\le t}
|\mu_\gamma(H_c,g)-\mu_0(H_c,g)|
\widehat K(dg\mid H_c,\pi(H_c)),
\label{eq:nonregular-outcome-modulus}\\
\omega_{K,C}(t)
&=
\sup_{\pi\in\Pclass}\frac1C\sum_{c=1}^C
\sup_{\|\eta-\eta_0\|\le t}
\left|
\int \mu_0(H_c,g)
\{K_\eta-K_{\eta_0}\}(dg\mid H_c,\pi(H_c))
\right|.
\label{eq:nonregular-formation-modulus}
\end{align}
If $\|\widehat\gamma-\gamma_0\|\le r_{\gamma,C}$ and $\|\widehat\eta-\eta_0\|\le r_{\eta,C}$, then substitution in \eqref{eq:nonregular-outcome-modulus} and the triangle inequality in \eqref{eq:nonregular-formation-modulus} give
\begin{equation*}
\mathcal E_{\mu,C}\le \omega_{\mu,C}(r_{\gamma,C}),
\qquad
\mathcal E_{K,C}\le \omega_{K,C}(r_{\eta,C}).
\end{equation*}
Under Assumption~\ref{ass:policy-relevant-sensitivity}, these moduli are bounded by linear functions of $t$. The cases below instead bound changes away from thresholds or singularities and control how often candidate policies approach them. The required margin conditions limit that probability or its sample analogue.

\paragraph{Formation models with link thresholds.}
First consider deterministic network formation, so the formation law assigns probability one to the selected network:
\[
K_\eta(dg\mid h,a)=\delta_{g_\eta(h,a)}(dg).
\]
Suppose the selected pairwise-stable network is determined by the signs of finitely many inequalities $s_q(\eta;h,a)$, indexed by $q\in\mathcal Q(h,a)$. These inequalities describe incentives to form or remove links and, when needed, the choice among stable networks. Assume
\[
|s_q(\eta;h,a)-s_q(\eta_0;h,a)|
\le B_q(h,a)\|\eta-\eta_0\|
\]
locally, and that preserving all signs preserves the selected network. Define the smallest absolute value of these inequalities and the largest bound on their local changes:
\[
m_K(h,a)=\min_{q\in\mathcal Q(h,a)}|s_q(\eta_0;h,a)|,
\qquad
B_K(h,a)=\max_{q\in\mathcal Q(h,a)}B_q(h,a).
\]
Then $g_\eta(h,a)=g_{\eta_0}(h,a)$ whenever $B_K(h,a)\|\eta-\eta_0\|<m_K(h,a)$. Let $\mathcal G(h,a)$ denote the union of network realizations that can be selected in the parameter neighborhood and define
\[
\Delta_\mu(h,a)
=
\sup_{g,g'\in\mathcal G(h,a)}
|\mu_0(h,g)-\mu_0(h,g')|,
\]
which is at most $2\bar B$ under the bounded-value condition. It follows that
\begin{equation}
\sup_{\|\eta-\eta_0\|\le t}
\left|
\int\mu_0(h,g)\{K_\eta-K_{\eta_0}\}(dg\mid h,a)
\right|
\le
\Delta_\mu(h,a)
\ind[m_K(h,a)\le B_K(h,a)t].
\label{eq:threshold-formation-pointwise}
\end{equation}
Consequently,
\begin{equation}
\label{eq:threshold-formation-sample}
\mathcal E_{K,C}
\le
\sup_{\pi\in\Pclass}\frac1C\sum_{c=1}^C
\Delta_\mu(H_c,\pi(H_c))
\ind\!
\left[
 m_K(H_c,\pi(H_c))
 \le B_K(H_c,\pi(H_c))r_{\eta,C}
\right].
\end{equation}
The bound depends on how often a link or selection inequality is close to zero, where a small parameter error can change the selected network. If $\Delta_\mu\le\bar\Delta$, $B_K\le\bar B_K$, and, for some $\alpha_K>0$,
\begin{equation}
\label{eq:formation-margin-condition}
\sup_{\pi\in\Pclass}\frac1C\sum_{c=1}^C
\ind[m_K(H_c,\pi(H_c))\le q]
=O_p(q^{\alpha_K})
\end{equation}
for the relevant sequences $q\downarrow0$, then
\[
\mathcal E_{K,C}=O_p(r_{\eta,C}^{\alpha_K}).
\]
If every relevant inequality stays uniformly away from zero, so $m_K(h,a)\ge\underline m>0$, the formation-error term is zero once $\bar B_Kr_{\eta,C}<\underline m$. If a nonvanishing fraction of cases under candidate policies instead has $m_K=0$ and the competing networks have different values, the bound in \eqref{eq:threshold-formation-sample} need not converge to zero.

The two-agent specialization of the connections model in \citet{JacksonWolinsky1996} gives the simplest example. Let $g\in\{0,1\}$ denote the link and let its net surplus be $s_\eta=b-c$. Under a deterministic tie-breaking rule,
\[
g_\eta=\ind[b-c\ge0].
\]
If the link changes cluster value by $\Delta(h)$, then
\[
\int\mu_0(h,g)K_\eta(dg)-
\int\mu_0(h,g)K_{\eta_0}(dg)
=
\Delta(h)
\{\ind[b-c\ge0]-\ind[b_0-c_0\ge0]\}.
\]
At $b_0=c_0$, this difference can equal $\Delta(h)$ for an arbitrarily small parameter perturbation, so no finite local $L_K$ exists. Away from the threshold it is locally zero. With continuously distributed formation shocks, the same calculation averages the indicator in \eqref{eq:threshold-formation-pointwise} over the shocks; a bounded density near zero gives the margin exponent $\alpha_K=1$.

\paragraph{Network responses near the boundary of equilibrium stability.}
Consider the standard linear-quadratic network game underlying \citet{BallesterCalvoArmengolZenou2006}. For a symmetric nonnegative adjacency or interaction matrix $G(g)$, write
\[
x_{\beta,\theta}(h,g)
=
R_\beta(g)b_\theta(h),
\qquad
R_\beta(g)=\{I-\beta G(g)\}^{-1},
\]
and let the cluster response be
\[
\mu_{\beta,\theta}(h,g)
=
\frac1{M_h}\sum_{i=1}^{M_h}\ell\{x_{\beta,\theta,i}(h,g)\},
\]
where $\ell$ is $L_\ell$-Lipschitz. Taking $\ell(u)=u$ gives the usual average equilibrium action; a bounded $\ell$, or truncation to the feasible outcome range, preserves the bounded-value condition used in the regret theorem. Define
\[
\kappa_0(g)=1-\beta_0\rho\{G(g)\},
\qquad
\widehat\kappa(g)=1-\widehat\beta\rho\{G(g)\},
\]
where $\rho(G)$ is the spectral radius and $\|G\|_{op}$ is the operator norm; they coincide for the symmetric matrices considered. The true and estimated interaction parameters are nonnegative. Stable equilibria require $\kappa_0(g)>0$, and values near zero allow the matrix inverse, and hence the outcome response, to be large.

The matrix inverse identity gives
\begin{equation}
\widehat x(h,g)-x_0(h,g)
=
R_{\widehat\beta}(g)
\{\widehat b(h)-b_0(h)\}
+
(\widehat\beta-\beta_0)
R_{\widehat\beta}(g)G(g)R_{\beta_0}(g)b_0(h).
\label{eq:lq-resolvent-expansion}
\end{equation}
Suppose, on the observed environments,
\[
\max_c\frac{\|\widehat b(H_c)-b_0(H_c)\|_2}{\sqrt{M_c}}
\le r_{b,C},
\qquad
|\widehat\beta-\beta_0|\le r_{\beta,C},
\]
$M_c^{-1/2}\|b_0(H_c)\|_2\le\bar b$, and $\rho\{G(g)\}\le\bar\rho$ on the domain relevant to the policy class. If $\kappa_0(g)>q$ and $q>2\bar\rho r_{\beta,C}$, then $\widehat\kappa(g)>q/2$. Applying \eqref{eq:lq-resolvent-expansion}, the operator-norm bounds for the two matrix inverses, and Cauchy--Schwarz to the cluster average yields
\begin{equation*}
|\widehat\mu(H_c,g)-\mu_0(H_c,g)|
\le
2L_\ell
\left\{
\frac{r_{b,C}}q
+
\frac{\bar b\bar\rho r_{\beta,C}}{q^2}
\right\}.
\end{equation*}
On the remaining networks, where $\kappa_0(g)\le q$, suppose both fitted and true responses are bounded by $\bar B$. Their largest average probability under the fitted formation laws and candidate policies is
\begin{equation*}
\mathcal A_{\kappa,C}(q)
=
\sup_{\pi\in\Pclass}\frac1C\sum_{c=1}^C
\widehat K
\left(
\kappa_0(g)\le q
\mid H_c,\pi(H_c)
\right).
\end{equation*}
Splitting the integral in $\mathcal E_{\mu,C}$ at $\kappa_0(g)=q$ gives
\begin{equation}
\label{eq:lq-outcome-regret-bound}
\mathcal E_{\mu,C}
\le
2L_\ell
\left\{
\frac{r_{b,C}}q
+
\frac{\bar b\bar\rho r_{\beta,C}}{q^2}
\right\}
+2\bar B\mathcal A_{\kappa,C}(q),
\qquad
q>2\bar\rho r_{\beta,C}.
\end{equation}
The first two terms bound response error away from the stability boundary; the last term bounds the contribution of networks close to it. If
\[
\mathcal A_{\kappa,C}(q)=O_p(q^{\alpha_\mu})
\]
uniformly over the relevant sequences $q\downarrow0$, and $r_{b,C}$ and $r_{\beta,C}$ are both $O_p(r_C)$, choosing $q$ of order $r_C^{1/(\alpha_\mu+2)}$ gives
\[
\mathcal E_{\mu,C}
=O_p\left(r_C^{\alpha_\mu/(\alpha_\mu+2)}\right).
\]
When $\beta_0$ is known and only $b_\theta$ is estimated, balancing the first and third terms in \eqref{eq:lq-outcome-regret-bound} gives the faster order $r_{b,C}^{\alpha_\mu/(\alpha_\mu+1)}$. Nonvanishing probability at $\kappa_0=0$ can prevent the modulus from converging to zero. Networks with $\kappa_0<0$ are outside the maintained stable-equilibrium region; candidate policies must exclude them or the response specification must be modified to remain well defined.

The same stability threshold governs the effect of formation error. Let $g'\sim\widehat K(\cdot\mid h,a)$ and $g\sim K_0(\cdot\mid h,a)$ have joint distribution $\Gamma$ with the specified marginals. Holding $(\beta_0,b_0)$ fixed, the matrix inverse identity gives
\[
R_{\beta_0}(g')-R_{\beta_0}(g)
=
\beta_0R_{\beta_0}(g')
\{G(g')-G(g)\}
R_{\beta_0}(g).
\]
On the event $\min\{\kappa_0(g'),\kappa_0(g)\}>q$,
\[
|\mu_0(h,g')-\mu_0(h,g)|
\le
\frac{L_\ell\beta_0\bar b}{q^2}
\|G(g')-G(g)\|_{op}.
\]
Therefore, for bounded value contributions and any collection of couplings $\Gamma_{c,a}$,
\begin{align}
\mathcal E_{K,C}
&\le
\sup_{\pi\in\Pclass}\frac1C\sum_{c=1}^C
\E_{\Gamma_{c,\pi(H_c)}}
\Bigg[
\frac{L_\ell\beta_0\bar b}{q^2}
\|G(g')-G(g)\|_{op}
\ind[\min\{\kappa_0(g'),\kappa_0(g)\}>q]
\notag\\
&\hspace{2.35in}
+2\bar B
\ind[\min\{\kappa_0(g'),\kappa_0(g)\}\le q]
\Bigg].
\label{eq:lq-formation-regret-bound}
\end{align}
The smallest bound is obtained by minimizing over joint distributions with the required marginals. Suppose the expected matrix difference on $\min\{\kappa_0(g'),\kappa_0(g)\}>q$ is $O_p(\delta_{G,C})$ and the probability of its complement is $O_p(q^{\alpha_\mu})$, uniformly over policies. Then
\[
\mathcal E_{K,C}
=O_p\left(\delta_{G,C}q^{-2}+q^{\alpha_\mu}\right).
\]
Balancing the terms yields order $\delta_{G,C}^{\alpha_\mu/(\alpha_\mu+2)}$. Formation errors on networks close to the stability boundary can therefore have a larger welfare effect than equally sized errors elsewhere.

\paragraph{Changes in equilibrium selection.}
A similar argument applies when a model admits multiple equilibria. Let $e_0(h,g)$ be the equilibrium selected at the true parameter, and suppose a criterion $S_\gamma(e;h,g)$ is maximized over a finite set of equilibria. The gap between the selected equilibrium's criterion value and the best alternative is
\[
m_e(h,g)
=
S_{\gamma_0}\{e_0(h,g);h,g\}
-
\max_{e\ne e_0(h,g)}S_{\gamma_0}(e;h,g).
\]
If
\[
\sup_e
|S_\gamma(e;h,g)-S_{\gamma_0}(e;h,g)|
\le B_e(h,g)\|\gamma-\gamma_0\|,
\]
then the equilibrium selection can change only if
\[
m_e(h,g)\le 2B_e(h,g)\|\gamma-\gamma_0\|.
\]
Let $\omega_{\mu,C}^{br}(t)$ denote the outcome-error modulus with the selected equilibrium held fixed. With bounded outcomes, the total error is bounded by this term plus the effect of changes in equilibrium selection:
\begin{equation}
\mathcal E_{\mu,C}
\le
\omega_{\mu,C}^{br}(r_{\gamma,C})
+
2\bar B
\sup_{\pi\in\Pclass}\frac1C\sum_{c=1}^C
\int
\ind\!
\left[
 m_e(H_c,g)
 \le 2B_e(H_c,g)r_{\gamma,C}
\right]
\widehat K(dg\mid H_c,\pi(H_c)).
\label{eq:equilibrium-selection-regret-bound}
\end{equation}
A margin condition bounds the probability of a change in equilibrium selection, giving a rate as in \eqref{eq:formation-margin-condition}. If ties have positive probability under a candidate policy and the tied equilibria yield different values, uniformly learning the value associated with a particular selection rule can require further restrictions. An alternative is to maximize the lowest policy value across admissible equilibria. The difference between the upper and lower values then measures uncertainty about equilibrium selection.

Each bound combines estimation error where the model responds smoothly with a bounded loss weighted by the frequency of threshold crossings, near-singular responses, or changes in equilibrium selection. These calculations control the same first-stage terms as the main regret theorem without imposing a uniform Lipschitz bound.

\clearpage
\section{Empirical Applications: Estimation and Supplementary Results}
\label{app:application-details}

This section supplements the empirical applications in Section~\ref{sec:applications}. It first describes simultaneous inference for policy gains, then reports sample construction, estimation, policy computation, and supplementary results for classroom seating and business groups.

\subsection{Simultaneous Inference for Welfare Gains}
\label{app:empirical-inference}

The reported policy is selected from several candidate rules, so confidence intervals are constructed simultaneously over the candidate class. Fix a learning objective $o$ and evaluation outcome $r$, and let $\mathcal J_{o,r}$ index the candidate rules for objective $o$. For rule $j$, $\Delta_j$ is its population welfare gain on outcome $r$ relative to random assignment. The classes contain 66 and 192 rules for friendship count and academic access, respectively, and 26 and 76 rules for all-firm and small-firm welfare. Unrestricted policies and implementation-cost comparisons are reported as point estimates.

\paragraph{Cluster bootstrap.}
Let $\widehat\Delta_j$ be the estimated gain for rule $j$. We use 1,999 cluster-bootstrap replications, resampling the 40 schools for seating and the 21 counties for business groups. In replication $b$, the resampling counts satisfy $M^{(b)}\sim\mathrm{Multinomial}(K;1/K,\ldots,1/K)$, where $K$ is the number of independent clusters. Each replication re-estimates the formation or outcome model and evaluates every candidate relative to random assignment under that model. Business regressions retain the original randomization-stratum fixed effects, and seating imputation means are re-estimated. Score coefficients, feature-standardization constants, and tie-breaking rules are held fixed.

Let $\widehat\Delta_{c,j}^{*(b)}$ be the gain for rule $j$ in county or classroom $c$, evaluated using the model fitted in replication $b$. For business groups, equal county weights give the bootstrap gain $K^{-1}\sum_c M_c^{(b)}\widehat\Delta_{c,j}^{*(b)}$. For seating, let $C_k$ be the set of classrooms in school $k$. Equal classroom weights give
\[
 \widehat\Delta_j^{*(b)}=
 \frac{\sum_k M_k^{(b)}\sum_{c\in C_k}\widehat\Delta_{c,j}^{*(b)}}
 {\sum_k M_k^{(b)}|C_k|}.
\]
The seating population gain is therefore the expected sum of classroom gains within a school divided by the expected number of classrooms per school. Building on the uniform bootstrap inference approach of
\citet[Supplementary Appendix B]{KitagawaTetenov2018},
we construct simultaneous confidence intervals using the
maximum bootstrap deviation across candidate rules, standardized by each rule's bootstrap standard error. Let $\widehat s_j$ be the bootstrap standard deviation of $\widehat\Delta_j^{*(b)}$, and let $q_{o,r}$ be the empirical 90th percentile of
\[
 T_{b,o,r}=\max_{j\in\mathcal J_{o,r}}
 \frac{|\widehat\Delta_j^{*(b)}-\widehat\Delta_j|}{\widehat s_j}.
\]
The interval for candidate $j$ is
\[
 [L_j,U_j]=[\widehat\Delta_j-q_{o,r}\widehat s_j,
             \widehat\Delta_j+q_{o,r}\widehat s_j].
\]
Gains that are identically zero by construction are excluded from the maximum and receive $[0,0]$. For business groups, gains and interval endpoints on the mean log scale are transformed using $100(\exp(x)-1)$.

\paragraph{Asymptotic justification.}
For a fixed finite class of candidate rules, an influence-function expansion describes the joint sampling variation that the bootstrap must reproduce. Let $\eta$ collect the common first-stage parameters, including imputation parameters. Suppose the independent clusters admit regular estimating equations $U_k(\eta)$, with nonsingular $A=-E\partial_\eta U_k(\eta_0)$. For business groups, let $Z_k$ denote the county data, $h_j(Z_k,\eta)$ the county gain for rule $j$, and $G_j=E\partial_\eta h_j(Z_k,\eta_0)$. Regular estimation and a uniformly negligible first-order remainder yield
\[
 \widehat\Delta_j-\Delta_j
 =\frac1K\sum_k\left\{
 h_j(Z_k,\eta_0)-\Delta_j+G_jA^{-1}U_k(\eta_0)
 \right\}+o_p(K^{-1/2}),
\]
uniformly over the finite class. The first term reflects variation in county gains; the second reflects first-stage estimation. The original-stratum fixed effects are removed within counties when estimating the common slopes. For seating, let $T_{k,j}$ sum classroom gains in school $k$ and let $N_k=|C_k|$. The influence function for the ratio defining the seating gain is
\[
 \frac{T_{k,j}-\Delta_jN_k}{EN_k}
 +\frac{E\partial_\eta T_{k,j}}{EN_k}A^{-1}U_k.
\]
Re-estimating the first-stage models in each bootstrap sample captures both components and their covariance when the joint bootstrap approximation is consistent. With positive limiting variances for nondegenerate gains and continuity of the limiting maximum's quantile, this yields asymptotic simultaneous coverage over the fixed candidate class, including a rule selected from that class. The sensitivity tables report the clipped linear-probability and gradient-boosting specifications as point estimates because inference for them requires additional regularity conditions.

\subsection{Classroom Seating}
\suppressfloats[t]
\label{app:hungary-application-details}

\paragraph{Sample and measurement.}
The sample contains 3,109 students in 182 classrooms across 40 schools, forming 26,840 unordered within-class pairs. Baseline achievement is the average of the available spring-2017 teacher-assigned grades. Grades are observed for 2,979 students and missing for 130. The 1,711 students at or below their classroom's median observed grade are the target group for academic access. Students with missing grades remain eligible for matching but contribute zero to the access objective. Formation regressions replace missing grade differences and pair-mean grades with their sample means and include missing-value indicators.

Recorded partner identifiers recover 1,273 deskmate pairs. Of the remaining students, 262 have no partner identifier and 301 name a partner outside the retained sample. The implementation-cost analysis preserves this recorded pair count in each classroom; the main policy comparison uses the maximum feasible number of pairs.

\paragraph{Friendship formation.}
Table~\ref{tab:hungary-formation-coefs} reports the main logit estimates. Counterfactual probabilities are computed by changing seating status and its interactions while holding baseline characteristics fixed.

\begin{table}[htbp]\centering
\caption{Friendship-formation coefficient estimates}
\label{tab:hungary-formation-coefs}
\begin{minipage}{\textwidth}
\renewcommand{\baselinestretch}{1.0}\small
\setlength{\tabcolsep}{4pt}
\captionsetup[subtable]{font=small,justification=centering,singlelinecheck=false,skip=3pt}
\begin{subtable}[t]{\linewidth}\centering
\begin{tabularx}{\linewidth}{@{}>{\raggedright\arraybackslash}Xcc@{}}
\toprule
Regressor & Coefficient & School-clustered SE \\
\midrule
Intercept & 0.09 & 0.38 \\
Deskmate indicator & 0.94 & 0.22 \\
Gender agreement & 3.21 & 0.12 \\
Mixed Roma/non-Roma pair & -0.08 & 0.07 \\
Both students Roma & 0.10 & 0.11 \\
Absolute difference in baseline grades & -0.46 & 0.06 \\
Mean baseline grade & 0.21 & 0.04 \\
Log classroom size & -1.61 & 0.13 \\
Deskmate $\times$ gender agreement & -0.46 & 0.23 \\
Deskmate $\times$ absolute baseline grade difference & 0.03 & 0.11 \\
Pair ethnicity information missing & -0.11 & 0.10 \\
Baseline grade information missing & 0.02 & 0.11 \\
\bottomrule
\end{tabularx}
\end{subtable}\par\vspace{5pt}
{\footnotesize\raggedright\noindent Notes: Logit estimates use 26,840 within-class pairs; standard errors are clustered across 40 schools.\par}
\end{minipage}\end{table}

\paragraph{Score-based policies and random assignment.}
The friendship-count score uses gender agreement, mixed ethnicity, and the negative absolute grade difference. The academic-access score uses the access weight and its interactions with these characteristics. For $d=3$ friendship-count features and $d=4$ academic-access features, the coefficient grids are
\[
\Theta_d=\left\{\frac{z}{\|z\|_2}:z\in\mathbb Z^d,\ \sum_{j=1}^d|z_j|=4\right\}.
\]
The grids contain 66 and 192 coefficient vectors, respectively. Features are standardized using their sample means and standard deviations. Exact maximum-weight matching with fixed tie-breaking implements each rule, producing 1,504 deskmate pairs across classrooms. Under uniform random matching, each pair in classroom $c$ is selected with probability $\lfloor n_c/2\rfloor/\binom{n_c}{2}$. Substituting this probability into the pair-additive welfare expression gives the random-assignment benchmark without simulation.

\paragraph{Welfare across objectives.}
Table~\ref{tab:seating-cross-improvements} evaluates each selected score-based policy on both friendship count and academic access. Holding assignments fixed shows how a policy chosen for one objective affects the other.

\begin{table}[!t]\centering
\caption{Welfare gains by learning objective and evaluation outcome}
\label{tab:seating-cross-improvements}
\begin{minipage}{\textwidth}
\renewcommand{\baselinestretch}{1.0}\small
\setlength{\tabcolsep}{4pt}
\captionsetup[subtable]{font=small,justification=centering,singlelinecheck=false,skip=3pt}
\begin{subtable}[t]{\linewidth}\centering
\begin{tabularx}{\linewidth}{@{}
>{\raggedright\arraybackslash\hsize=1.20\hsize}X
>{\centering\arraybackslash\hsize=1.50\hsize}X
>{\centering\arraybackslash\hsize=0.55\hsize}X
>{\centering\arraybackslash\hsize=0.75\hsize}X@{}}
\toprule
Learning objective & Evaluation outcome & Gain & 90\% interval \\
\midrule
Friendship count & Friendship count & 3.93 & [-0.51, 8.38] \\
Friendship count & Academic access & -2.24 & [-3.94, -0.54] \\
Academic access & Friendship count & 2.61 & [0.10, 5.12] \\
Academic access & Academic access & 5.99 & [2.33, 9.64] \\
\bottomrule
\end{tabularx}
\end{subtable}\par\vspace{5pt}
{\footnotesize\raggedright\noindent Notes: Each score-based rule is evaluated on both outcomes without changing its assignments. Gains relative to random assignment are mutual friends per 100 students or grade-gap units per 100 target students. Brackets are nominal 90\% bootstrap intervals; see Appendix~\ref{app:empirical-inference}.\par}
\end{minipage}\end{table}

\paragraph{Sensitivity to the formation specification.}
The probit specification changes only the link function of the main logit. The quadratic-grade logit adds squared grade difference, squared pair-mean grade, their interactions with seating status, and the interaction of seating status with pair-mean grade. The linear-probability model uses the same regressors as the main logit and clips predictions to $[0,1]$ before calculating the difference between seating together and apart. The gradient-boosting model uses depth-three trees, a learning rate of 0.06, and 300 iterations. All models use the same sample, score classes, welfare objectives, and assignment constraints.

Panel A of Table~\ref{tab:seating-functional-form} reports the area under the receiver-operating-characteristic curve (AUC), the mean squared probability-prediction error (Brier score), and the mean predicted increase in friendship probability from seating a pair together. Fit is evaluated under the recorded assignments. Panel B evaluates the rule selected under the main logit and the rule reselected under each alternative model, using that model for both welfare calculations. Their difference shows the fitted gain from changing the rule, rather than only changing the model used to evaluate it.

\begin{table}[!t]\centering
\caption{Sensitivity to the friendship-formation specification}
\label{tab:seating-functional-form}
\begin{minipage}{\textwidth}
\renewcommand{\baselinestretch}{1.0}\small
\setlength{\tabcolsep}{4pt}
\captionsetup[subtable]{font=small,justification=centering,singlelinecheck=false,skip=3pt}
\begin{subtable}[t]{\linewidth}\centering
\caption*{\textit{Panel A: In-sample fit and seating increments}}
\begin{tabularx}{\linewidth}{@{}>{\raggedright\arraybackslash}Xccc@{}}
\toprule
Model & AUC & Brier $\times100$ & Increment (pp) \\
\midrule
Main logit & 0.82 & 10.71 & 6.62 \\
Probit & 0.82 & 10.72 & 6.77 \\
Quadratic-grade logit & 0.82 & 10.69 & 6.56 \\
Linear probability, clipped & 0.82 & 10.88 & 6.72 \\
Gradient boosting & 0.83 & 10.45 & 6.42 \\
\bottomrule
\end{tabularx}
\end{subtable}\par\vspace{5pt}
\begin{subtable}[t]{\linewidth}\centering
\caption*{\textit{Panel B: Welfare under alternative specifications}}
\begin{tabularx}{\linewidth}{@{}>{\raggedright\arraybackslash}Xccccc@{}}
\toprule
Model & Objective & \shortstack{Main-rule\\gain} & \shortstack{90\%\\interval} & \shortstack{Reselected\\gain} & \shortstack{Pair overlap\\(\%)} \\
\midrule
Main logit & Friendship count & 3.93 & [-0.51, 8.38] & 3.93 & 100.00 \\
Main logit & Academic access & 5.99 & [2.33, 9.64] & 5.99 & 100.00 \\
Probit & Friendship count & 4.25 & [0.17, 8.33] & 4.28 & 81.74 \\
Probit & Academic access & 5.85 & [2.34, 9.36] & 5.85 & 100.00 \\
Quadratic-grade logit & Friendship count & 4.04 & [-1.15, 9.24] & 4.06 & 81.74 \\
Quadratic-grade logit & Academic access & 6.01 & [2.10, 9.92] & 6.10 & 47.20 \\
Linear probability, clipped & Friendship count & 5.05 & -- & 5.12 & 81.74 \\
Linear probability, clipped & Academic access & 6.38 & -- & 6.39 & 72.06 \\
Gradient boosting & Friendship count & 3.54 & -- & 4.15 & 11.42 \\
Gradient boosting & Academic access & 6.21 & -- & 6.36 & 58.08 \\
\bottomrule
\end{tabularx}
\end{subtable}\par\vspace{5pt}
{\footnotesize\raggedright\noindent Notes: Increment is the mean predicted friendship-probability change from seating a pair together, in percentage points. Main-rule and reselected gains use the model in each row and its random benchmark; units are mutual friends per 100 students or grade-gap units per 100 target students. Intervals are nominal 90\% bootstrap intervals for the main rule; dashes mean not reported. Pair overlap is the percentage of selected pairs shared with the main rule.\par}
\end{minipage}\end{table}

\paragraph{Pair characteristics and the nomination limit.}
We compare the baseline characteristics of pairs selected by each policy with those of recorded deskmates, and compare pairs left apart with recorded non-deskmates. Every pair selected by the learned rules has an absolute grade difference within the observed deskmate range. The friendship-count rule selects one pair whose mean grade lies outside the observed range of deskmate pair-mean grades; the academic-access rule selects none. The corresponding expected counts under random assignment are also small.

The survey allows each student to nominate at most five friends. A mutual friendship requires both students to nominate each other, so a student's mutual-friendship count cannot exceed five. Table~\ref{tab:hungary-marginal-coherence} reports small exceedances in fitted expected counts under the learned policies. These fitted marginal probabilities therefore cannot all arise from a joint distribution that respects the nomination limit.

\begin{table}[!t]\centering
\caption{Fitted friendship counts and the nomination limit}
\label{tab:hungary-marginal-coherence}
\begin{minipage}{\textwidth}
\renewcommand{\baselinestretch}{1.0}\small
\setlength{\tabcolsep}{4pt}
\captionsetup[subtable]{font=small,justification=centering,singlelinecheck=false,skip=3pt}
\begin{subtable}[t]{\linewidth}\centering
\begin{tabularx}{\linewidth}{@{}>{\raggedright\arraybackslash}Xcc@{}}
\toprule
Assignment & \shortstack{Students with expected\\count above five} & \shortstack{Maximum\\expected count} \\
\midrule
Random assignment & 0 & 4.99 \\
Score-based policy (friendship count) & 12 & 5.02 \\
Score-based policy (academic access) & 8 & 5.02 \\
Unrestricted policy (friendship count) & 12 & 5.02 \\
Unrestricted policy (academic access) & 8 & 5.02 \\
\bottomrule
\end{tabularx}
\end{subtable}\par\vspace{5pt}
{\footnotesize\raggedright\noindent Notes: Counts are fitted expected mutual friendships per student. The survey allows at most five nominations per student.\par}
\end{minipage}\end{table}

\paragraph{Implementation costs.}
Let $a_c^0$ be the recorded chart, $m_c^0$ its number of deskmate pairs, and $n_c$ the number of students in classroom $c$. To account for the cost of changing partners, we add the following term to each pair's score:
\[
\frac{n_c\lambda}{2m_c^0}\,\mathbf1\{\{i,j\}\in a_c^0\}.
\]
The factor $n_c/(2m_c^0)$ reconciles welfare per student with the fraction of recorded pairs broken. The coefficient $\lambda$ rewards retaining a recorded pair and is chosen jointly with the other score coefficients. By contrast, $\kappa$ is the cost per unit of seating disruption in the welfare objective. The class contains 1,123 candidate rules, including the rule that preserves the recorded chart, with the combined score coefficients normalized to unit Euclidean norm. Tables~\ref{tab:hungary-frontier} and~\ref{tab:hungary-retention-coefficients} report gains, disruption, and coefficients as $\kappa$ varies; every assignment preserves the recorded pair count in each classroom.

\begin{table}[!t]\centering
\caption{Welfare gains and seating disruption}
\label{tab:hungary-frontier}
\begin{minipage}{\textwidth}
\renewcommand{\baselinestretch}{1.0}\small
\setlength{\tabcolsep}{4pt}
\captionsetup[subtable]{font=small,justification=centering,singlelinecheck=false,skip=3pt}
\begin{subtable}[t]{\linewidth}\centering
\begin{tabularx}{\linewidth}{@{}>{\raggedright\arraybackslash}Xcccc@{}}
\toprule
$\kappa$ & Policy & Gross gain & \shortstack{Pairs broken\\(\%)} & Net gain \\
\midrule
0.00 & Unrestricted policy & 3.65 & 93.38 & 3.65 \\
0.00 & Score-based policy & 3.40 & 85.72 & 3.40 \\
0.005 & Unrestricted policy & 3.60 & 64.93 & 3.28 \\
0.005 & Score-based policy & 3.35 & 54.86 & 3.08 \\
0.01 & Unrestricted policy & 3.54 & 55.88 & 2.98 \\
0.01 & Score-based policy & 3.35 & 54.86 & 2.81 \\
0.02 & Unrestricted policy & 3.46 & 50.57 & 2.45 \\
0.02 & Score-based policy & 3.35 & 54.86 & 2.26 \\
0.05 & Unrestricted policy & 3.24 & 44.49 & 1.01 \\
0.05 & Score-based policy & 3.17 & 46.12 & 0.87 \\
0.10 & Unrestricted policy & 0.00 & 0.00 & 0.00 \\
0.10 & Score-based policy & 0.00 & 0.00 & 0.00 \\
0.20 & Unrestricted policy & 0.00 & 0.00 & 0.00 \\
0.20 & Score-based policy & 0.00 & 0.00 & 0.00 \\
1.00 & Unrestricted policy & 0.00 & 0.00 & 0.00 \\
1.00 & Score-based policy & 0.00 & 0.00 & 0.00 \\
\bottomrule
\end{tabularx}
\end{subtable}\par\vspace{5pt}
{\footnotesize\raggedright\noindent Notes: Gains are mutual friends per 100 students relative to the recorded chart, with equal classroom weights. Net gain equals gross gain minus $\kappa$ times the reported percentage of pairs broken. Each classroom retains its recorded pair count.\par}
\end{minipage}\end{table}

\begin{table}[!t]\centering
\caption{Selected score coefficients by implementation cost}
\label{tab:hungary-retention-coefficients}
\begin{minipage}{\textwidth}
\renewcommand{\baselinestretch}{1.0}\small
\setlength{\tabcolsep}{4pt}
\captionsetup[subtable]{font=small,justification=centering,singlelinecheck=false,skip=3pt}
\begin{subtable}[t]{\linewidth}\centering
\begin{tabularx}{\linewidth}{@{}>{\raggedright\arraybackslash}Xcccc@{}}
\toprule
$\kappa$ & $\lambda$ & Gender agreement & Mixed ethnicity & Grade proximity \\
\midrule
0.00 & 0.16 & 0.81 & -0.40 & 0.40 \\
0.005 & 0.53 & 0.80 & 0.00 & 0.27 \\
0.01 & 0.53 & 0.80 & 0.00 & 0.27 \\
0.02 & 0.53 & 0.80 & 0.00 & 0.27 \\
0.05 & 0.78 & 0.59 & 0.00 & 0.20 \\
0.10 & -- & -- & -- & -- \\
0.20 & -- & -- & -- & -- \\
1.00 & -- & -- & -- & -- \\
\bottomrule
\end{tabularx}
\end{subtable}\par\vspace{5pt}
{\footnotesize\raggedright\noindent Notes: Score coefficients have unit Euclidean norm. The cost $\kappa$ enters welfare; $\lambda$ rewards retaining recorded pairs. Dashes indicate that the recorded chart is preserved. Grade proximity is the negative absolute grade difference.\par}
\end{minipage}\end{table}
\FloatBarrier

\subsection{Business Groups}
\label{app:cs-application-details}

\paragraph{Sample and measurement.}
The assignment sample contains 1,409 firms in 141 groups across 21 counties. The response model uses the 1,365 firms with observed follow-up outcomes. The other 44 firms remain in policy evaluation under the same maintained response model. Assignments preserve the observed group sizes within each county. Small and large firms are defined using fixed baseline county employment medians; these definitions differ from the size categories used in the original randomization.

Because original administrative capacities are unavailable, the feasible assignments preserve group sizes in the available sample. The released attrition and shutdown records cannot be linked to firm identifiers, and five groups have inconsistent experimental-design labels.

\paragraph{Response model and welfare.}
For each firm, peer characteristics are calculated using the other members of its group. They include the shares of large firms, firms in the same sector, and owners with government work experience, together with the log of mean peer employment. Employment dispersion is the standard deviation of log employment across all group members. Group size is the number of members in the available sample.

Table~\ref{tab:cs-response-slopes} reports the linear response coefficients. The model includes firm-level controls and fixed effects for the original county-by-size-category-by-sector randomization strata. These terms affect welfare levels but cancel from comparisons of assignments because firm characteristics and stratum membership remain fixed. Let $x_i(a)$ be firm $i$'s exposure vector, $w_{i,o}$ its welfare weight for evaluation group $o$, and $E_R$ expectation under random assignment $A_c^R$. The within-county change in average exposure relative to random assignment and the resulting fitted log-revenue gain are
\[
d_{c,o}(\pi)=\frac{\sum_iw_{i,o}[x_i\{\pi(H_c)\}-E_Rx_i(A_c^R)]}{\sum_iw_{i,o}},\qquad
\widehat\Delta_o(\pi)=\frac1{21}\sum_c d_{c,o}(\pi)'\widehat\beta.
\]
Welfare is averaged first within each county and then equally across counties. Within a county, all-firm welfare is the average of small- and large-firm welfare weighted by their numbers of firms. The exposure differences in $d_{c,o}(\pi)$, multiplied by the estimated response coefficients $\widehat\beta$, determine the fitted gain.

\begin{table}[!t]\centering
\caption{Business outcome-response coefficients}
\label{tab:cs-response-slopes}
\begin{minipage}{\textwidth}
\renewcommand{\baselinestretch}{1.0}\small
\setlength{\tabcolsep}{4pt}
\captionsetup[subtable]{font=small,justification=centering,singlelinecheck=false,skip=3pt}
\begin{subtable}[t]{\linewidth}\centering
\begin{tabularx}{\linewidth}{@{}>{\raggedright\arraybackslash}Xc@{}}
\toprule
Regressor & Coefficient \\
\midrule
Log baseline employment & 0.26 \\
Male firm owner & 0.19 \\
Firm-owner age (centered) & 0.00 \\
Firm owner has a college degree or higher & 0.15 \\
Firm owner has government work experience & -0.12 \\
Log baseline revenue & 0.66 \\
Large-peer share & -0.51 \\
Same-sector peer share & -0.17 \\
Standard deviation of log employment & -0.39 \\
Log mean peer employment & 0.36 \\
Government-experience peer share & -0.51 \\
Group size & 0.08 \\
\bottomrule
\end{tabularx}
\end{subtable}\par\vspace{5pt}
{\footnotesize\raggedright\noindent Notes: The outcome is follow-up log revenue for 1,365 firms. The model includes the listed controls and fixed effects for the original county-by-size-category-by-sector strata.\par}
\end{minipage}\end{table}

\paragraph{Score-based policies and optimization.}
For all-firm welfare, the grid contains the 26 normalized nonzero vectors in $\{-1,0,1\}^3$, applied to the three exposure coordinates whose county averages can change with assignment. For small-firm welfare, it contains the 12 signed coordinate axes in six dimensions and 64 fixed unit vectors generated from Gaussian draws. Features are divided by their baseline exposure standard deviations. Writing $s$ for these standard deviations and $\odot$ for elementwise multiplication, coefficients proportional to $s\odot\widehat\beta$ rank assignments by fitted linear-model welfare after assignment-invariant terms are removed. If the grid contains this vector, exact score maximization attains the same fitted welfare as unrestricted maximization.

For each county and coefficient vector, optimization considers partitions with the observed group sizes. Small two-group problems are solved by enumeration; the others use mixed-integer nonlinear optimization with a fixed time limit. We report the best feasible assignment found, with global optimality certified for some problems. The assignments found by score maximization are also supplied as feasible candidates when maximizing fitted welfare directly. Differences between the reported unrestricted and score-based policies can therefore reflect both the score-class restriction and the numerical solutions found.

\paragraph{Random-assignment welfare.}
The random benchmark averages the fitted response over uniform within-county assignments with the observed group sizes. Expected exposures are calculated analytically where possible and by Monte Carlo integration otherwise. For nonlinear responses, the response is evaluated under each simulated assignment before averaging. Across the parametric specifications, Monte Carlo standard errors are small relative to the bootstrap standard errors of policy gains.

\paragraph{Welfare across firm groups.}
Table~\ref{tab:business-cross-improvements} evaluates each selected score-based policy for all firms, small firms, and large firms, holding its assignments fixed. This shows the distribution of fitted gains across firm sizes.

\begin{table}[!t]\centering
\caption{Welfare gains by learning objective and firm size}
\label{tab:business-cross-improvements}
\begin{minipage}{\textwidth}
\renewcommand{\baselinestretch}{1.0}\small
\setlength{\tabcolsep}{4pt}
\captionsetup[subtable]{font=small,justification=centering,singlelinecheck=false,skip=3pt}
\begin{subtable}[t]{\linewidth}\centering
\begin{tabularx}{\linewidth}{@{}>{\raggedright\arraybackslash}Xccc@{}}
\toprule
Learning objective & Evaluation group & Gain & 90\% interval \\
\midrule
All firms & All firms & 10.65 & [-15.53, 44.93] \\
All firms & Small firms & -1.87 & [-38.16, 55.70] \\
All firms & Large firms & 26.79 & [-15.54, 90.35] \\
Small firms & All firms & 4.89 & [-1.07, 11.21] \\
Small firms & Small firms & 21.80 & [-7.24, 59.95] \\
Small firms & Large firms & -11.44 & [-31.94, 15.24] \\
\bottomrule
\end{tabularx}
\end{subtable}\par\vspace{5pt}
{\footnotesize\raggedright\noindent Notes: Each score-based rule is evaluated on all three firm groups without changing its assignments. Gains are percentage changes in geometric-mean revenue over random assignment, with equal county weights. Brackets are nominal 90\% bootstrap intervals; see Appendix~\ref{app:empirical-inference}.\par}
\end{minipage}\end{table}

\paragraph{Exposure contributions to welfare gains.}
Under the linear response, each exposure's contribution is its change relative to random assignment multiplied by its estimated coefficient. Table~\ref{tab:cs-score-decomposition} reports these contributions for each rule's learning objective. They sum to the gain in mean log revenue; applying $100(\exp(x)-1)$ gives the reported percentage revenue gain.

\begin{table}[!t]\centering
\caption{Exposure changes and fitted welfare gains}
\label{tab:cs-score-decomposition}
\begin{minipage}{\textwidth}
\renewcommand{\baselinestretch}{1.0}\small
\setlength{\tabcolsep}{4pt}
\captionsetup[subtable]{font=small,justification=centering,singlelinecheck=false,skip=3pt}
\begin{subtable}[t]{\linewidth}\centering
\begin{tabularx}{\linewidth}{@{}>{\raggedright\arraybackslash}Xccc@{}}
\toprule
Objective & Exposure & Mean change & \shortstack{Contribution to\\log-revenue gain} \\
\midrule
All firms & Large-peer share & 0.00 & 0.00 \\
All firms & Same-sector peer share & 0.01 & 0.00 \\
All firms & Standard deviation of log employment & -0.72 & 0.28 \\
All firms & Log mean peer employment & -0.49 & -0.18 \\
All firms & Government-experience peer share & 0.00 & 0.00 \\
All firms & Group size & 0.00 & 0.00 \\
All firms & Total & -- & 0.10 \\
Small firms & Large-peer share & -0.15 & 0.07 \\
Small firms & Same-sector peer share & -0.03 & 0.01 \\
Small firms & Standard deviation of log employment & 0.02 & -0.01 \\
Small firms & Log mean peer employment & 0.10 & 0.04 \\
Small firms & Government-experience peer share & -0.13 & 0.06 \\
Small firms & Group size & 0.35 & 0.03 \\
Small firms & Total & -- & 0.20 \\
\bottomrule
\end{tabularx}
\end{subtable}\par\vspace{5pt}
{\footnotesize\raggedright\noindent Notes: Each score-based rule is evaluated on its learning objective relative to random assignment, with equal county weights. Contributions equal exposure changes times estimated response coefficients and sum to the log-revenue gain. \par}
\end{minipage}\end{table}

\paragraph{Sensitivity to the response specification.}
The additive quadratic model adds the square of each exposure to the linear response. The exposure-interaction model adds interactions of employment dispersion with log mean peer employment and group size. The boosted model replaces the linear exposure response with a fitted tree-based function while retaining additive firm-level controls and original-stratum fixed effects without penalizing their coefficients. All specifications use the same estimation sample, score classes, welfare objectives, and assignment constraints.

Panel A of Table~\ref{tab:business-functional-form} reports in-sample root mean squared error (RMSE) and within-stratum $R^2$. The latter compares residual variation with outcome variation around the original-stratum means. Panel B evaluates the rule selected under the main linear model and the rule reselected under each alternative model, using the corresponding model's random-assignment benchmark. Pair overlap is measured by the Jaccard index: the number of firm pairs assigned together under both rules divided by the number assigned together under either rule.

\begin{table}[!t]\centering
\caption{Sensitivity to the business outcome specification}
\label{tab:business-functional-form}
\begin{minipage}{\textwidth}
\renewcommand{\baselinestretch}{1.0}\small
\setlength{\tabcolsep}{4pt}
\captionsetup[subtable]{font=small,justification=centering,singlelinecheck=false,skip=3pt}
\begin{subtable}[t]{\linewidth}\centering
\caption*{\textit{Panel A: In-sample model fit}}
\begin{tabularx}{\linewidth}{@{}>{\raggedright\arraybackslash}Xcc@{}}
\toprule
Model & RMSE & Within-stratum $R^2$ \\
\midrule
Main linear & 1.09 & 0.63 \\
Additive quadratic & 1.08 & 0.63 \\
Exposure interactions & 1.09 & 0.63 \\
Boosted exposure response & 1.01 & 0.68 \\
\bottomrule
\end{tabularx}
\end{subtable}\par\vspace{5pt}
\begin{subtable}[t]{\linewidth}\centering
\caption*{\textit{Panel B: Welfare under alternative specifications}}
\begin{tabularx}{\linewidth}{@{}>{\raggedright\arraybackslash}Xccccc@{}}
\toprule
Model & Objective & \shortstack{Main-rule\\gain} & \shortstack{90\%\\interval} & \shortstack{Reselected\\gain} & \shortstack{Pair overlap\\(\%)} \\
\midrule
Main linear & All firms & 10.65 & [-15.53, 44.93] & 10.65 & 100.00 \\
Main linear & Small firms & 21.80 & [-7.24, 59.95] & 21.80 & 100.00 \\
Additive quadratic & All firms & 43.53 & [-3.02, 112.41] & 43.53 & 100.00 \\
Additive quadratic & Small firms & 25.34 & [-3.57, 62.92] & 36.13 & 17.32 \\
Exposure interactions & All firms & 16.63 & [-12.49, 55.45] & 16.63 & 100.00 \\
Exposure interactions & Small firms & 20.32 & [-11.95, 64.41] & 20.32 & 100.00 \\
Boosted exposure response & All firms & 14.80 & -- & 14.86 & 42.71 \\
Boosted exposure response & Small firms & 14.91 & -- & 19.25 & 27.15 \\
\bottomrule
\end{tabularx}
\end{subtable}\par\vspace{5pt}
{\footnotesize\raggedright\noindent Notes: Main-rule and reselected gains use the model in each row and its random benchmark. Gains are percentage changes in geometric-mean revenue. Intervals are nominal 90\% bootstrap intervals for the main rule; dashes mean not reported. Pair overlap is the percentage of firm pairs assigned together under both rules relative to those assigned together under either rule.\par}
\end{minipage}\end{table}

\paragraph{Policy-induced exposures.}
Table~\ref{tab:cs-full-support} compares policy-induced exposure vectors with exposures in the estimation sample. The range check identifies firms with at least one exposure outside its observed range. The nearest-neighbor check measures distance to the closest observed exposure vector, while Mahalanobis distance accounts for exposure scales and correlations when measuring joint differences. Nearest-neighbor calibration excludes all firms in the same original group. Relative to random assignment, the score-based all-firm policy has fewer firms outside individual exposure ranges but more firms flagged by the two joint-distance checks. Thus its exposure combinations can differ more from those observed even when their individual coordinates remain within range.

\begin{table}[!t]\centering
\caption{Business exposure diagnostics}
\label{tab:cs-full-support}
\begin{minipage}{\textwidth}
\renewcommand{\baselinestretch}{1.0}\small
\setlength{\tabcolsep}{4pt}
\captionsetup[subtable]{font=small,justification=centering,singlelinecheck=false,skip=3pt}
\begin{subtable}[t]{\linewidth}\centering
\begin{tabularx}{\linewidth}{@{}>{\raggedright\arraybackslash}Xccccc@{}}
\toprule
Assignment & Evaluation group & \shortstack{Outside\\range} & \shortstack{Nearest-\\neighbor} & Mahalanobis & \shortstack{At least\\one} \\
\midrule
Random assignment & All firms & 2.94 & 5.68 & 9.27 & 11.37 \\
Random assignment & Small firms & 2.88 & 5.79 & 9.69 & 11.73 \\
Random assignment & Large firms & 3.02 & 5.70 & 8.93 & 11.11 \\
Score-based policy (all firms) & All firms & 1.41 & 21.55 & 35.98 & 40.92 \\
Score-based policy (all firms) & Small firms & 0.17 & 20.22 & 33.13 & 37.06 \\
Score-based policy (all firms) & Large firms & 2.65 & 22.76 & 38.70 & 44.52 \\
Score-based policy (small firms) & All firms & 6.53 & 14.18 & 20.61 & 23.05 \\
Score-based policy (small firms) & Small firms & 1.51 & 10.41 & 16.44 & 18.48 \\
Score-based policy (small firms) & Large firms & 12.22 & 18.86 & 26.05 & 29.01 \\
Unrestricted policy (all firms) & All firms & 2.00 & 27.17 & 40.17 & 43.72 \\
Unrestricted policy (all firms) & Small firms & 3.62 & 22.74 & 32.27 & 35.99 \\
Unrestricted policy (all firms) & Large firms & 0.37 & 33.01 & 49.91 & 53.30 \\
Unrestricted policy (small firms) & All firms & 6.47 & 16.57 & 26.73 & 29.40 \\
Unrestricted policy (small firms) & Small firms & 1.17 & 15.19 & 24.50 & 25.81 \\
Unrestricted policy (small firms) & Large firms & 12.40 & 18.13 & 29.73 & 33.77 \\
\bottomrule
\end{tabularx}
\end{subtable}\par\vspace{5pt}
{\footnotesize\raggedright\noindent Notes: Entries are percentages of firms flagged, with equal county weights within each evaluation group. Outside range means at least one exposure lies outside its observed range; the two distance checks assess joint exposures. At least one is the union of the three flags. Parentheses give the learning objective.\par}
\end{minipage}\end{table}
\FloatBarrier

\clearpage
\end{appendix}

\begingroup
\renewcommand{\refname}{Appendix References}
\putbib
\endgroup

\end{bibunit}


\begin{thebibliography}{44}
\newcommand{\enquote}[1]{``#1''}
\expandafter\ifx\csname natexlab\endcsname\relax\def\natexlab#1{#1}\fi

\bibitem[\protect\citeauthoryear{Aronow and Samii}{Aronow and Samii}{2017}]{AronowSamii2017}
\textsc{Aronow, P.~M. and C.~Samii} (2017): \enquote{Estimating Average Causal Effects under General Interference, with Application to a Social Network Experiment,} \emph{The Annals of Applied Statistics}, 11, 1912--1947.

\bibitem[\protect\citeauthoryear{Athey and Wager}{Athey and Wager}{2021}]{AtheyWager2021}
\textsc{Athey, S. and S.~Wager} (2021): \enquote{Policy Learning with Observational Data,} \emph{Econometrica}, 89, 133--161.

\bibitem[\protect\citeauthoryear{Auerbach, Guo, and Tabord-Meehan}{Auerbach et~al.}{2026}]{AuerbachGuoTabordMeehan2026}
\textsc{Auerbach, E., H.~Guo, and M.~Tabord-Meehan} (2026): \enquote{The Local Approach to Causal Inference under Network Interference,} \emph{Quantitative Economics}, 17, 173--199.

\bibitem[\protect\citeauthoryear{Bala and Goyal}{Bala and Goyal}{2000}]{BalaGoyal2000}
\textsc{Bala, V. and S.~Goyal} (2000): \enquote{A Noncooperative Model of Network Formation,} \emph{Econometrica}, 68, 1181--1229.

\bibitem[\protect\citeauthoryear{Balas and Padberg}{Balas and Padberg}{1976}]{BalasPadberg1976}
\textsc{Balas, E. and M.~W. Padberg} (1976): \enquote{Set Partitioning: A Survey,} \emph{SIAM Review}, 18, 710--760.

\bibitem[\protect\citeauthoryear{Ballester, Calv{\'o}-Armengol, and Zenou}{Ballester et~al.}{2006}]{BallesterCalvoArmengolZenou2006}
\textsc{Ballester, C., A.~Calv{\'o}-Armengol, and Y.~Zenou} (2006): \enquote{Who's Who in Networks. Wanted: The Key Player,} \emph{Econometrica}, 74, 1403--1417.

\bibitem[\protect\citeauthoryear{Barnhart, Johnson, Nemhauser, Savelsbergh, and Vance}{Barnhart et~al.}{1998}]{BarnhartJohnsonNemhauserSavelsberghVance1998}
\textsc{Barnhart, C., E.~L. Johnson, G.~L. Nemhauser, M.~W.~P. Savelsbergh, and P.~H. Vance} (1998): \enquote{Branch-and-Price: Column Generation for Solving Huge Integer Programs,} \emph{Operations Research}, 46, 316--329.

\bibitem[\protect\citeauthoryear{Basse and Feller}{Basse and Feller}{2018}]{BasseFeller2018}
\textsc{Basse, G.~W. and A.~Feller} (2018): \enquote{Analyzing Two-Stage Experiments in the Presence of Interference,} \emph{Journal of the American Statistical Association}, 113, 41--55.

\bibitem[\protect\citeauthoryear{Bloch and Jackson}{Bloch and Jackson}{2006}]{BlochJackson2006}
\textsc{Bloch, F. and M.~O. Jackson} (2006): \enquote{Definitions of Equilibrium in Network Formation Games,} \emph{International Journal of Game Theory}, 34, 305--318.

\bibitem[\protect\citeauthoryear{Bramoull{\'e}, Djebbari, and Fortin}{Bramoull{\'e} et~al.}{2009}]{BramoulleDjebbariFortin2009}
\textsc{Bramoull{\'e}, Y., H.~Djebbari, and B.~Fortin} (2009): \enquote{Identification of Peer Effects through Social Networks,} \emph{Journal of Econometrics}, 150, 41--55.

\bibitem[\protect\citeauthoryear{Bramoull{\'e}, Djebbari, and Fortin}{Bramoull{\'e} et~al.}{2020}]{BramoulleDjebbariFortin2020}
---\hspace{-.1pt}---\hspace{-.1pt}--- (2020): \enquote{Peer Effects in Networks: A Survey,} \emph{Annual Review of Economics}, 12, 603--629.

\bibitem[\protect\citeauthoryear{Cai and Szeidl}{Cai and Szeidl}{2018}]{CaiSzeidl2018}
\textsc{Cai, J. and A.~Szeidl} (2018): \enquote{Interfirm Relationships and Business Performance,} \emph{The Quarterly Journal of Economics}, 133, 1229--1282.

\bibitem[\protect\citeauthoryear{Carrell, Sacerdote, and West}{Carrell et~al.}{2013}]{CarrellSacerdoteWest2013}
\textsc{Carrell, S.~E., B.~I. Sacerdote, and J.~E. West} (2013): \enquote{From Natural Variation to Optimal Policy? The Importance of Endogenous Peer Group Formation,} \emph{Econometrica}, 81, 855--882.

\bibitem[\protect\citeauthoryear{Christensen, Moon, and Schorfheide}{Christensen et~al.}{2026}]{ChristensenMoonSchorfheide2026}
\textsc{Christensen, T., H.~R. Moon, and F.~Schorfheide} (2026): \enquote{Optimal decision rules when payoffs are partially identified,} \emph{Review of Economic Studies}, rdag017.

\bibitem[\protect\citeauthoryear{de~Paula}{de~Paula}{2020}]{dePaula2020}
\textsc{de~Paula, {\'A}.} (2020): \enquote{Econometric Models of Network Formation,} \emph{Annual Review of Economics}, 12, 775--799.

\bibitem[\protect\citeauthoryear{de~Paula, Richards-Shubik, and Tamer}{de~Paula et~al.}{2018}]{dePaulaRichardsShubikTamer2018}
\textsc{de~Paula, {\'A}., S.~Richards-Shubik, and E.~Tamer} (2018): \enquote{Identifying Preferences in Networks With Bounded Degree,} \emph{Econometrica}, 86, 263--288.

\bibitem[\protect\citeauthoryear{Edmonds}{Edmonds}{1965}]{Edmonds1965}
\textsc{Edmonds, J.} (1965): \enquote{Paths, Trees, and Flowers,} \emph{Canadian Journal of Mathematics}, 17, 449--467.

\bibitem[\protect\citeauthoryear{Galil}{Galil}{1986}]{Galil1986}
\textsc{Galil, Z.} (1986): \enquote{Efficient Algorithms for Finding Maximum Matching in Graphs,} \emph{ACM Computing Surveys}, 18, 23--38.

\bibitem[\protect\citeauthoryear{Garlick}{Garlick}{2018}]{Garlick2018}
\textsc{Garlick, R.} (2018): \enquote{Academic Peer Effects with Different Group Assignment Policies: Residential Tracking versus Random Assignment,} \emph{American Economic Journal: Applied Economics}, 10, 345--369.

\bibitem[\protect\citeauthoryear{Graham}{Graham}{2017}]{Graham2017}
\textsc{Graham, B.~S.} (2017): \enquote{An Econometric Model of Network Formation with Degree Heterogeneity,} \emph{Econometrica}, 85, 1033--1063.

\bibitem[\protect\citeauthoryear{Han, Owusu, and Shin}{Han et~al.}{2022}]{HanOwusuShin2022}
\textsc{Han, S., J.~Owusu, and Y.~Shin} (2022): \enquote{Statistical Treatment Rules under Social Interaction,} \emph{ArXiv Working Paper}, arXiv:2209.09077, revised November 2022.

\bibitem[\protect\citeauthoryear{Hazard and Kitagawa}{Hazard and Kitagawa}{2025}]{HazardKitagawa2025}
\textsc{Hazard, Y. and T.~Kitagawa} (2025): \enquote{Who With Whom? Learning Optimal Matching Policies,} \emph{ArXiv Working Paper}, arXiv:2507.13567.

\bibitem[\protect\citeauthoryear{Hirano and Porter}{Hirano and Porter}{2009}]{HiranoPorter2009}
\textsc{Hirano, K. and J.~R. Porter} (2009): \enquote{Asymptotics for Statistical Treatment Rules,} \emph{Econometrica}, 77, 1683--1701.

\bibitem[\protect\citeauthoryear{Hoff, Raftery, and Handcock}{Hoff et~al.}{2002}]{HoffRafteryHandcock2002}
\textsc{Hoff, P.~D., A.~E. Raftery, and M.~S. Handcock} (2002): \enquote{Latent Space Approaches to Social Network Analysis,} \emph{Journal of the American Statistical Association}, 97, 1090--1098.

\bibitem[\protect\citeauthoryear{Hudgens and Halloran}{Hudgens and Halloran}{2008}]{HudgensHalloran2008}
\textsc{Hudgens, M.~G. and M.~E. Halloran} (2008): \enquote{Toward Causal Inference with Interference,} \emph{Journal of the American Statistical Association}, 103, 832--842.

\bibitem[\protect\citeauthoryear{Imai, Keele, and Yamamoto}{Imai et~al.}{2010}]{ImaiKeeleYamamoto2010}
\textsc{Imai, K., L.~Keele, and T.~Yamamoto} (2010): \enquote{Identification, Inference and Sensitivity Analysis for Causal Mediation Effects,} \emph{Statistical Science}, 25, 51--71.

\bibitem[\protect\citeauthoryear{Jackson and Wolinsky}{Jackson and Wolinsky}{1996}]{JacksonWolinsky1996}
\textsc{Jackson, M.~O. and A.~Wolinsky} (1996): \enquote{A Strategic Model of Social and Economic Networks,} \emph{Journal of Economic Theory}, 71, 44--74.

\bibitem[\protect\citeauthoryear{Keller and Elwert}{Keller and Elwert}{2023}]{KellerElwert2023}
\textsc{Keller, T. and F.~Elwert} (2023): \enquote{Feasible Peer Effects: Experimental Evidence for Deskmate Effects on Educational Achievement and Inequality,} \emph{Sociological Science}, 10, 806--829.

\bibitem[\protect\citeauthoryear{Kitagawa and Tetenov}{Kitagawa and Tetenov}{2018}]{KitagawaTetenov2018}
\textsc{Kitagawa, T. and A.~Tetenov} (2018): \enquote{Who Should Be Treated? Empirical Welfare Maximization Methods for Treatment Choice,} \emph{Econometrica}, 86, 591--616.

\bibitem[\protect\citeauthoryear{Kitagawa and Wang}{Kitagawa and Wang}{2023}]{KitagawaWang2023}
\textsc{Kitagawa, T. and G.~Wang} (2023): \enquote{Who Should Get Vaccinated? Individualized Allocation of Vaccines Over {SIR} Network,} \emph{Journal of Econometrics}, 232, 109--131.

\bibitem[\protect\citeauthoryear{Kitagawa and Wang}{Kitagawa and Wang}{2026}]{KitagawaWang2026}
---\hspace{-.1pt}---\hspace{-.1pt}--- (2026): \enquote{Individualized Treatment Allocation in Sequential Network Games,} \emph{ArXiv Working Paper}, arXiv:2302.05747, revised May 2026.

\bibitem[\protect\citeauthoryear{Leung}{Leung}{2022}]{Leung2022}
\textsc{Leung, M.~P.} (2022): \enquote{Causal Inference under Approximate Neighborhood Interference,} \emph{Econometrica}, 90, 267--293.

\bibitem[\protect\citeauthoryear{Manski}{Manski}{1993}]{Manski1993}
\textsc{Manski, C.~F.} (1993): \enquote{Identification of Endogenous Social Effects: The Reflection Problem,} \emph{The Review of Economic Studies}, 60, 531--542.

\bibitem[\protect\citeauthoryear{Manski}{Manski}{2004}]{Manski2004}
---\hspace{-.1pt}---\hspace{-.1pt}--- (2004): \enquote{Statistical Treatment Rules for Heterogeneous Populations,} \emph{Econometrica}, 72, 1221--1246.

\bibitem[\protect\citeauthoryear{Mele}{Mele}{2017}]{Mele2017}
\textsc{Mele, A.} (2017): \enquote{A Structural Model of Dense Network Formation,} \emph{Econometrica}, 85, 825--850.

\bibitem[\protect\citeauthoryear{Rohrer, Keller, and Elwert}{Rohrer et~al.}{2021}]{RohrerKellerElwert2021}
\textsc{Rohrer, J.~M., T.~Keller, and F.~Elwert} (2021): \enquote{Proximity Can Induce Diverse Friendships: A Large Randomized Classroom Experiment,} \emph{PLOS ONE}, 16, e0255097.

\bibitem[\protect\citeauthoryear{Sacerdote}{Sacerdote}{2011}]{Sacerdote2011}
\textsc{Sacerdote, B.} (2011): \enquote{Peer Effects in Education: How Might They Work, How Big Are They and How Much Do We Know Thus Far?} in \emph{Handbook of the Economics of Education}, ed. by E.~A. Hanushek, S.~Machin, and L.~Woessmann, Elsevier, vol.~3, 249--277.

\bibitem[\protect\citeauthoryear{Sheng}{Sheng}{2020}]{Sheng2020}
\textsc{Sheng, S.} (2020): \enquote{A Structural Econometric Analysis of Network Formation Games Through Subnetworks,} \emph{Econometrica}, 88, 1829--1858.

\bibitem[\protect\citeauthoryear{Snijders}{Snijders}{2001}]{Snijders2001}
\textsc{Snijders, T. A.~B.} (2001): \enquote{The Statistical Evaluation of Social Network Dynamics,} \emph{Sociological Methodology}, 31, 361--395.

\bibitem[\protect\citeauthoryear{Viviano}{Viviano}{2025}]{Viviano2025}
\textsc{Viviano, D.} (2025): \enquote{Policy Targeting under Network Interference,} \emph{The Review of Economic Studies}, 92, 1257--1292.

\bibitem[\protect\citeauthoryear{Wu, Zhang, and Wang}{Wu et~al.}{2023}]{WuZhangWang2023}
\textsc{Wu, J., J.~Zhang, and C.~Wang} (2023): \enquote{Student Performance, Peer Effects, and Friend Networks: Evidence from a Randomized Peer Intervention,} \emph{American Economic Journal: Economic Policy}, 15, 510--542.

\bibitem[\protect\citeauthoryear{Xu and Zhou}{Xu and Zhou}{2025}]{XuZhou2025}
\textsc{Xu, Y. and B.~Zhou} (2025): \enquote{Batched Adaptive Network Formation,} \emph{ArXiv Working Paper}, arXiv:2507.18961.

\bibitem[\protect\citeauthoryear{You}{You}{2026}]{You2026}
\textsc{You, Z.} (2026): \enquote{Using Prior Studies to Design Experiments: An Empirical Bayes Approach,} \emph{ArXiv Working Paper}, arXiv:2602.20581.

\bibitem[\protect\citeauthoryear{Zhang and Imai}{Zhang and Imai}{2023}]{ZhangImai2023}
\textsc{Zhang, Y. and K.~Imai} (2023): \enquote{Individualized Policy Evaluation and Learning under Clustered Network Interference,} \emph{ArXiv Working Paper}, arXiv:2311.02467, revised April 2025.

\end{thebibliography}


\begin{thebibliography}{8}
\newcommand{\enquote}[1]{``#1''}
\expandafter\ifx\csname natexlab\endcsname\relax\def\natexlab#1{#1}\fi

\bibitem[\protect\citeauthoryear{Ballester, Calv{\'o}-Armengol, and Zenou}{Ballester et~al.}{2006}]{BallesterCalvoArmengolZenou2006}
\textsc{Ballester, C., A.~Calv{\'o}-Armengol, and Y.~Zenou} (2006): \enquote{Who's Who in Networks. Wanted: The Key Player,} \emph{Econometrica}, 74, 1403--1417.

\bibitem[\protect\citeauthoryear{Cai and Szeidl}{Cai and Szeidl}{2018}]{CaiSzeidl2018}
\textsc{Cai, J. and A.~Szeidl} (2018): \enquote{Interfirm Relationships and Business Performance,} \emph{The Quarterly Journal of Economics}, 133, 1229--1282.

\bibitem[\protect\citeauthoryear{Fafchamps and Quinn}{Fafchamps and Quinn}{2018}]{FafchampsQuinn2018}
\textsc{Fafchamps, M. and S.~Quinn} (2018): \enquote{Networks and Manufacturing Firms in Africa: Results from a Randomized Field Experiment,} \emph{The World Bank Economic Review}, 32, 656--675.

\bibitem[\protect\citeauthoryear{Feigenberg, Field, and Pande}{Feigenberg et~al.}{2013}]{FeigenbergFieldPande2013}
\textsc{Feigenberg, B., E.~Field, and R.~Pande} (2013): \enquote{The Economic Returns to Social Interaction: Experimental Evidence from Microfinance,} \emph{The Review of Economic Studies}, 80, 1459--1483.

\bibitem[\protect\citeauthoryear{Jackson and Wolinsky}{Jackson and Wolinsky}{1996}]{JacksonWolinsky1996}
\textsc{Jackson, M.~O. and A.~Wolinsky} (1996): \enquote{A Strategic Model of Social and Economic Networks,} \emph{Journal of Economic Theory}, 71, 44--74.

\bibitem[\protect\citeauthoryear{Kitagawa and Tetenov}{Kitagawa and Tetenov}{2018}]{KitagawaTetenov2018}
\textsc{Kitagawa, T. and A.~Tetenov} (2018): \enquote{Who Should Be Treated? Empirical Welfare Maximization Methods for Treatment Choice,} \emph{Econometrica}, 86, 591--616.

\bibitem[\protect\citeauthoryear{van~der Vaart and Wellner}{van~der Vaart and Wellner}{1996}]{vanDerVaartWellner1996}
\textsc{van~der Vaart, A.~W. and J.~A. Wellner} (1996): \emph{Weak Convergence and Empirical Processes}, New York: Springer.

\bibitem[\protect\citeauthoryear{Vasilaky and Leonard}{Vasilaky and Leonard}{2018}]{VasilakyLeonard2018}
\textsc{Vasilaky, K.~N. and K.~L. Leonard} (2018): \enquote{As Good as the Networks They Keep? Improving Outcomes through Weak Ties in Rural Uganda,} \emph{Economic Development and Cultural Change}, 66, 755--792.

\end{thebibliography}
\end{document}